\documentclass[preprints,article,accept,moreauthors,pdftex,10pt,a4paper]{mdpi}

\firstpage{1}
\pubvolume{xx}
\issuenum{1}
\articlenumber{1}
\pubyear{xxxx}
\copyrightyear{xxxx}
\history{
    Preprint uploaded XXXXXX
    Received: date;
    Accepted: date;
    Published: date
}

\pdfoutput=1

\Title{Wake structures and performance of wind turbine rotor with harmonic surging motions under laminar and turbulent inflows}

\Author{
    YuanTso Li $^{1, 2}$\orcidA{},
    Wei Yu $^{2}$*\orcidB{}
    and Hamid Sarlak $^{1,}$\orcidC{}
}

\AuthorNames{YuanTso Li , Wei Yu and Hamid Sarlak}

\address{%
$^{1}$ \quad Technical University of Denmark, Denmark\\
$^{2}$ \quad Delft University of Technology, The Netherlands}

\corres{Correspondence: W.Yu@tudelft.nl}

\abstract{This study presents a comprehensive numerical analysis of a full-scale horizontal-axis Floating Offshore Wind Turbine (FOWT) rotor subjected to harmonic surging motions under both laminar and turbulent inflow conditions. Utilizing high-fidelity Computational Fluid Dynamics (CFD) simulations, namely Large-Eddy Simulation (LES) with Actuator Line Model (ALM), this research investigates the rotor performance, wake characteristics, and wake structures of a surging FOWT in detail. The study delves into the influence of varying inflow turbulence intensities, surging settings, and their interplay on the aerodynamic performance and the wake aerodynamics of a FOWT rotor. The results show that, through employing the phase-averaging technique, Surge-Induced Periodic Coherent Structures (SIPeCS) can be identified in the wake of all the surging cases studied, irrespective of the inflow conditions and the surging settings. Additionally, the findings show that the faster wake recovery observed in the surging cases is not caused by enhancing the instability-induced turbulence level, a previously accepted hypothesis. Instead, the results indicate that it is due to the enhanced advection process resulting from the induction fields of SIPeCS that causes the wake to recover faster. The analysis of rotor performance shows that the time-averaged rotor performances are affected by the intricate aerodynamics arising from the surging motions. With certain surging settings, the time-averaged thrust and the time-averaged power of a surging rotor are found to be simultaneously lower and higher compared to those of a fixed rotor. Furthermore, the study underscores the importance of considering both the magnitude of surging and the rate of surging simultaneously to fully characterize the hysteresis load on a surging rotor.
}


\begin{document}



\section{Introduction}

The rapid development in offshore wind energy has led the industry to explore sites further away from coasts, where more spaces and better wind resources are available. This trend indicates that future offshore wind farms will likely be situated in deeper waters ($> 60$~m), where floating concepts offer economic advantages over traditional bottom-fixed designs. However, several aspects of floating concepts, such as the effects of unsteady aerodynamics caused by platform motions, are under-explored \cite{butterfield2007engineering, van2016long, veers2019grand, micallef2021floating}. Several experimental and numerical studies have indicated that the additional Degrees of Freedom (DoF) introduced with platform motions, such as surging and pitching, heavily affect power performances and wake characteristics of Floating Offshore Wind Turbines (FOWTs) \cite{farrugia2016study, tran2016cfd, johlas2021floating}. However, a complete understanding of these effects is still lacking \cite{micallef2021floating, fontanella2021unaflow}.

Previous numerical studies indicate that FOWTs subjected to motions may exhibit accelerated wake recovery rates, suggesting the potential for reduced spacings in floating wind farms compared to bottom-fixed counterparts \cite{rezaeiha2021wake, arabgolarcheh2022modeling}. For instance, Kopperstad et al. \cite{kopperstad2020aerodynamic} had found faster wake recovery rates for a FOWT in motion using LES (large-eddy simulation) with ADM (actuator disk model) under both laminar and turbulent inflow conditions. Chen et al. \cite{chen2022modelling} also found faster wake recovery rates using IDDES (improved delayed detached eddy simulation) with geometric resolved FOWT rotor under laminar inflow conditions. To this point, it has been widely accepted that the enhanced wake recovery for a surging wind turbine rotor is mainly attributed to the increased turbulence level in its wake~\cite{micallef2021floating, chen2022modelling, kleine2022stability, ramos2022investigationI}. However, most of the previous numerical research about FOWT in motion employed CFD (computational fluid dynamic) models that are not able to capture fine flow structures well (such as tip vortices) and/or imposed laminar inflow conditions, which are unrealistic in the fields \cite{micallef2021floating, tran2016cfd, chen2022modelling}. These shortcomings are relevant since the wind turbine wake aerodynamics modeled with turbulent inflow conditions differ markedly from those modeled with laminar inflow conditions \cite{troldborg2009actuator, sarlak2014large, li2022onset}. Also, different dynamics of the released tip vortices are the main cause for a FOWT in motion to have different wake characteristics and wake structures from the bottom-fixed ones \cite{chen2022modelling, kleine2022stability}. Note that the mentioned shortcomings are also applicable when modeling the rotor aerodynamic performance, since inflow conditions directly affect rotor performance. Lower fidelity CFD models, such as ADM, cannot reliably capture the effects of tip/root vortices and shed vortices properly \cite{micallef2021floating, troldborg2009actuator}. Regarding the above-mentioned, this work studies the wakes of a FOWT in motion both under laminar and turbulent inflow conditions with a high-fidelity CFD model that can capture the fine flow structures, namely LES with ALM (actuator line model), intending to provide deeper insights into the wake aerodynamics of a FOWT in motion.

The motion type focused on in this study is harmonic surging, and the motions are prescribed. Surging is a prevalent type of motion studied in previous research since the apparent inflow velocity seen by the FOWT rotor is directly influenced \cite{micallef2021floating, ramos2022investigationI}.

The primary objectives of this work are to investigate the impacts of varying inflow conditions, surging settings, and their combined effects on the rotor performance and wake characteristics of a FOWT rotor. To the authors' knowledge, this is the first numerical study that uses LES with ALM to comprehensively investigate surging turbine rotors under a variety of conditions, where the simulation cases cover the scenarios of various inflow turbulence intensities (TI), surging amplitudes ($A_S$), and surging frequencies ($\omega_S$). This holistic approach is crucial due to the significant interplay of these factors on both the rotor performance and the wake characteristics, which we extensively explore in this paper.

\section{Methodology}

\label{Sec:Methodologies}

This section details the methodologies used in our study and is divided into four subsections. In Section~\ref{sec:phaseLock}, we define harmonic surging motions and discuss the application of the phase-averaging technique. Next, Section~\ref{sec:setup} details the setups of the simulations, including the software, algorithm, governing equations, discretization schemes, mesh layouts, boundary conditions, and parameterization of the FOWT rotor using the ALM. Later, Section~\ref{sec:Statistics} describes how the statistics are obtained. Lastly, Section~\ref{sec:testMatrices} introduces the test matrix and explains the rationale for grouping the cases. 

\subsection{Defining surging motions and phase-averaging techniques}

\label{sec:phaseLock}

In this paper, the surging motions are prescribed to be harmonic. The streamwise position ($x$-position) of the rotor $p_R(t)$ in surging motion is defined in Equation~\ref{eq:surgeMotion}, where $A_S$ represents the surging amplitude, $\omega_S$ is the surging frequency, $\phi_S$ is the phase angle of surging with a phase shift of $\phi_{S_0}$, and $p_{R_0}$ denotes the neutral position of the rotor. $\phi_{S_0} = 0.0 \pi$ and $p_{R_0} = 0$ is maintained throughout this study, as written on the right-hand side of Equation~\ref{eq:surgeMotion}. The surging velocity of the rotor, $V_{S}$, is expressed on the left-hand side of Equation~\ref{eq:V_WTEquation}. It is important to note that non-zero values of $V_{S}$ alter the apparent inflow velocity $V_{0, \mathrm{app}}$ seen by the surging rotor, as depicted on the right-hand side of Equation~\ref{eq:V_WTEquation}.

\begin{multline}
    p_R(t) = A_S \hspace{1pt} \sin \left( \omega_S t + \phi_{S_0} \right) + p_{R_0} = A_S \hspace{1pt} \sin \phi_{S} + p_{R_0}, \\
     p_R(t) \Big{|}_{\phi_{S_0} = 0.0\pi, \hspace{2pt} p_{R_0} = 0} = A_S \hspace{1pt} \sin \omega_S t = A_S \hspace{1pt} \sin \phi_{S}
    \label{eq:surgeMotion}
\end{multline}

\begin{equation}
    V_{S} = \frac{\mathrm{d} \hspace{1pt} p_R(t)}{\mathrm{d} t} = A_S \omega_S \hspace{1pt} \cos \omega_S t , \qquad
    V_{0, \mathrm{app}} = V_0 - V_{S}
    \label{eq:V_WTEquation}
\end{equation}

Two crucial non-dimensional parameters for a FOWT rotor in harmonic surging motions are the ratio of the maximum surging velocity ($V_{S, \mathrm{max}}$) to inflow velocity (denoted as $v_{\mathrm{max}}$) and the reduced frequency based on the rotor diameter $D$ (denoted as $k_D$), as presented in the study of Ferreira et al. \cite{ferreira2022dynamic}. Their definitions are given in Equation~\ref{eq:VandW}. For instance, applying a surging motion with $A_S = 4$~m and $\omega_S = 0.63$~rad/s to an NREL 5MW baseline wind turbine under its rated conditions results in $v_{\mathrm{max}} = 0.22$ and $k_D = 7.00$  ($D = 126$~m, $V_{0, \mathrm{rated}} = 11.4$~m/s). The surging velocity of the rotor $V_{S}$ can be expressed using $v_{\mathrm{max}}$, as written on the right-hand side of Equation~\ref{eq:VandW}. In this work, we interpret $v_{\mathrm{max}}$ as the \emph{magnitude of surging} and $k_D$ as the \emph{rate of surging}.

\begin{equation}
    v_{\mathrm{max}} \overset{\Delta}{=} \frac{V_{S, \mathrm{max}}}{V_0} = \frac{A_S \omega_S}{V_0}, \qquad
    k_D \overset{\Delta}{=} \frac{\omega_S D}{V_0}, \qquad V_{S} 
    = V_{\mathrm{0}} \hspace{1.5pt} v_{\mathrm{max}} \hspace{1pt} \cos  \phi_{S}
    \label{eq:VandW}
\end{equation}

Besides surging, another important periodicity in our study is the rotor's rotation frequency $\Omega$, denoted by Equation~\ref{eq:rotatephi}. In this equation, $\phi_{\Omega}$ represents the rotational phase angle of the rotor, with $\Omega$ being the rotational frequency and $\phi_{\Omega_0}$ being the phase shift. $\phi_{\Omega_0} = 0.0\pi$ is maintained for all cases in this paper, and $\phi_{\Omega} = 0.0\pi$ corresponds to a blade pointing upward in the positive $z$-direction. It is important to note that the phase-averaging technique, which will be explained later, involves $\phi_{\Omega}$.

\begin{equation}
    \phi_{\Omega} = \Omega t + \phi_{\Omega_0}
    \label{eq:rotatephi}
\end{equation}

In wind energy science or flow physics, when time-varying properties oscillate at specific frequencies, it is common to match the sampling rates with these frequencies and study the statistics based on the acquired data \cite{lignarolo2015tip}. This approach, commonly known as the \emph{phase-averaging technique}, is used to understand the periodicities of the system. In this work, we refer to the data obtained at these specific rates as \emph{phase-locking} quantities, and the statistics obtained during averaging the phase-locking data are termed \emph{phase-averaging} quantities.

In our study of a harmonic surging FOWT rotor, we encounter two major periodicities, which are the surging frequency $\omega_S$ and the rotational frequency $\Omega$ (periodicity of $\omega_S$ is more focused in this work). To make the phase-averaging technique effective, we synchronize $\omega_S$ with $\Omega$ by ensuring $\Omega$ is an integer multiple of $\omega_S$. This synchronization ensures that each specific $\phi_S$ in the surging cycle corresponds consistently to the same $\phi_{\Omega}$ for the acquired phase-locking data. Implementing this synchronization across all cases is a crucial aspect of our study since it minimizes the impacts of $\phi_{\Omega}$ on the periodicity of $\omega_S$.

When applying the phase-averaging technique on quantities such as $u$ at a given $\phi_{S}$, we denote the sampled phase-locking data as $u_{\phi_{S}}$. For instance, phase-locking $u$ collected as $\phi_S = 0\pi$ are represented as $u_{0 \pi}$. In this paper, the operator $<\cdot>$ represents quantities that have been phase-averaged. For example, the phase-averaged $u$, denoted as $<u>_{0\pi}$, is the average of $u_{0\pi}$ over a given number of cycles (time interval), while $<\sigma_u>_{0\pi}$ denotes its standard deviation, as described in Equation~\ref{eq:stdPhaseLock}. Note that $u'_{0\pi}$ represents the fluctuation part of phase-averaging. Similarly, the cycle-averaged $C_T$, denoted as $<C_T>$, is obtained based on the phase-averaged $C_T$ across a full surging cycle (see Figure~\ref{fig:paper_1_TIcomparecyc_CTCPArea}(b) for examples). Moreover, analogous to turbulence kinetic energy (TKE), we introduce phase-averaged TKE ($<$TKE$>_{\phi_S}$) to better understand the velocity fluctuations after removing the effects of phase differences due to surging and rotation ($\phi_S$ and $\phi_{\Omega}$). The definition of TKE and $<$TKE$>_{\phi_S = 0\pi}$ used in this study is presented in Equation~\ref{eq:TKEs}.

\begin{equation}
    <u>_{0\pi} = \frac{\sum^N_{n = 1} u_{0\pi, n} }{N}, \qquad
    u_{0\pi} = <u>_{0\pi} + u'_{0\pi}, \qquad
    <\sigma_u>_{0\pi} = \sqrt{ \frac{\sum^N_{n=1} \left( u'_{0\pi, n} \right)^2 }{N} }
    \label{eq:stdPhaseLock}
\end{equation}

\begin{equation}
    \mathrm{TKE} = \frac{1}{2} \left( \sigma_{u}^2 + \sigma_{v}^2 + \sigma_{w}^2 \right)
    \qquad
    <\mathrm{TKE}>_{0 \pi} = \frac{1}{2} \left( <\sigma_u>_{0\pi}^2 + <\sigma_v>_{0\pi}^2 + <\sigma_w>_{0\pi}^2 \right)
    \label{eq:TKEs}
\end{equation}

\subsection{Simulation setups}
\label{sec:setup}

The NREL 5MW baseline turbine (Jonkman et al. \cite{jonkman2009definition}) is selected as the rotor model for our simulations due to the extensive validation database \cite{micallef2021floating, otter2022review}. The three-bladed turbine has a rotor diameter of $D=126$ m, a rated wind speed of $V_{0, \mathrm{rated}} = 11.4$~m/s, and a rated tip speed ratio of $\lambda_{\mathrm{rated}} = 7.0$. We neglect the tower, tilt angles, pre-coning, floor effects, wind shear, controller, and aeroelasticity for simplicity. Unless stated otherwise, the operational conditions are set to the turbine's rated condition.

Large-eddy simulations of the surging turbine are performed using the open source toolbox {\emph{OpenFOAM v2106}} \cite{weller1998tensorial}, and a modified version of \emph{turbinesFoam} (an ALM module originally developed by Bachant et al. \cite{turbinesFoam}) is used to parameterize the surging rotors. We consider the flow (air) to be incompressible and Newtonian, with a density of $\rho = 1.225$~kg/m$^3$ and kinematic viscosity $\nu = 1.5 \times 10^{-5}$~m$^2$/s. Thermal effects and Coriolis force are neglected. The (LES filtered) incompressible Navier-Stokes equations, Equation~\ref{eq:NSE_LES_mom}, are solved using eddy-viscosity closure, with the subgrid-scale (SGS) stress tensor modeled through eddy-viscosity $\nu_T$. Since as long as the resolutions are adequate, the choice of the SGS model is not considered a deterministic factor for wind turbine modeling using LES with ALM \cite{sarlak2015role}. Thus, the standard Smagorinsky model \cite{smagorinsky1963general}, one of the simplest and most used SGS model, is chosen, where $\nu_T$ is modeled using Equation~\ref{eq:sma_nut} with $C_k = 0.094$ and $C_{\varepsilon} = 1.048$, and $\Delta$ corresponds to the grid size. Second-order central differencing (\texttt{Gauss linear}) is used for spatial interpolations, and Crank-Nicolson scheme \cite{crank1947practical} (\texttt{CrankNicolson}, with a coefficient of $0.9$) is selected for temporal interpolations. PISO (Pressure-Implicit with Splitting of Operators) algorithm is used for pressure-velocity coupling. The simulations are performed on the high-performance clusters of DTU Computing Center \cite{DTU_DCC_resource}, where a 600~s simulation requires about 66 hours on 64 processors.

\begin{gather}
    \frac{\partial{u}_i}{\partial x_i} = 0, \qquad
    \frac{\partial {u}_i}{\partial t} + 
    {u}_j \frac{\partial {u}_i }{\partial x_j} =
    - \frac{1}{\rho} \frac{\partial {p}}{\partial x_i} + 
    \frac{\partial}{\partial x_j} \left[ (\nu + \nu_T ) \left( \frac{\partial {u}_i}{\partial x_j} + \frac{\partial{u}_j}{\partial x_i} \right) \right] +
    \frac{{f}_{\mathrm{body},i}}{\rho}
    \label{eq:NSE_LES_mom}\\
    \nu_T = C_k \Delta \sqrt{k_{\mathrm{sgs}}} = C_k \sqrt{\frac{C_k}{C_{\varepsilon}}} \Delta^2 \sqrt{2 {S}_{pq} {S}_{pq}}, \qquad
    {S}_{pq} = \frac{1}{2} \left( \frac{\partial {u}_p}{\partial x_q} + \frac{\partial {u}_q}{\partial x_p} \right)
    \label{eq:sma_nut}
\end{gather}

\begin{figure}[t]
\centering
\includegraphics[width=400pt]{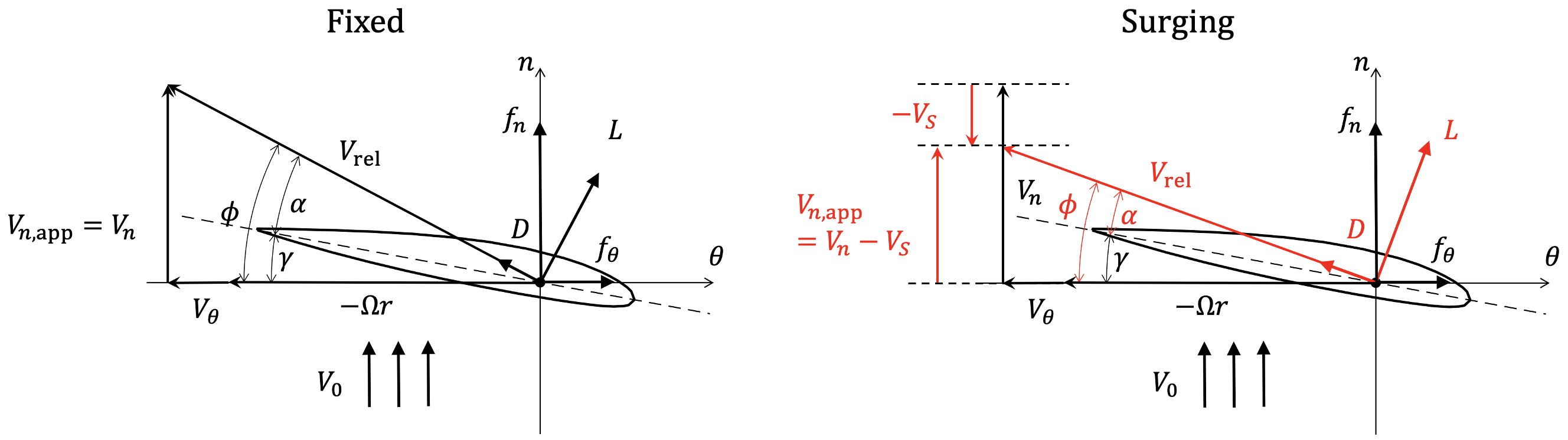}
\caption{Velocity triangles of a blade element for a fixed rotor (left) or a surging rotor (right).}
\label{fig:Vel_tri_paper}
\end{figure}

\begin{figure}[t]
\centering
\includegraphics[width=430pt]{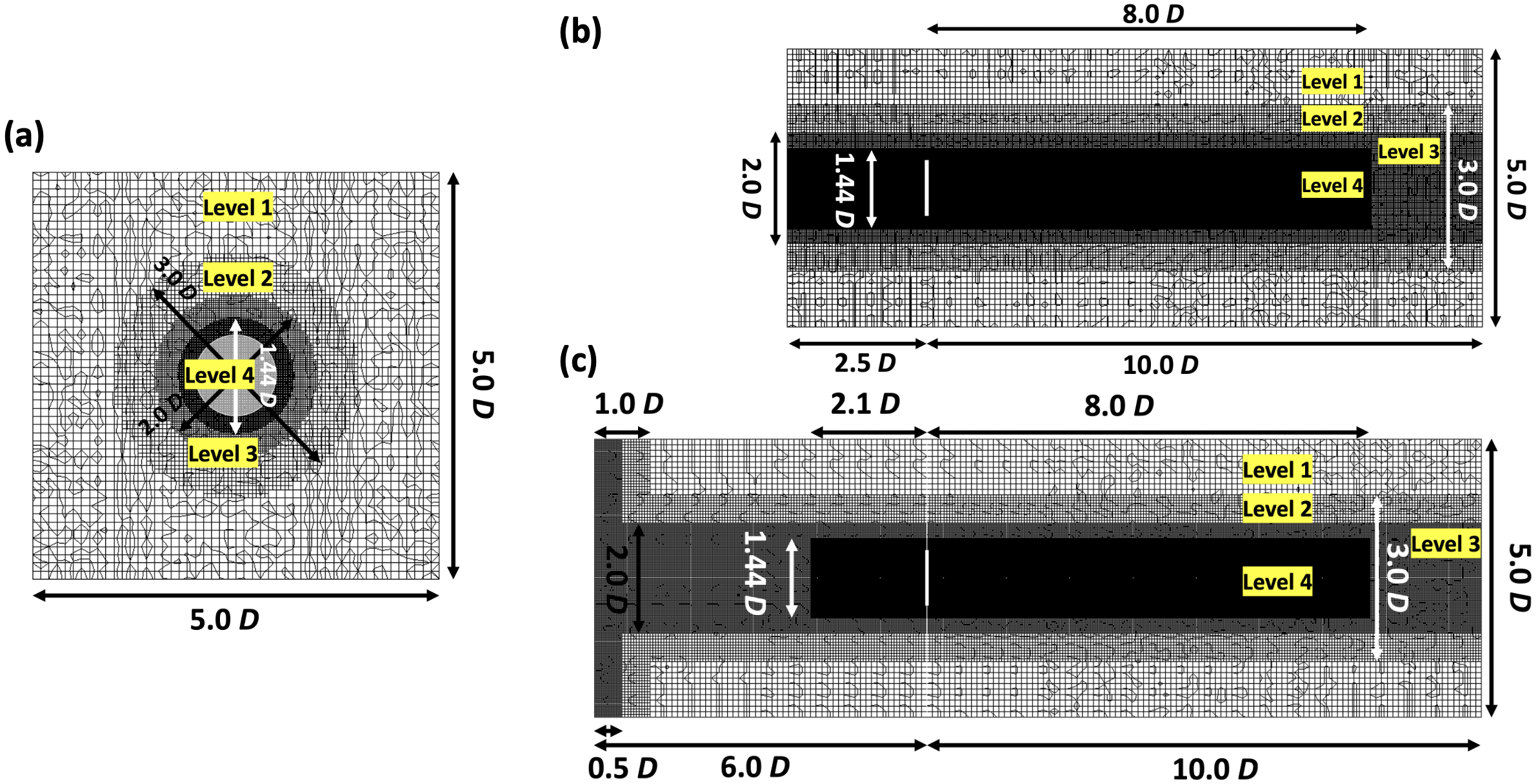}
\caption{The mesh layouts for the simulation cases. (a) is the cross-section of plane $x/D = 0$, and it is shared by both laminar and turbulent cases. (b) and (c) are the cross sections of plane $y/D = 0$, where (b) is for the laminar cases ($10.4$M cells) and (c) is for the turbulent ($10.9$M cells). \emph{Level} indicates the refinement levels, where $\Delta$ (grid size) doubles as \emph{Level} decreases by one.}
\label{fig:paper_mesh}
\end{figure}

The surging FOWT rotor in our simulations is parameterized using the actuator line model (ALM) \cite{sorensen2002numerical}. In the simulations, each blade is represented by 40 actuator line points with equidistant spacing, $\Delta_r$. For hub modeling, an additional actuator line element with a drag coefficient $C_d = 0.3$ and a reference area of $\pi r_{\mathrm{hub}}^2$ (where $r_{\mathrm{hub}} = 1.5$~m is the hub radius) is introduced \cite{hoerner1965fluid, naderi2017numerical}. The term ${\boldsymbol{f}}_{\mathrm{body}}$ in Equation~\ref{eq:NSE_LES_mom} refers to the body forces that the actuator lines exert on the flow. They are first calculated based on the lift and drag forces ($\boldsymbol{L}$ and $\boldsymbol{D}$) obtained through the blade element approach and then projected onto the CFD grid through the Gaussian regularization kernel $\eta_{\varepsilon}$, as detailed from Equation~\ref{eq:f_2D} to \ref{eq:tip_corr}. In these equations, $R$ and $r$ stand for rotor radius and the radial distance from the rotor center, respectively. $c$ is cord length, $\gamma$ is blade twist angle, and $B$ is the number of blades. $f_{\mathrm{tip}}$ is a tip correction factor based on the Glauert model to ensure that loads at the tips/roots drop to zero \cite{nathan2017comparison, daug2020new}. $\varepsilon$ is the smoothing factor for the Gaussian regularization kernel in Equation\ref{eq:ALM}, which is set to be twice the size of the grid near the rotor ($2 \Delta$ at \emph{Level} 4 in Figure~\ref{fig:paper_mesh}), following the recommendations of previous work using ALM with LES \cite{troldborg2009actuator, sarlak2015role, martinez2015large}. For a visual representation of the velocity triangles of a blade element (actuator line point), see Figure~\ref{fig:Vel_tri_paper}. Note that $V_{S}$ (surging velocity) directly influences $V_{\mathrm{rel}}$ (relative velocity seen by the actuator line point), $\alpha$ (angle of attack), and $\phi$ (inflow angle) through altering $V_{n, \mathrm{app}}$ (apparent normal velocity seen by the actuator line point). Naturally, the locations to project ${\boldsymbol{f}}_{\mathrm{body}}$ are influenced by the surging motions as shown in Equation~\ref{eq:ALM}, where $\boldsymbol{x}$ denotes the position vector. Furthermore, no additional dynamic stall model is implemented as the chord-based reduced frequency is rather low under the given conditions \cite{leishman2002challenges, leishman2006principles}. Thus, the term \emph{stalling} in this work is confined to \emph{static (quasi-steady) stalling}. 

\begin{gather}
    {\boldsymbol{f}}_{2D}(r) = \left(  \boldsymbol{L}, \boldsymbol{D} \right) = \frac{1}{2} \rho V_{\mathrm{rel}}^2 c \Big( C_l(Re_c, \alpha) \hat{\boldsymbol{e}}_L, C_d(Re_c, \alpha) \hat{\boldsymbol{e}}_D \Big) = f_n \hat{\boldsymbol{e}}_n + f_{\theta} \hat{\boldsymbol{e}}_{\theta}
    \label{eq:f_2D}\\
    V_{\mathrm{rel}} = \sqrt{V_{n, \mathrm{app}}^2 + (-\Omega r + V_{\theta})^2}, \qquad
    \phi = \arctan \left( \frac{ V_{n, \mathrm{app}} }{-\Omega r + V_{\theta}} \right) = \alpha + \gamma, \qquad  V_{n, \mathrm{app}} = V_{n} - V_{S}
    \label{eq:f_2D_mod}\\
    {\boldsymbol{f}}_{\mathrm{body}}(\boldsymbol{x}) = \sum^{B}_{i=1} \int^R_0 f_{\mathrm{tip}}(r_i)
    {\boldsymbol{f}}_{2D}(r_i) \eta_{\varepsilon} ( \lVert \boldsymbol{x} - (r_i \hat{\boldsymbol{e}}_i + \boldsymbol{p}_R ) \rVert ) \mathrm{d}r_i, \qquad
    \eta_{\varepsilon}(d) = \frac{1}{\varepsilon^3 \pi^{3/2}} \exp \left[ -\left( \frac{d}{\varepsilon} \right)^2 \right]
    \label{eq:ALM}\\
    f_{\mathrm{tip}}(r) = \frac{2}{\pi} \arccos \left[ \exp \left( - \frac{B(R - r)}{2r \sin \phi)} \right) \right]
    \label{eq:tip_corr}
\end{gather}

The computational domains and meshes for the simulations are illustrated in Figure~\ref{fig:paper_mesh}. Specifically, Figure~\ref{fig:paper_mesh}(b) depicts the mesh with section plane of $y/D = 0$ for laminar inflow cases (comprising 10.4M cells), while Figure~\ref{fig:paper_mesh}(c) shows the mesh for turbulent inflow cases (10.9M cells). The slight differences between these meshes are primarily for mitigating the undesired pressure fluctuations introduced by the synthetic turbulent inlet conditions. \emph{Level} in Figure~\ref{fig:paper_mesh} refers to the mesh refinement levels. Both meshes share the same $\Delta$ value for corresponding \emph{Level}s. Near the rotor (in the wake region), $\Delta$ is similar to the spacing of the actuator line points $\Delta_r$. That is, at \emph{Level} 4, $\Delta = D/80$. A Cartesian coordinate system is used, with the positive $x$-axis pointing downstream. The neutral position of the rotor center is at the origin, rotating clockwise when viewed from upstream. Temporally, there are 360 time steps per rotor revolution for the rated condition of NREL 5MW ($\Delta t = 0.0138$ s), ensuring that the rotor tip travels less than $1\Delta$ per time step (which is $<0.7\Delta$ for our cases), in line with recommendations by previous studies \cite{troldborg2009actuator, martinez2015large}.

For the inlet boundary conditions, the laminar cases utilize a velocity inlet with a uniform fixed value, while turbulent cases employ the divergence-free synthetic eddy method (DFSEM) \cite{poletto2013new}. DFSEM introduces inflow with the desired turbulence intensities, length scales, and anisotropy, and one of its key features is the requirement of much less computational resources than the precursor method. Additionally, DFSEM can reproduce identical inflow in both space and time if the settings are the same, enabling meaningful comparisons of instantaneous fields between cases. Laminar and turbulent cases share the same boundary conditions besides the velocity inlet conditions. The velocity boundary conditions on all four sides are treated as slip walls, and an advective boundary condition ($\mathrm{D}/\mathrm{D}t = 0$) is used for the outlet. For the pressure fields, symmetry boundary conditions are applied to all four sides and the inlet, while the outlet is set to a uniform fixed value under the assumption that the pressure fields have returned to ambient levels at the outlet ($x/D = 10$). Detailed simulation setups are provided in Li \cite{Li2023Numerical}.

\subsection{Statistics} 
\label{sec:Statistics}

The averaging windows for the simulations are set at 20 and 50 rotational periods at the rated conditions ($T_{\Omega, \mathrm{rated}}$) for the laminar and turbulent cases, respectively. These windows have been determined to ensure the convergence of the statistics of interest, including time-averaged and cycle-averaged rotor performances (e.g., $\overline{C}_T$ and $<C_T>$), time-averaged and phase-averaged values for flow fields (e.g., $\overline{u}$ and $<u>_{0\pi}$), and their second-order statistics (e.g., $\sigma_u$ and $<\sigma_u>_{0\pi}$). In this work, quantities with an overline are time-averaged values, and the terms ``time-averaged'' and ``mean'' are used interchangeably. A brief summary of the convergence test results is presented in the Appendix~\ref{app:convergenceStudy} as well as Li \cite{Li2023Numerical}.

\subsection{Test matrix}

\label{sec:testMatrices}

A total of 16 simulation cases are conducted and are detailed in Table~\ref{tab:cases_Single_rotor}. Four of the cases feature a fixed rotor, and the remaining twelve involve a surging rotor, as separated by the horizontal line in the table. Case numbers and grouping indices are listed in the two leftmost columns. TI, $A_S$, and $\omega_S$ are the inflow turbulence intensity, surging amplitude, and surging frequency. For convenience, $A_S$ is set to $0$~m and $\omega_S$ is manually made to $0.63$~rad/s for the cases with a fixed rotor when analysis involves phase-averaging or cycle-averaging. $v_{\mathrm{max}}$ and $k_D$ are the two non-dimensional parameters introduced in Equation~\ref{eq:VandW}, and can be used to characterize $A_S$ and $\omega_S$. 

Cases subjected to both laminar and turbulent inflow conditions are included in our test matrix. Despite being somewhat unrealistic in nature, the cases with laminar inflows are introduced and analyzed because of their simpler context compared to those with turbulent inflows. The insights gained from the laminar cases form the basis for analyzing the more intricate dynamics found in the cases subjected to turbulent inflow conditions, facilitating easier and more effective interpretations.

The base settings for surging parameters $A_S$ and $\omega_S$ in the surging cases of Table~\ref{tab:cases_Single_rotor} are chosen as $4$~m and $0.63$~rad/s, respectively. The value of this $\omega_S$ is based on the typical wave frequencies for offshore environments \cite{stewart2016creation}. Although $A_S = 4$~m is larger than the typical values for the floating platform, which are around $1.5$-$2.5$~m for typical sea states (the floating platform is assumed to be a semi-submersible or tension leg platform and the significant wave height is around $4.0$~m) \cite{stewart2016creation, kyle2020propeller, ramachandran2013investigation, wen2018power}, it is still chosen. Choosing a larger $A_S$ gives advantages in identifying the impact of surging motions since it is expected to show stronger effects. Surging parameters for the other settings half/double $A_S$ or $\omega_S$ of the base settings one at a time (the synchronization between $\omega_S$ and $\Omega$ elaborated in Section~\ref{sec:phaseLock} is maintained), resulting in five distinct surging settings (excluding the fixed scenario). Although some of the surging settings represent extreme states or may even be impractical in real-world scenarios (e.g., the cases with $v_{\mathrm{max}} = 0.44$ and $k_D = 14.0$), they are still tested since the main focus of this work is not on simulating real-world cases, but on studying the impacts of surging settings and inflow conditions. Regarding the base settings for inflow TI, $5.3\%$ is chosen based on the typical inflow TI for offshore environments. Furthermore, all the three tested inflow TI, namely $2.7\%$, $5.3\%$, and $11.5\%$, fall within the practical range for offshore environments, which is around $2.5\%$-$12.0\%$ \cite{hansen2012impact}.

For a more concise analysis, the 16 cases outlined in Table~\ref{tab:cases_Single_rotor} are categorized into three groups, namely \textbf{Group}~$\boldsymbol{a}$, \textbf{Group}~$\boldsymbol{b}$, and \textbf{Group}~$\boldsymbol{c}$, as indicated in Table~\ref{tab:cases_Single_rotor}. Note that some cases are included in more than one group. \textbf{Group}~$\boldsymbol{c}$, comprising cases \textbf{1}, \textbf{5}, \textbf{3}, and \textbf{7}, represents the core scenarios of our analysis, which are fixed-laminar, surging-laminar, fixed-turbulent, and surging-turbulent. These cases provide an overview of both the impacts of surging motions and inflow conditions, as well as their interplay. This group is the foundational backbone of our study and is also included in the other two groups. \textbf{Group}~$\boldsymbol{a}$ consists of eight cases (cases \textbf{1}-\textbf{8}), which includes instances of a surging rotor with the same surging settings ($v_{\mathrm{max}} = 0.22$ and $k_D = 7.0$) subjected to four different inflow conditions (laminar, TI $=2.7\%$, $5.3\%$, and $11.5\%$), along with the other four cases featuring a fixed rotor subjected to the four same inflow conditions. The primary focus of this group is to examine how varying inflow TI influences the effects of surging motions. Finally, \textbf{Group}~$\boldsymbol{b}$, encompassing twelve cases (cases \textbf{1}, \textbf{3}, \textbf{5}, \textbf{7}, and \textbf{9}-\textbf{16}), is dedicated to examining the effects of different surging settings under both laminar and turbulent inflow conditions (TI = $5.3\%$). This group covers six surging settings, and each of them is tested with the two inflow conditions. The focus here is on assessing the influence of the surging parameters $A_S$ and $\omega_S$ on rotor performance and wake structures and understanding the interaction between the inflow turbulence and the surging settings.

\begin{table}[htb!]
\centering
\caption{Case settings conducted with a single fixed or surging NREL~5MW baseline wind turbine rotor at its rated condition ($V_0 = 11.4$~m/s, $\lambda = 7$). ``Laminar'' indicates the case has laminar inflow conditions. $\boldsymbol{a}$, $\boldsymbol{b}$, and $\boldsymbol{c}$ in the column of \textbf{Group} serve as indices for grouping. TI, $A_S$, and $\omega_S$ are the inflow turbulence intensity, surging amplitude, and surging frequency. $v_{\mathrm{max}}$ and $k_D$ are the two non-dimensional parameters derived from $A_S$ and $\omega_S$, where their definitions can be found in Equation~\ref{eq:VandW}.}
\label{tab:cases_Single_rotor}
\adjustbox{max width=\textwidth}{%
\begin{tabular}{lc|ccccc}

\textbf{Case} & \textbf{Group} & TI~[$\%$] & $A_S$~[m] & $\omega_S$~[rad/s] & $v_{\mathrm{max}}$ & $k_D$ \\
\hline

\textbf{1} & $\boldsymbol{a}$, $\boldsymbol{b}$, $\boldsymbol{c}$ & Laminar& $0$ & - & $0.00$ & - \\

\textbf{2} & $\boldsymbol{a}$ & $2.7$& $0$ & - & $0.00$ & - \\

\textbf{3} & $\boldsymbol{a}$, $\boldsymbol{b}$, $\boldsymbol{c}$ & $5.3$& $0$ & - & $0.00$ & -  \\

\textbf{4} & $\boldsymbol{a}$ & $11.5$& $0$ & - & $0.00$ & - \\

\hline

\textbf{5} & $\boldsymbol{a}$, $\boldsymbol{b}$, $\boldsymbol{c}$ & Laminar & $4$ & $0.63$ & $0.22$ & $7.0$  \\

\textbf{6} & $\boldsymbol{a}$ & $2.7$ & $4$ & $0.63$ & $0.22$ & $7.0$  \\

\textbf{7} & $\boldsymbol{a}$, $\boldsymbol{b}$, $\boldsymbol{c}$ & $5.3$ & $4$ & $0.63$ & $0.22$ & $7.0$ \\

\textbf{8} & $\boldsymbol{a}$ & $11.5$ & $4$ & $0.63$ & $0.22$ & $7.0$ \\

\textbf{9} & $\boldsymbol{b}$ & Laminar & $2$ & $0.63$ & $0.11$ & $7.0$  \\

\textbf{10} & $\boldsymbol{b}$ & Laminar & $8$ & $0.63$ & $0.44$ & $7.0$ \\

\textbf{11} & $\boldsymbol{b}$ & Laminar & $4$ & $0.32$ & $0.11$ & $3.5$ \\

\textbf{12} & $\boldsymbol{b}$ & Laminar & $4$ & $1.27$ & $0.44$ & $14.0$ \\

\textbf{13} & $\boldsymbol{b}$ & $5.3$ & $2$ & $0.63$ & $0.11$ & $7.0$ \\

\textbf{14} & $\boldsymbol{b}$ & $5.3$ & $8$ & $0.63$ & $0.44$ & $7.0$  \\


\textbf{15} & $\boldsymbol{b}$ & $5.3$ & $4$ & $0.32$ & $0.11$ & $3.5$ \\

\textbf{16} & $\boldsymbol{b}$ & $5.3$ & $4$ & $1.27$ & $0.44$ & $14.0$ \\

\hline

\end{tabular}
}
\end{table}

\section{Verification and validation}

\subsection{Inflow turbulence characterization}

\label{sec:inflow_turb}

\begin{figure}[t]
\centering
\includegraphics[width=430pt]{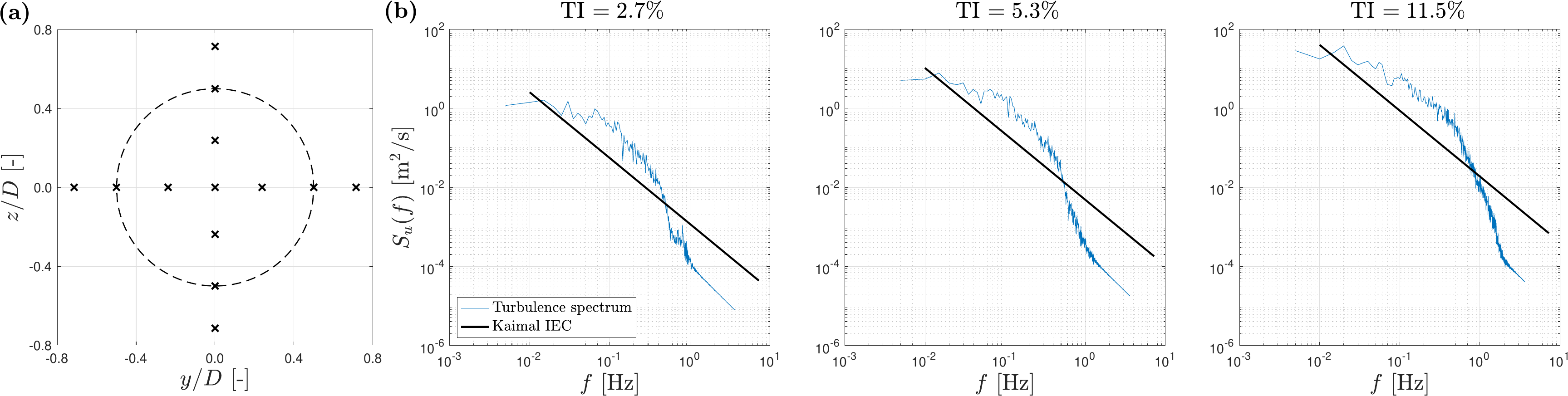}
\caption{(a): Positions of the probes for sampling the inflow properties, they are located at $x/D = -2$. (b): Turbulence spectra for different inflow conditions with their inflow turbulence intensities (TI) labeled at the top.}
\label{fig:paper1_V2_freqMoreProbe}
\end{figure}

In our simulations, inflow turbulence is analysed at $2.0D$ upstream the rotor, using the 13 probes shown in Figure~\ref{fig:paper1_V2_freqMoreProbe}(a). For various turbulent inflow conditions with different TI, several key parameters are measured, including mean streamwise velocity ($\overline{u}$), standard deviation of $u$, $v$, and $w$ ($\sigma_u$, $\sigma_v$, and $\sigma_w$), turbulence intensity (TI, as defined in Equation~\ref{eq:def_TI}), power spectrum of $u$ (denoted as $S_u(f)$), and the integral length scale of $u$ in the streamwise direction ($L_u$). $L_u$ is determined when the auto-correlation of $u$ first reaches zero. All the values for $\overline{u}$, $\sigma_u$, $\sigma_v$, $\sigma_w$, TI, $L_u$, and $S_u(f)$ are computed by averaging data from all probes with equal weighting. The reference inflow velocity, $V_0$, is set at $11.4$~m/s.

\begin{equation}
    \mathrm{TI} = \frac{\sqrt{\frac{1}{3}(\sigma_{u}^2 + \sigma_{v}^2 + \sigma_{w}^2)}}{V_0} \times 100\% = \frac{\sqrt{\frac{2}{3}\mathrm{TKE}}}{V_0} \times 100\%
    \label{eq:def_TI}
\end{equation}

Table~\ref{tab:InflowTurbInformation} details the turbulence-related quantities sampled using the probes depicted in Figure~\ref{fig:paper1_V2_freqMoreProbe}(a), where three different turbulent inflow conditions with different inflow TI are characterized. According to the IEC 61400-1 edition 4.0 (2019) \cite{IECStandard}, the integral length scale $L_{u}$ should be $42$~m for our cases, where the designed hub height, $z_{\mathrm{hub}}$, is $90$~m (see Equation\ref{eq:IEC_length}). Additionally, the standard specifies that the ratio $\sigma_v/\sigma_u$ should exceed $0.7$ (in scenarios considering floor effects, with $v$ representing the lateral direction). In our simulations, while $\sigma_v$ and $\sigma_w$ align with the IEC standard, we observe a larger $L_u$. This difference is reasonable as the wall-normal distance influences $L_u$ \cite{pena2019osterild}, and in our setup, the imposed (slip) wall is situated further than $90$~m from the rotor. Furthermore, civil engineering standards such as ASCE 7-16 and AIJ (2004) suggest that $L_u$ should range around $150$ to $250$~m for offshore conditions of a similar height \cite{nandi2021estimation}. Therefore, our simulations' higher $L_u$ values are considered realistic within these broader standards. Figure~\ref{fig:paper1_V2_freqMoreProbe}(b) illustrates the turbulence spectrum $S_u(f)$, averaged from the power spectra measured by the probes in Figure~\ref{fig:paper1_V2_freqMoreProbe}(a). These spectra align well with the Kaimal spectrum defined in IEC 61400-1  \cite{IECStandard} (see Equation~\ref{eq:Kaimal}).

\begin{table}[htb!]
\centering
\caption{Characteristics of the inflow turbulence measured at $2.0D$ upstream of the rotor, using the probes shown in Figure~\ref{fig:paper1_V2_freqMoreProbe}(a). Most of the numbers presented in this table are the averaged values based on all the thirteen probes, with the exception of $\sigma_{\mathrm{TI}}$ and $\sigma_{L_u}$, where they are are the standard deviations of TI and $L_u$ based on the measurements of the probes, respectively.}
\label{tab:InflowTurbInformation}
\begin{tabular}{c|cccccccc}
\hline
\textbf{TI}~[\%] & $\overline{u}/V_0$ & $\sigma_{\mathrm{TI}}/$TI & $\sigma_u/V_0$ & $\sigma_v/\sigma_u$ & $\sigma_w/\sigma_u$ & {$L_{u}$~[m]} & $\sigma_{L_u}/L_u$  \\
\hline
2.66 & $1.003$ & $0.079$ & $0.029$ & $0.93$ & $0.87$ & $108.2$ & $0.59$ \\
5.27 & $1.008$ & $0.083$ & $0.057$ & $0.91$ & $0.85$ & $104.9$ & $0.58$ \\
11.52 & $1.018$ & $0.061$ & $0.121$ & $0.97$ & $0.90$ & $130.3$ & $0.47$ \\

\hline
\end{tabular}
\end{table}

\begin{equation}
    L_u = 0.7 z_{\mathrm{hub}}, \quad z_{\mathrm{hub}} \leq 60~\mathrm{m}, \qquad 
    L_u = 42~\mathrm{m}, \quad z_{\mathrm{hub}} > 60~\mathrm{m}
    \label{eq:IEC_length}
\end{equation}

\begin{equation}
    S_{u}(f) = 0.05 \sigma_u^2 \left(\frac{L_{u}}{V_0}  \right)^{-2/3} f^{-5/3} 
    \label{eq:Kaimal}
\end{equation}

\subsection{Verification}

\begin{figure}[t]
\centering
\includegraphics[width=430pt]{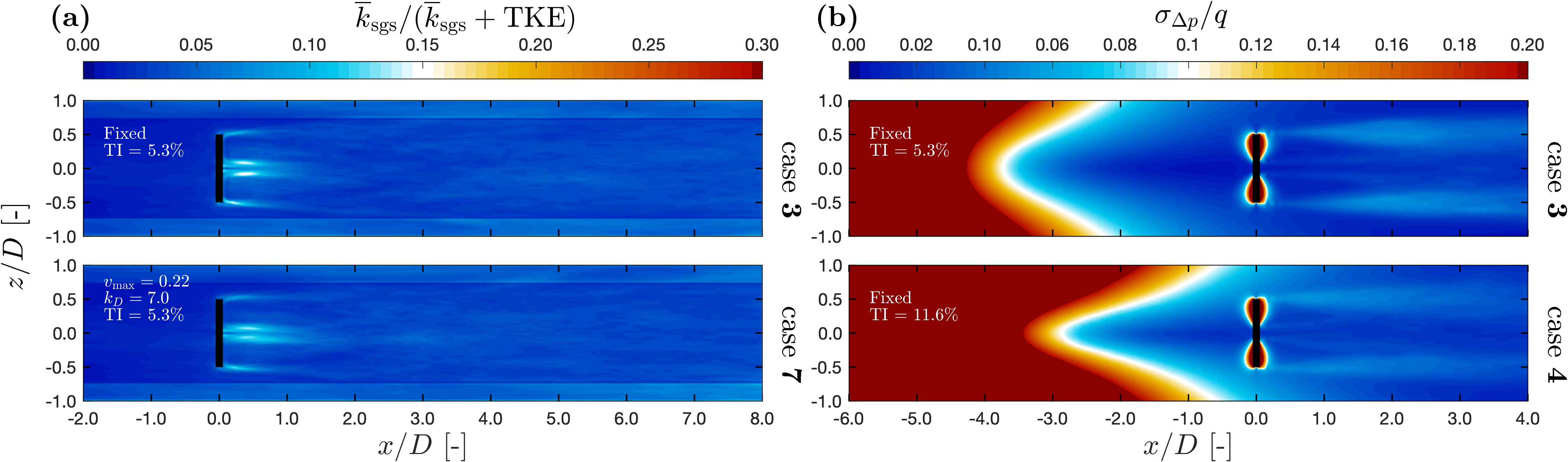}
\caption{(a): Ratios of modeled turbulent kinetic energy ($\overline{k}_{\mathrm{sgs}}$) and total turbulent kinetic energy for cases \textbf{3} (top) and cases \textbf{7} (bottom). (b): The fields of pressure fluctuations ($\sigma_{\Delta p}$, standard deviation of pressure) for cases \textbf{3} (top) and cases \textbf{4} (bottom), where $q = \rho V_{0}^2/2$.}
\label{fig:sgsTest}
\end{figure}

Several tests related to grid resolution are conducted to ensure the reliability of our simulations. A grid independence test demonstrates that our results are not significantly affected by variations in grid resolution (see Appendix \ref{app:grid_test}). Furthermore, to verify the adequacy of LES, we examined the ratio of subgrid-scale TKE ($k_{\mathrm{sgs}}$) to the total TKE, which is the sum of (resolved) TKE and $k_{\mathrm{sgs}}$. As Figure~\ref{fig:sgsTest}(a) demonstrates, more than $80\%$ of the TKE is resolved, aligning with best practices in turbulence modeling using LES \cite{pope2001turbulent}. Additionally, the pressure fluctuations stemming from DFSEM used at the inlet for synthetic turbulent inflow are assessed. The standard deviation of pressure fields ($\sigma_{\Delta p}$, where $\Delta p$ represents the difference between measured and ambient pressure), which is shown in Figure~\ref{fig:sgsTest}(b), indicates that pressure fluctuations are predominantly restricted to the vicinity of the inlet and have limited impacts on the solutions near and after the rotor.

\subsection{Validation}

Table~\ref{tab:NREL_bench} presents the values of the time-averaged thrust coefficient ($\overline{C}_T$) and power coefficient ($\overline{C}_P$) of a fixed NREL~5MW rotor operating in its rated condition (tip speed ratio $\lambda_{\mathrm{rated}} = 7.00$, inflow velocity $V_{0, \mathrm{rated}} = 11.4$~m/s) from our work and other previous studies. While our $\overline{C}_T$ and $\overline{C}_P$ values show some deviation from the original design specifications given by Jonkman et al. \cite{jonkman2009definition}, they fall within the range of results reported by other studies. This comparison suggests that our results are reasonable and reliable within the context of existing research.

\begin{table}[htb!]
\centering
\caption{Comparison of the $\overline{C}_T$ and $\overline{C}_P$ from our study of a fixed NREL~5MW rotor under its rated conditions with results from other previous numerical studies.}
\label{tab:NREL_bench}
\begin{tabular}{l|ccc|cc}
{Source} & Turbulence model & Force model & TI~[$\%$] & {$\overline{C}_T$} & {$\overline{C}_P$} \\
\hline
\textbf{Current work} & \textbf{LES} & \textbf{ALM} & \textbf{Laminar} & $\mathbf{0.728}$ & $\mathbf{0.518}$ \\
\textbf{Current work} & \textbf{LES} & \textbf{ALM} & $\mathbf{5.3}$ & $\mathbf{0.727}$ & $\mathbf{0.518}$ \\
Jonkman et al. \cite{jonkman2009definition} & - & BEM & - & $0.81$ & $0.47$\\
Johlas et al. \cite{johlas2019large} & LES & ALM & $4.1$ & $0.752$ & - \\
Xue et al. \cite{xue2022research} & LES & ALM & Laminar & $0.75$ & $0.52$ \\
Li et al. \cite{li2015numerical} & RANS & ALM & - & $0.77$ & $0.49$ \\
Yu et al. \cite{yu2018study} & RANS & ALM & - & $0.728$ & $0.472$ \\
Rezaeiha et al. \cite{rezaeiha2021wake} & RANS & ADM & $5$ & $0.715$ & $0.567$ \\

\hline
\end{tabular}
\end{table}

\section{Result and discussion on rotor performances}

This section examines the rotor performance for the cases listed in Table~\ref{tab:cases_Single_rotor}. In Section~\ref{sec:CT_CP_time_Avg}, we begin by presenting and analyzing the results of integral rotor performance, focusing on the effects of inflow TI and surging settings on the thrust and power coefficients ($C_T$ and $C_P$). Next, in Section~\ref{sec:AOA}, the angle of attack ($\alpha$) along the blade span for the surging cases is presented, specifically to confirm the occurrence of (static) stalling due to surging. Finally, Section~\ref{sec:Exp_act_CT} discusses the cycle-averaged rotor performance based on the analytical derivations and simulation results. The findings from Section~\ref{sec:AOA} are integrated into this discussion, providing deeper insights into the results observed in Section~\ref{sec:CT_CP_time_Avg}.

The definitions of $C_T$ and $C_P$ for this work are described in Equation~\ref{eq:CT_and_CP}. Except for the apparent thrust and power coefficients ($C_T^{\mathrm{app}}$ and $C_P^{\mathrm{app}}$) appear in Section~\ref{sec:Exp_act_CT}, the reference velocity for all $C_T$ and $C_P$ related quantities are set to be $V_{0, \mathrm{rated}}$ throughout this work, which is the inflow wind speed $V_0$ for all the $16$ cases in Table~\ref{tab:cases_Single_rotor}. This choice is made to facilitate a more straightforward comparison among the cases.

\begin{equation}
    C_T = \frac{T}{\frac{1}{2} \rho V^2_{0, \mathrm{rated}} \pi R^2}, \qquad   C_P = \frac{P}{\frac{1}{2} \rho V^3_{0, \mathrm{rated}} \pi R^2}
    \label{eq:CT_and_CP}
\end{equation}

\subsection{Integral rotor performances}

\label{sec:CT_CP_time_Avg}

This subsection delves into how inflow conditions and surging settings influence integral rotor performances. In the analysis, the 16 cases in Table~\ref{tab:cases_Single_rotor} are divided into \textbf{Group}~$\boldsymbol{a}$ and \textbf{Group}~$\boldsymbol{b}$ as previously mentioned in Section~\ref{sec:testMatrices}. Although the interplay between inflow conditions and surging settings can be observed in both groups, the first focuses more on the effects of inflow conditions, while the second focuses more on the effects of surging settings.

The integral rotor performances in this work refer to the quantities related to $C_T$ and $C_P$, including time-averaged thrust and power coefficients ($\overline{C}_T$ and $\overline{C}_P$) and cycle-averaged thrust and power coefficients ($<C_T>$ and $<C_P>$, introduced in Section~\ref{sec:phaseLock}). $\overline{C}_T$ and $\overline{C}_P$ quantify rotor performance for each case with single values, allowing quick comparisons. While $<C_T>$ and $<C_P>$ give more details of how the surging motions impact the rotor performance according to the surging phase angle $\phi_S$.

\begin{figure}[t]
\centering
\includegraphics[width=430pt]{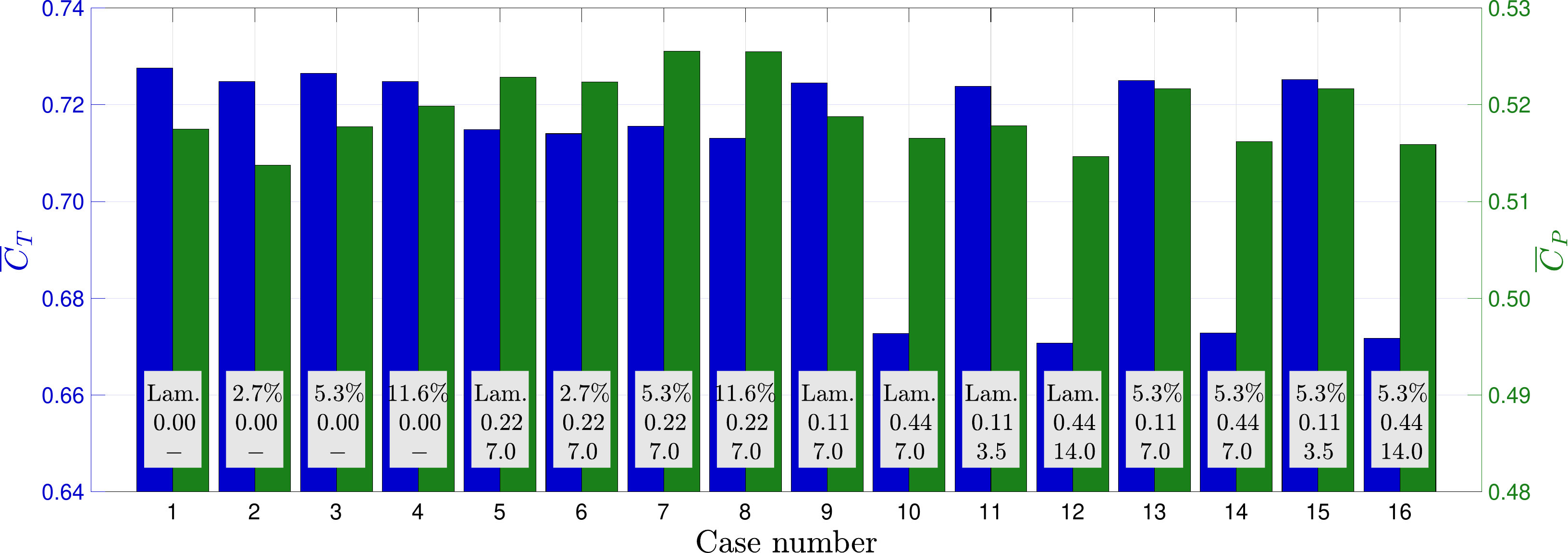}
\caption{Bar plots of $\overline{C}_T$ and $\overline{C}_P$ for the 16 cases in Table~\ref{tab:cases_Single_rotor}. Entries on each bar are the settings for each case. The top, middle, and bottom entries are for inflow turbulence intensity, the ratio between maximum surging velocity and inflow velocity, $v_{\mathrm{max}}$, and the rotor-based reduced frequency, $k_D$ (see Equation~\ref{eq:VandW}).}
\label{fig:singleHis_wide}
\end{figure}

\paragraph{Impact of surging effects with various inflow TI}

$\overline{C}_T$ and $\overline{C}_P$ for the eight cases in \textbf{Group}~$\boldsymbol{a}$ (cases \textbf{1}-\textbf{8}) are illustrated in Figure~\ref{fig:singleHis_wide} together with the other cases in Table~\ref{tab:cases_Single_rotor}. Examining the four fixed cases (cases \textbf{1}-\textbf{4}), it is observed that their $\overline{C}_T$ and $\overline{C}_P$ values are very similar across different inflow conditions. A similar pattern of consistency is observed with the four surging cases (cases \textbf{5}-\textbf{8}) as well. Moreover, it can be seen that $\overline{C}_T$ for the four surging cases are consistently lower than the four fixed cases, and $\overline{C}_P$ for the four surging cases are consistently higher than the four fixed cases. These indicate that the integral time-averaged rotor performances, regardless of the rotor being fixed or surging, are not significantly affected by inflow turbulence, and the surging settings have more impacts on $\overline{C}_T$ and $\overline{C}_P$ compared to the inflow conditions.

\begin{figure}[t]
\centering
\includegraphics[width=380pt]{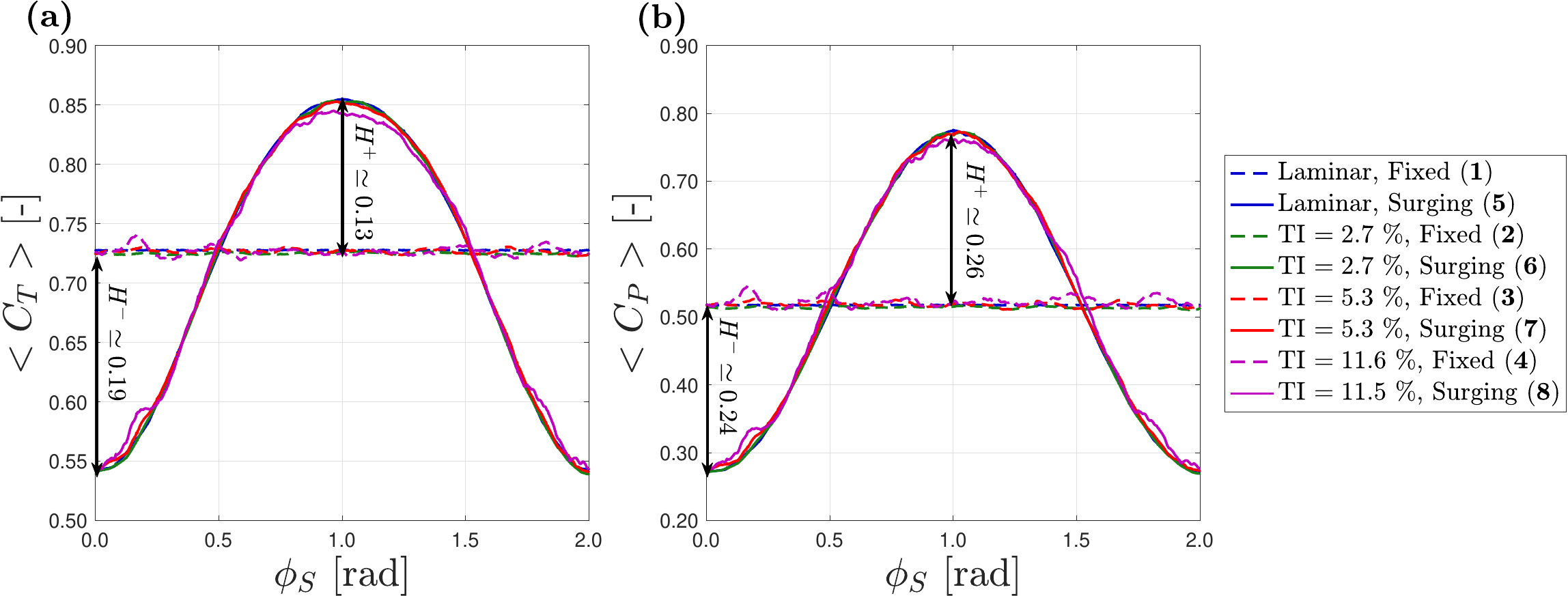}
\caption{(a) \& (b): $<C_T>$ and $<C_P>$ for cases of \textbf{Group}~$\boldsymbol{a}$, which they have a fixed or a surging rotor and subjected to different inflow conditions. Numbers enclosed by parentheses correspond to the case numbers in Table~\ref{tab:cases_Single_rotor}. The overshoots and undershoots for $<C_T>$ or $<C_P>$ of the surging cases are indicated with $H^+$ and $H^-$, respectively. These undershoots and overshoots are evaluated based on the neutral values ($\overline{C}_T$ and $\overline{C}_P$ of the fixed case).}
\label{fig:paper_1_TIcomparecyc_CTCPArea}
\end{figure}

Figures~\ref{fig:paper_1_TIcomparecyc_CTCPArea}(a) and \ref{fig:paper_1_TIcomparecyc_CTCPArea}(b) present the cycle-averaged thrust coefficient ($<C_T>$) and power coefficient ($<C_P>$) for the cases in \textbf{Group}~$\boldsymbol{a}$. For the fixed rotor cases, the reference frequency for cycle-averaging is aligned with the $\omega_S$ of the surging cases within \textbf{Group}~$\boldsymbol{a}$, as described in Section~\ref{sec:phaseLock}. These figures reveal that both $<C_T>$ and $<C_P>$ display a similar insensitivity to inflow TI, akin to $\overline{C}_T$ and $\overline{C}_P$. Consistent with previous research \cite{tran2016cfd, de2014effect}, $<C_T>$ and $<C_P>$ fluctuate in response to the surging velocity $V_S$. Specifically, when $V_S$ opposes the inflow velocity $V_0$ ($\phi_S = 1.0 \pi$), the values of $<C_T>$ and $<C_P>$ are at their highest. Conversely, when $V_S$ aligns with $V_0$ ($\phi_S = 0.0 \pi$), the values are at their lowest.

\paragraph{Impact of surging settings under laminar or turbulent inflow conditions}

$\overline{C}_T$ and $\overline{C}_P$ for the twelve cases in \textbf{Group}~$\boldsymbol{b}$ (cases \textbf{1}, \textbf{3}, \textbf{5}, \textbf{7}, and \textbf{9}-\textbf{16}) are also plotted in Figure~\ref{fig:singleHis_wide}. It is clear that $\overline{C}_T$ and $\overline{C}_P$ are more sensitive to $v_{\mathrm{max}}$ (magnitudes of surging) than to $k_D$ (rates of surging). For example, the results of cases \textbf{10} and \textbf{12} are relatively similar (same $v_{\mathrm{max}}$ different $k_D$), but the results of cases \textbf{5} and \textbf{10} are significantly different (different $v_{\mathrm{max}}$ same $k_D$). This finding is in line with what has been reported by Ferreira et al. \cite{ferreira2022dynamic}, where they conducted a comprehensive literature review on existing studies regarding surging FOWTs. Additionally, in agreement with the observations in \textbf{Group}~$\boldsymbol{a}$, the values of $\overline{C}_T$ and $\overline{C}_P$ show little variations with different inflow conditions.

Upon further examining the values of $\overline{C}_T$ and $\overline{C}_P$, we observe that, with the exception of cases exhibiting larger $v_{\mathrm{max}}$ values (specifically, those with $v_{\mathrm{max}} = 0.44$ where severe stalling occurs during the surging cycle as shown in Section~\ref{sec:AOA}), surging cases typically exhibit lower $\overline{C}_T$ and slightly higher $\overline{C}_P$ compared to the fixed case. This trend becomes more pronounced with larger $v_{\mathrm{max}}$. Although simultaneously having lower $\overline{C}_T$ and higher $\overline{C}_P$ may seem contradictory according to one-dimensional momentum theory \cite{manwell2010wind}, this phenomenon can be attributed to the nonlinear response of $f_n$ and $f_{\theta}$ (normal and tangential forces of blade elements) to the changes of $V_S$ due to surging (see Figure~\ref{fig:Vel_tri_paper}). Detailed analysis and discussions of this phenomenon are provided in Section~\ref{sec:Exp_act_CT}.

\begin{figure}[t]
\centering
\includegraphics[width=430pt]{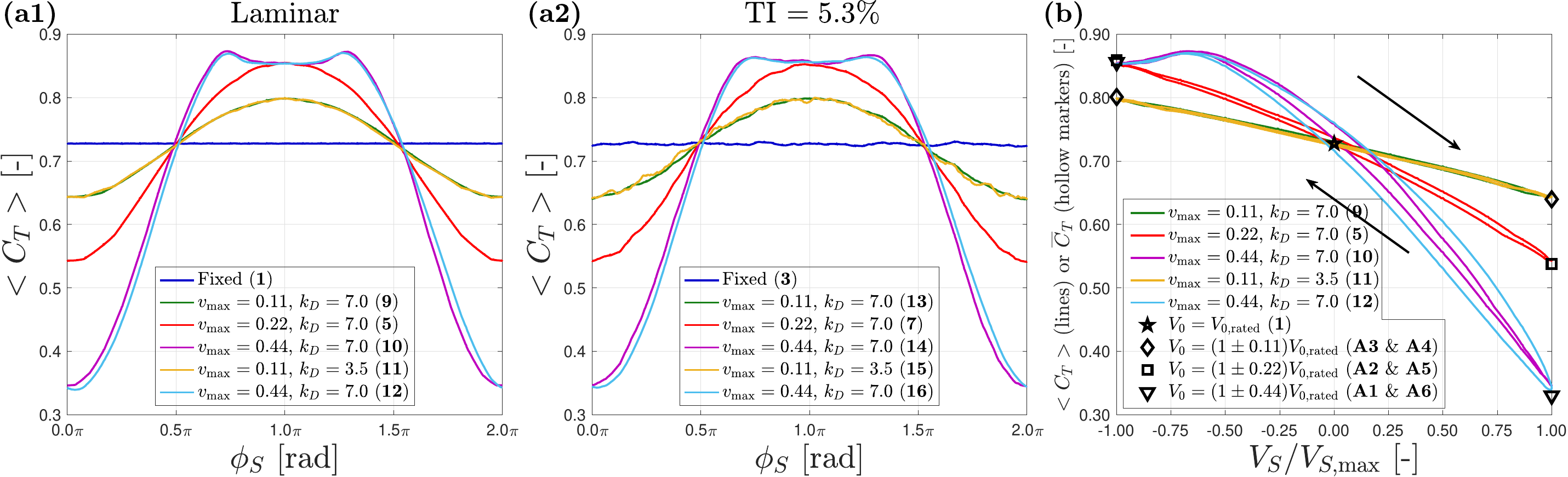}
\caption{$<C_T>$ for cases of \textbf{Group}~$\boldsymbol{b}$, where the cases have different surging settings and subjected to laminar or turbulent inflow conditions. (a1): Cases subjected to laminar inflows. (a2) Cases subjected to turbulent inflows (TI $= 5.3\%$). (b): $<C_T>$ against $V_{S}$ for the laminar cases. $V_{S, \mathrm{max}}$ is the maximum value of $V_S$ for a case, which is $A_S$ multiplied by $\omega_S$. The loops go in the clockwise direction with respect to time advancement, as indicated by the arrows in the plot. The hollow markers indicate the quasi-steady predictions for $<C_T>$ with certain $V_S$, which corresponds to the values of $\overline{C}_T$ for case \textbf{1} or the auxiliary cases in Appendix~\ref{app:auxiliary}. The auxiliary cases are simulated with different $V_0$, laminar inflow conditions, and a fixed rotor. The numbers enclosed by parentheses correspond to the case numbers in Table~\ref{tab:cases_Single_rotor} or \ref{tab:aux_cases_Single_rotor}.}
\label{fig:paper_1_ampFre_concyc_CT}
\end{figure}

In Figures~\ref{fig:paper_1_ampFre_concyc_CT}(a1) and \ref{fig:paper_1_ampFre_concyc_CT}(a2), $<C_T>$ for the twelve cases in \textbf{Group}~$\boldsymbol{b}$ is shown. These figures clearly demonstrate that $<C_T>$ fluctuates throughout the surging cycle in the surging cases, with larger $v_{\mathrm{max}}$ leading to more pronounced variations. Furthermore, $<C_T>$ is observed to be close to identical in both laminar and turbulent cases with identical surging settings, reaffirming that cycle-averaged rotor performance is almost unaffected by inflow turbulence. Notably, an upper limit for $<C_T>$ is observed, and $<C_T>$ starts to decrease after reaching this limit during a surging cycle even though the apparent inflow velocity seen by the rotor ($V_{0, \mathrm{app}}$) continues to increase. This is particularly in the cases with higher $v_{\mathrm{max}}$ ($v_{\mathrm{max}} = 0.44$). This pattern suggests the occurrence of (static) stalling, a phenomenon that will be confirmed and discussed in more detail in Section~\ref{sec:AOA}.

Further observing the curves of $<C_T>$ in Figures~\ref{fig:paper_1_ampFre_concyc_CT}(a1) and \ref{fig:paper_1_ampFre_concyc_CT}(a2), it is found that they are almost in sync with the surging $V_S$, where phase differences between $<C_T>$ and $V_S$ are minimal (note that $V_S = V_0v_{\mathrm{max}} \cos \phi_S$, see Equation~\ref{eq:VandW}). This finding is consistent with most of the existing literature \cite{tran2016cfd, arabgolarcheh2022modeling, ferreira2022dynamic}. However, subtle hysteresis (unsteady aerodynamic) effects are observed, especially when comparing the results of the two surging settings with $v_{\mathrm{max}} = 0.44$. To further explore this hysteresis phenomenon, we plotted $<C_T>$ against $V_{S}$ for the laminar cases in \textbf{Group}$~\boldsymbol{b}$ with the forms of hysteresis loops in Figure\ref{fig:paper_1_ampFre_concyc_CT}(b). Moreover, this figure is labeled with hollow markers indicating the values of quasi-steady predictions (QSP) for $<C_T>$. These QSP are based on values of $\overline{C}_T$ from case \textbf{1} in Table~\ref{tab:cases_Single_rotor} and the auxiliary cases (\textbf{A1}-\textbf{A6}) documented in Appendix~\ref{app:auxiliary}. The auxiliary cases are additional LES-ALM simulations with the setups same as case \textbf{1} (fixed-laminar) but with several different inflow velocities ($V_0$) differ from $V_{0, \mathrm{rated}}$. Specifically, the values of these $V_0$ are set to $(1 \pm v_{\mathrm{max}})V_{0, \mathrm{rated}}$ with $v_{\mathrm{max}}$ being $0.11$, $0.22$, or $0.44$, matching the maximum and minimum $V_{0, \mathrm{app}}$ of a surging FOWT rotor would encounter for the cases in Table~\ref{tab:cases_Single_rotor}.

In Figure~\ref{fig:paper_1_ampFre_concyc_CT}(b), the markers of QSP effectively outline the general shapes and the fluctuation amplitudes of the $<C_T>$ loops. However, subtle hysteresis effects are observed, especially prominent around $V_S = 0.0$~m/s, when a surging FOWT rotor experiences maximum acceleration. These hysteresis effects become more pronounced with larger $v_{\mathrm{max}}$ and $k_D$. And for the cases with the same values of $v_{\mathrm{max}}$, stronger hysteresis effects are observed with the cases having higher $k_D$. This observation highlights that the aerodynamic behavior of FOWTs under harmonic surging cannot be solely attributed to $v_{\mathrm{max}}$. Instead, both $v_{\mathrm{max}}$ and $k_D$ influences the dynamics. To the best of our knowledge, we are the first to explicitly show that the hysteresis loading ($<C_T>$) of an (uncontrolled) surging FOWT depends on both $v_{\mathrm{max}}$ and $k_D$.

Further examining the values of $<C_T>$ as the curves pass through the point where $V_S = 0.0$~m/s, we notice that the neutral value (values for the fixed case, marked by a star) does not fall consistently inside the $<C_T>$ loops. Our current study, based on the test matrix used, does not completely explain the observed phenomena. However, several hypotheses are proposed to explain the behavior in order to provide foundations for future research. Specifically, we believe that this behavior stems from the interplay between the effects of rotor-level dynamic inflow and blade-level unsteady aerodynamics \cite{leishman2002challenges}. As demonstrated by de Vaal et al. \cite{de2014effect}, the rotor-level dynamic inflow effect causes $<C_T>$ to lead ahead of $V_S$ due to axial induction lagging. In contrast, blade-level unsteady aerodynamic effects, as demonstrated by Wen et al. \cite{wen2018power} (referred to as blade-wake interaction by them), result in $<C_T>$ lagging behind $V_S$ due to the presence of shed vortices, which delay the changes of $\alpha$ at the blade (actuator line) elements. Note that the two mentioned effects have different characteristic timescales and respond nonlinearly to $V_{0, \mathrm{app}}$, $\mathrm{d} V_{0, \mathrm{app}}/\mathrm{d}t$, and the evolution history of $V_{0, \mathrm{app}}$ \cite{leishman2002challenges}, adding layers of complexity to the already intricate system. In view of this, to further understand the complex dynamics at play, future research with dedicated efforts is recommended.

\subsection{Angle of attack}

\label{sec:AOA}

The main objective of this subsection is to confirm that the dips of the $<C_T>$ curves in Figure~\ref{fig:paper_1_ampFre_concyc_CT} are due to stalling and to pave the way for the analysis for $\overline{C}_T$ and $\overline{C}_P$ in the next subsection. Thus, even though a similar analysis has already been carried out in previous work \cite{rezaeiha2021wake, ramos2022investigationI}, the analysis of angle of attack $\alpha$ is still presented both for the reasons mentioned and for the completeness of the current study.

The stalling effects in this work are modeled using the static airfoil polar provided in the report of NREL~5MW turbine \cite{jonkman2009definition}. That is, \emph{stalling} in this context specifically refers to \emph{static stalling}. Additional dynamic stall models are not implemented, as the chord-based reduced frequencies based on the surging motions mainly considered ($k_c = 0.5 \hspace{1pt} c \hspace{1pt} \omega_S / V_{\mathrm{rel}}$) are generally very low for the outward blade span. Specifically, as $\omega_S = 0.63$~rad/s, $k_c < 0.05$ is satisfied from $r/R > 0.38$ to the blade tip, which is sufficiently low to neglect the dynamic effects \cite{leishman2002challenges, leishman2006principles}. Despite the absence of a dynamic stall model, most of the effects of unsteady rotor aerodynamics, excluding the effects of the airfoil boundary layer, are resolved explicitly using ALM with LES.

As the apparent inflow velocity, $V_{0, \mathrm{app}}$, is sufficiently high to cause $\alpha$ of a blade section to surpass the stalling angle $\alpha_{\mathrm{stall}}$ (see Equation~\ref{eq:f_2D_mod}), that section will experience stalling. Generally speaking, stalling will result in decreasing the lift coefficient ($C_l$) and increasing the drag coefficient ($C_d$) of a blade section, consequently reducing $<C_T>$ and $<C_P>$ and eventually lowering the values of $\overline{C}_T$ and $\overline{C}_P$. It should be noted that, in this study, stalling is modeled via the input airfoil polar data within the ALM framework. Therefore, only changes in $C_l$ and $C_d$ are considered, without accounting for additional complexities such as enhanced turbulence due to the boundary layer separation and the formation of leading-edge vortices.

\begin{figure}[t]
\centering
\includegraphics[width=430pt]{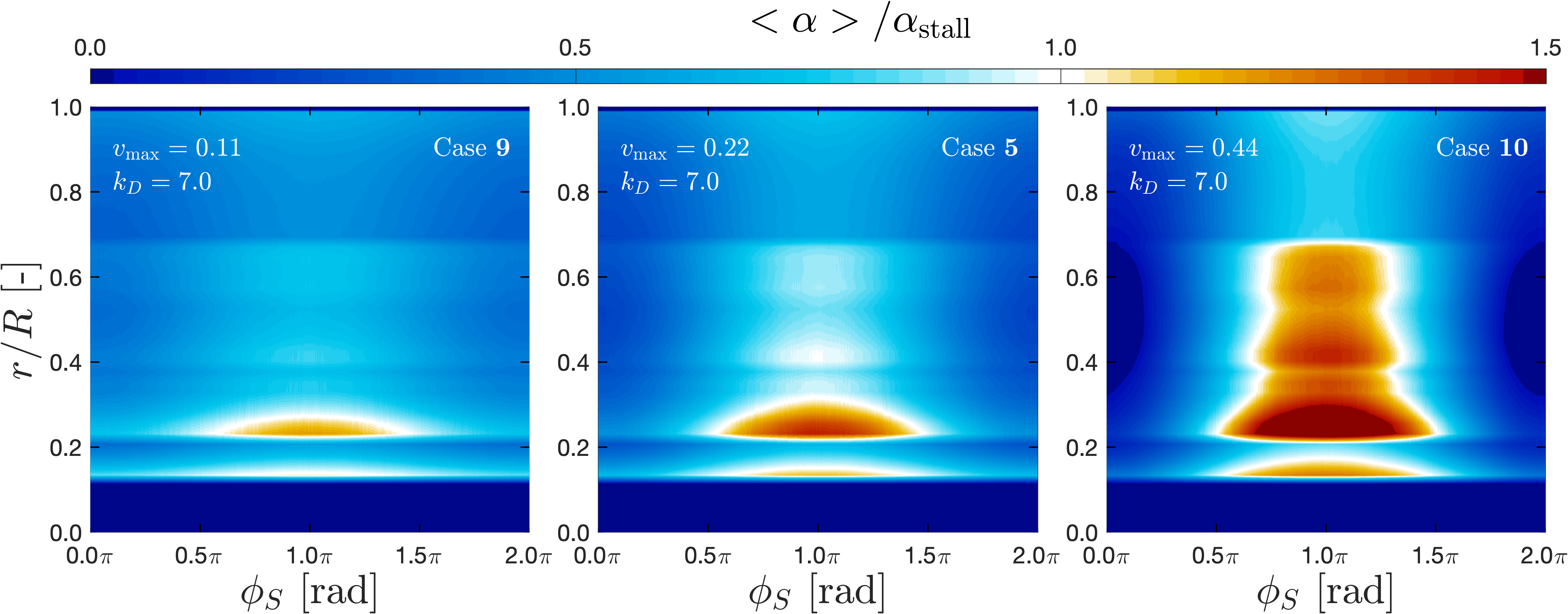}
\caption{Cycle-averaged angle of attack $<\alpha>$ for cases with different surging settings under laminar inflow conditions. $\alpha_{\mathrm{stall}}$ refers to the angle of attack at which stalling starts to occur, and its profile for NREL~5MW baseline turbine is in Appendix~\ref{app:alpha_stall}.}
\label{fig:AOA_all}
\end{figure}

Figure~\ref{fig:AOA_all} illustrates the cycle-averaged angle of attack, $<\alpha>$, along the blade during a surging cycle based on $\phi_S$. The figure includes three surging cases subjected to laminar inflow conditions with varying values of $v_{\mathrm{max}}$ but identical $k_D$ (cases \textbf{5}, \textbf{9}, and \textbf{10}). The general behavior is depicted by presenting these cases, as $\alpha$ during a surging cycle is predominantly influenced by $v_{\mathrm{max}}$ rather than $k_D$ or inflow conditions. The fixed cases are not shown as their $<\alpha>$ values are constant, and stalling is not observed. In Figure~\ref{fig:AOA_all}, $\alpha_{\mathrm{stall}}$ refers to the angle of attack at which the (first local) maximum of $C_l$ is reached, and the profile of $\alpha_{\mathrm{stall}}$ for NREL~5MW along its blade span can be found in Appendix~\ref{app:alpha_stall}. Note that the variation in airfoil geometries along the blade span leads to abrupt changes in $<\alpha>/\alpha_{\mathrm{stall}}$ along the blade, as seen in the figure. The values of $<\alpha>$ are observed to be larger around $\phi_S = 1.0\pi$, coinciding when $V_{0, \mathrm{app}}$ are higher. Moreover, the variability of $<\alpha>$ increases with larger $v_{\mathrm{max}}$, indicating more severe stalling in cases with higher $v_{\mathrm{max}}$. Furthermore, this analysis confirms that stalling predominantly occurs near the blade root, as expected, because varying $V_{n, \mathrm{app}}$ has more profound impacts on inflow angle $\phi$ close to the root as $\Omega r$ is smaller there (see Equation~\ref{eq:f_2D_mod}). The alignment of the stalling timings with the dips of $<C_T>$ for the surging cases with $v_{\mathrm{max}} = 0.44$ at $\phi_S = 1.0\pi$ (as shown in Figure~\ref{fig:paper_1_ampFre_concyc_CT}(a1) and \ref{fig:paper_1_ampFre_concyc_CT}(a2)) solidifies the statement that (static) stalling is responsible for these reductions.

\subsection{Discussions about the curves of \texorpdfstring{$<C_T>$}{text} and \texorpdfstring{$<C_P>$}{text}}

\label{sec:Exp_act_CT}

While $<C_T>$ and $<C_P>$ for the surging cases that do not experience severe stalling (when $v_{\mathrm{max}} \leq 0.22$) may superficially appear to follow simple harmonic curves, as shown in Section~\ref{sec:CT_CP_time_Avg}, their actual dynamics are much more complex, preventing straightforward analytical interpretations. However, this complexity makes the surging cases simultaneously exhibit lower $\overline{C}_T$ and higher $\overline{C}_P$ compared to the fixed cases. Note that both lower $\overline{C}_T$ and higher $\overline{C}_P$ are desirable for a wind turbine rotor \cite{porte2020wind}, while achieving the two at the same time is contrary to common perception. This trend has also been previously found in Rezaeiha et al. \cite{rezaeiha2021wake} and Chen et al. \cite{chen2022modelling}, but no previous work has explicitly addressed this particular finding. Here, analytical expressions are derived for how $<C_T>$ and $<C_P>$ would vary against the surging motions under certain assumptions. We also examine how $<C_T>$ and $<C_P>$ are influenced by surging motions throughout the entire surging cycle based on the CFD simulations and seek to elucidate the interplay of the surging velocity $V_S$, apparent normal velocity $V_{n, \mathrm{app}}$, relative velocity $V_{\mathrm{rel}}$, inflow angle $\phi$, and angle of attack $\alpha$ (see Figure~\ref{fig:Vel_tri_paper}).

\paragraph{Analytical derivation of \texorpdfstring{$<C_T>$}{text} and \texorpdfstring{$<C_P>$}{text} under surging motion}

To initiate the analysis, we introduce the apparent thrust and power coefficients, denoted by $C_{T}^{\mathrm{app}}$ and $C_{P}^{\mathrm{app}}$, respectively. These coefficients are linked to the instantaneous rotor thrust ($T$) and power ($P$) through the apparent inflow velocity seen by the rotor ($V_{0, \mathrm{app}}$), as defined in Equation~\ref{eq:def_ins_ct_cp}. 

\begin{equation}
    C_{T}^{\mathrm{app}} \overset{\Delta}{=} \frac{T}{\frac{1}{2} \rho V_{0, \mathrm{app}}^2 \pi R^2} = C_T \left( \frac{ V_{0, \mathrm{rated}}^2}{V_{0, \mathrm{app}}^2} \right)
    , \qquad   
    C_{P}^{\mathrm{app}} \overset{\Delta}{=} \frac{P}{\frac{1}{2} \rho V_{0, \mathrm{app}}^3 \pi R^2} = C_P \left( \frac{ V_{0, \mathrm{rated}}^3}{V_{0, \mathrm{app}}^3} \right)
    \label{eq:def_ins_ct_cp}
\end{equation}

If $C_{T}^{\mathrm{app}}$ and $C_{P}^{\mathrm{app}}$ match the values of $\overline{C}_T$ and $\overline{C}_P$ for the fixed cases and were constant throughout the surging cycle, the corresponding $\overline{C}_T$ and $\overline{C}_P$ for the surging cases would be higher than those of the fixed rotor case, as analyzed by Johlas et al. \cite{johlas2021floating} (see Equations~\ref{eq:const_CT} and \ref{eq:const_CP}, where $T_S$ denotes the period of harmonic surging motion). Furthermore, the hypothesis states that for surging cases with a larger $v_{\mathrm{max}}$, the values of $\overline{C}_T$ and $\overline{C}_P$ will be more elevated. However, this hypothesis does not align entirely with our observed data, as illustrated in Figure~\ref{fig:singleHis_wide}. In our study, $\overline{C}_T$ for all the surging cases are found to be lower than those for the fixed cases, and the cases with larger $v_{\mathrm{max}}$ have further lower $\overline{C}_T$, contradicting the predictions of Equation~\ref{eq:const_CT}. This discrepancy implies that the assumption of $C_{T}^{\mathrm{app}}$ and $C_{P}^{\mathrm{app}}$ being constants throughout the surging cycle does not hold. Therefore, the procedures for calculating $f_n$ and $f_{\theta}$ at actuator line points (as outlined in Equations~\ref{eq:f_2D} and \ref{eq:f_2D_mod}) should be closely reviewed. Nevertheless, as $V_{0, \mathrm{app}}$ are influenced by surging motions in the simulations, the operational conditions (such as $\Omega$) do not adjust accordingly to maintain the tip speed ratio $\lambda$, which causes the $\lambda$ of the rotor to drift away from $\lambda_{\mathrm{rated}}$, and thus affects the values of $C_{T}^{\mathrm{app}}$ and $C_{P}^{\mathrm{app}}$ \cite{manwell2010wind}.

\begin{gather}
    \mathrm{For}~v_{\mathrm{max}} > 0~\mathrm{:} \qquad
    \overline{C}_T
    = \frac{1}{T_S}\int_0^{T_S} \underbrace{ C_{T}^{\mathrm{app}} \left( \frac{V_{0, \mathrm{app}}^2}{ V_{0, \mathrm{rated}}^2} \right) }_{C_T} \hspace{2pt} \mathrm{d}t 
    = 
    \frac{1}{T_S}\int_0^{T_S}  C_{T}^{\mathrm{app}} (1 - v_{\mathrm{max}}\cos \omega_S t)^2 \hspace{2pt} \mathrm{d}t
    \label{eq:int_1PlusEta}\\
    \mathrm{If}~C_{T}^{\mathrm{app}}~\mathrm{is~constant:} \qquad
    \overline{C}_T = C_{T}^{\mathrm{app}} \left( 1 + \frac{v_{\mathrm{max}}^2}{2} \right)
    {\huge >} \hspace{5pt} C_{T}^{\mathrm{app}}
    \label{eq:const_CT}
\end{gather}

\begin{gather}
    \mathrm{For}~v_{\mathrm{max}} > 0~\mathrm{:} \qquad
    \overline{C}_P
    = \frac{1}{T_S}\int_0^{T_S} \underbrace{ C_{P}^{\mathrm{app}} \left( \frac{V_{0, \mathrm{app}}^3}{ V_{0, \mathrm{rated}}^3} \right)  }_{C_P}\hspace{2pt} \mathrm{d}t 
    = 
    \frac{1}{T_S}\int_0^{T_S} C_{P}^{\mathrm{app}} (1 - v_{\mathrm{max}}\cos \omega_S t)^3 \hspace{2pt} \mathrm{d}t
    \label{eq:int_1PlusEta2}\\
    \mathrm{If}~C_{P}^{\mathrm{app}}~\mathrm{is~constant:} \qquad
    \overline{C}_P = C_{P}^{\mathrm{app}} \left(1 + \frac{3 v_{\mathrm{max}}^2}{2} \right) 
    {\huge >} \hspace{5pt} C_{P}^{\mathrm{app}}
    \label{eq:const_CP}
\end{gather}

\paragraph{Actual behaviors of \texorpdfstring{$<C_T>$}{text} and \texorpdfstring{$<C_P>$}{text} under surging motion based on the simulation results}

Here, we analyze how $T$ and $P$ are influenced by the surging motion based on the setups and the results of the simulations. For a surging rotor, an increase in $V_{n, \mathrm{app}}$ leads to a larger $\phi$, which in turn results in a greater $\alpha$, as indicated in Equation~\ref{eq:f_2D_mod}. A combination of higher $V_{n, \mathrm{app}}$, and larger $\alpha$ would result in a stronger lift $\boldsymbol{L}$ therefore a higher thrust $T$ and power $P$ if severe stalling does not occur. 
However, changes in $V_{n, \mathrm{app}}$ will also alter $\phi$, therefore the lift force vector, as described in Equation~\ref{eq:phiandCt}. Basic trigonometry shows that an increase in $\phi$ reduces the normal component contributing to $T$ through the $\cos{\phi}$ term. On the other hand, an increase in $\phi$ enhances $\sin{\phi}$, thus increasing the tangential component contributing to $P$. Consequently, when $V_{n, \mathrm{app}} > V_{{n}}$ (e.g., $V_{0, \mathrm{app}} > V_{{0}}$), the increase of $T$ is less aggressive than the increase of $P$, since vector $\boldsymbol{L}$ is projected less to the normal component but more to the tangential component. On the contrary, when $V_{n, \mathrm{app}} < V_{{n}}$, the thrust force decreases less aggressively than $P$, since the lift force vector is projected more to the normal component. Although these effects may initially appear to offset each other for a complete surging cycle, the simulation results suggest that the period when $V_{n, \mathrm{app}}$ exceeds $V_{{n}}$ plays a more significant role in time-averaging $T$ and $P$. This is indicated in Figure~\ref{fig:paper_1_TIcomparecyc_CTCPArea}, which shows that $H^+$ for $<C_T>$ is smaller than that of $H^-$, while $H^+$ for $<C_P>$ is larger than its $H^-$. This is related to the fact that the magnitude of $\boldsymbol{L}$ has a quadratic relation to $V_{n, \mathrm{app}}$, as shown in Equations~\ref{eq:f_2D} and \ref{eq:f_2D_mod}. Up to this point, the reason for a surging rotor to simultaneously have lower $\overline{C}_T$ and higher $\overline{C}_P$ compared to a fixed rotor has been explained.

\begin{equation}
    \boldsymbol{L}  = \lvert \boldsymbol{L} \rvert \cos{\phi} \hspace{3pt} \hat{\boldsymbol{e}}_{n} + \lvert \boldsymbol{L} \rvert \sin{\phi} \hspace{3pt}  \hat{\boldsymbol{e}}_{\theta},
    \qquad
    \mathrm{where:} \hspace{8pt}
    \phi = \arctan \left( \frac{V_{n, \mathrm{app}}}{- \Omega r + V_\theta} \right)
    \label{eq:phiandCt}
\end{equation}

For simulations that experience severe stalling (the cases with $v_{\mathrm{max}} = 0.44$), the magnitude of the lift force, $\lvert \boldsymbol{L} \rvert$, is reduced based on the inputted airfoil polar, leading to a decrease in both $T$ and $P$, and thus lowering both $\overline{C}_T$ and $\overline{C}_P$. Note that the values of $\overline{C}_T$ and $\overline{C}_P$ for simulations that experience severe stalling is the lowest among all the cases considered in this study, as demonstrated in Figure~\ref{fig:singleHis_wide}. Therefore, we conclude that the surging motion could favor both the thrust and power if severe stalling does not occur.


\section{Result and discussion on wake characteristics and wake structures}

This section explores the wake characteristics and wake structures of the simulation cases listed in Table~\ref{tab:cases_Single_rotor}. In Section~\ref{sec:wake_vel}, we present and discuss the results of the time-averaged streamwise velocity $\overline{u}$ and its disk-averaged $\overline{u}_{\mathrm{Disk}}$ (the latter being the area-averaged of $\overline{u}$ within a radius $r<R$) to outline the general wake characteristics. The detailed analysis of $\overline{u}$ profiles at specific $x$-positions is confined to the cases in \textbf{Group}~$\boldsymbol{a}$, since the cases in \textbf{Group}~$\boldsymbol{b}$ exhibit broadly similar characteristics. Conversely, the examination of $\overline{u}_{\mathrm{Disk}}$ encompasses all the 16 cases listed in Table~\ref{tab:cases_Single_rotor}, which provides insights into the effects of inflow turbulence intensity (TI), surging settings, and their interplay on wake recovery. Analysis of $\overline{u}_{\mathrm{Disk}}$ is divided into two parts based on \textbf{Group}~$\boldsymbol{a}$ and \textbf{Group}~$\boldsymbol{b}$, paralleling the approach taken for the analysis of ${C}_T$ and ${C}_P$ in Section~\ref{sec:CT_CP_time_Avg}. Subsequently, in Section~\ref{sec:wake_structure}, wake structures are depicted through contour plots of selected field quantities for the four cases in \textbf{Group}~$\boldsymbol{c}$, where surge-induced periodic coherent structures (SIPeCS) are identified. The general impacts of inflow TI, surging settings, and their interplay on wake structures are presented and discussed based on the contour plots. Next, Section~\ref{sec:wake_SIPeCS} delves into SIPeCS, where they are analyzed in detail with qualitative and quantitative analysis. Significantly, the findings in this subsection imply that the primary mechanisms that drive faster wake (energy) recoveries in surging cases are attributed to the enhanced advection process. This discovery challenges long-believed hypotheses. Traditionally, it has been suggested that the faster wake recovery of surging FOWT rotors is attributed to enhanced turbulent mixing, and this mixing was thought to stem from the quicker wake breakdown triggered by the instabilities introduced by the surging motion \cite{micallef2021floating, chen2022modelling, kleine2022stability, ramos2022investigationI}. Lastly, to provide evidence that could challenge the common belief mentioned before, a term-by-term analysis of the flow kinetic energy entrains into the wakes is conducted in Section~\ref{sec:phase_energy} with the cases of \textbf{Group}~$\boldsymbol{c}$. The findings in the subsection support the claim, showing that the enhanced advection process is the main cause of the faster wake recovery for the surging FOWT.

\subsection{General wake characteristics with time-averaged streamwise velocity profiles}

\label{sec:wake_vel}

\subsubsection{Time-averaged streamwise velocity profiles \texorpdfstring{$\overline{u}$}{text}}

\begin{figure}[t]
\centering
\includegraphics[width=430pt]{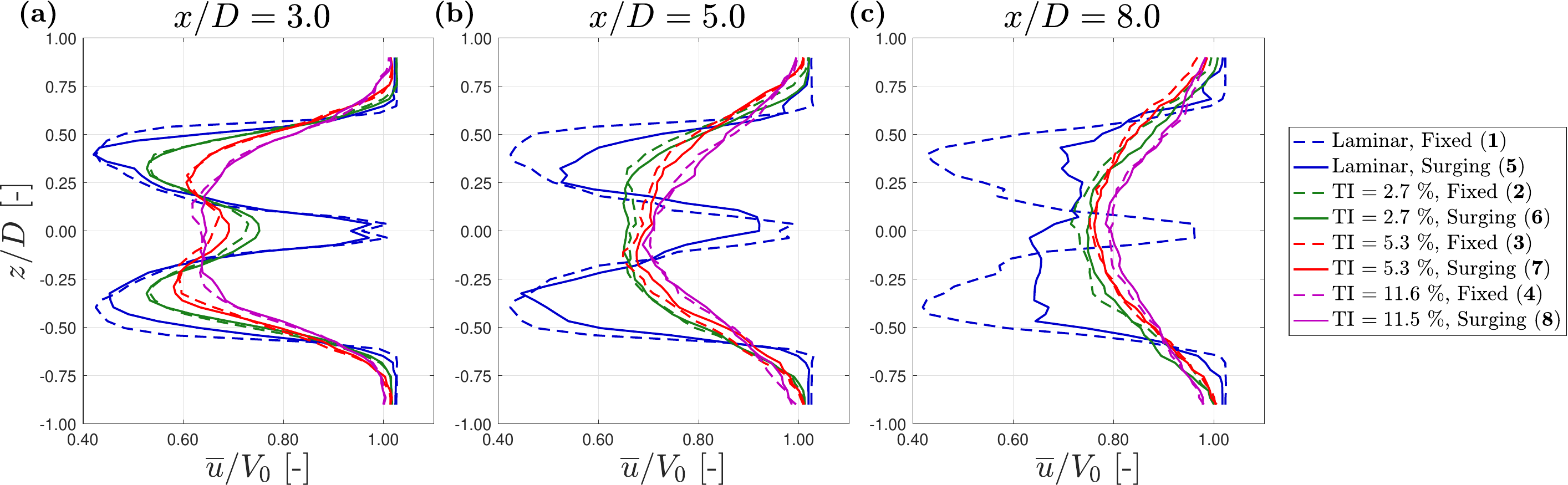}
\caption{Profiles of $\overline{u}$ at different $x/D$ for fixed or surging rotor with different inflow TI. (a) $x/D = 3$, (b) $x/D = 5$, (c) $x/D = 8$. The case numbers are labeled in parentheses.}
\label{fig:paper_1_TIcompare_UmeanProfile_}
\end{figure}

The time-averaged streamwise velocity profiles, denoted as $\overline{u}$, for the eight cases within \textbf{Group}~$\boldsymbol{a}$ are plotted at several downstream sections in Figure~\ref{fig:paper_1_TIcompare_UmeanProfile_}. The analysis begins by examining the two laminar cases (cases \textbf{1} and \textbf{5}), specifically to highlight the impact of surging motions under laminar inflow conditions. It is then followed by an evaluation of the turbulent cases (cases \textbf{2}-\textbf{4}, and \textbf{6}-\textbf{8}) to explore the impact of surging motions with the presence of inflow turbulence. The analysis also includes a comparative study between the laminar and turbulent cases, aiming to understand the combined effects of inflow conditions and surging settings on profiles of $\overline{u}$.

The analysis indicates that the two laminar cases, one with a fixed rotor and the other with a surging one, display relatively similar $\overline{u}$ profiles at $x/D = 3$. However, there are notable contrasts at $x/D = 8$. At this downstream point, the $\overline{u}$ profile for the surging-laminar case exhibits significant recovery, diverging markedly from the fixed-laminar case, which continues to show a profile similar to those at upstream locations.

Moving to the other six turbulent cases, similar to the previous research \cite{troldborg2009actuator, sarlak2014large}, it is clear that these cases exhibit more pronounced recoveries in their $\overline{u}$ profiles compared to the laminar cases, and the cases having higher inflow TI recover faster. Furthermore, the $\overline{u}$ profiles of the fixed and surging cases with identical inflow TI are very similar for all downstream sections, and all $\overline{u}$ profiles eventually appear in Gaussian shapes by $x/D = 8$. This similarity suggests that the wake aerodynamics of the turbulent cases are less affected by the surging motion than that of the laminar cases. However, a closer inspection of Figure~\ref{fig:paper_1_TIcompare_UmeanProfile_} reveals that the $\overline{u}$ profiles for the surging cases (cases \textbf{6}-\textbf{8}), in general, are marginally higher than those for the fixed cases (cases \textbf{2}-\textbf{4}). Nevertheless, these differences are less pronounced than those attributed to changing inflow TI. This observation suggests that when subjected to turbulent inflow conditions, the strengths of inflow TI dominate in shaping the $\overline{u}$ profiles, and the effects of surging motion are much reduced by the ambient turbulence.

\subsubsection{Mean disk-averaged streamwise velocity \texorpdfstring{$\overline{u}_{\mathrm{Disk}}$}{text}}

\paragraph{Impact of surging effects with various inflow TI}

\begin{figure}[t]
\centering
\includegraphics[width=430pt]{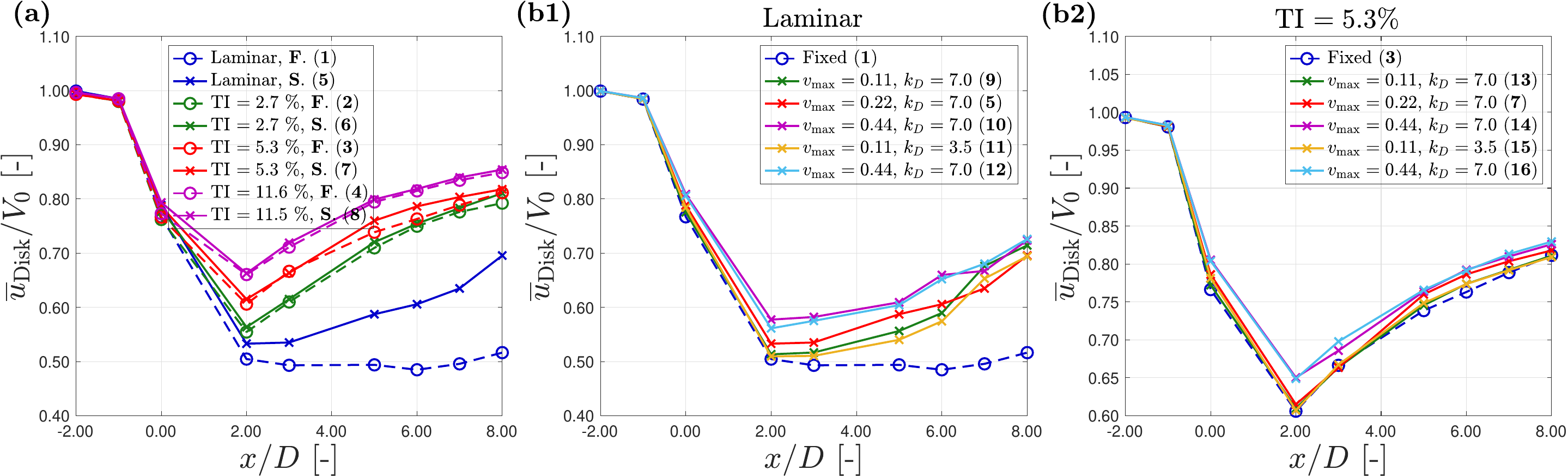}
\caption{$\overline{u}_{\mathrm{Disk}}$ along $x$-direction for cases in Table~\ref{tab:cases_Single_rotor}. (a): Cases of \textbf{Group}~$\boldsymbol{a}$, where \textbf{F.} and \textbf{S.} appear in the labels that stand for fixed or surging. (b1): Cases of \textbf{Group}~$\boldsymbol{b}$ with laminar inflow conditions. (b2): Cases of \textbf{Group}~$\boldsymbol{b}$ with turbulent inflow conditions (TI $= 5.3\%$). The case numbers are labeled in parentheses.}
\label{fig:paper_1_v2_ampFre_con_UAreaAvg}
\end{figure}

Figure~\ref{fig:paper_1_v2_ampFre_con_UAreaAvg}(a) displays the profiles of the mean disk-averaged streamwise velocity, $\overline{u}_{\mathrm{Disk}}$, plotted along the $x$-direction for the eight cases in \textbf{Group}~$\boldsymbol{a}$. The analysis begins with the two laminar cases, followed by the other six turbulent cases, and concludes with a comprehensive investigation encompassing the eight cases together.

First, the comparison between the two laminar cases, one with a surging rotor (case \textbf{5}) and the other with a fixed rotor (case \textbf{1}), reveals profound discrepancies in the $\overline{u}_{\mathrm{Disk}}$ profiles. For the surging case, $\overline{u}_{\mathrm{Disk}}$ increases with larger $x/D$. While for the fixed case, $\overline{u}_{\mathrm{Disk}}$ remains relatively constant. Consequently, the value of $\overline{u}_{\mathrm{Disk}}$ at $x/D = 8$ for the surging case is significantly higher compared to that for the fixed case, and this is consistent with the analysis of the $\overline{u}$ profiles performed previously. The results here indicate that surging motion triggers substantial wake recovery when inflow turbulence is absent, contrasting with the minimal recovery observed in the fixed-laminar case.

Next, the other six turbulent cases are analyzed, covering both the fixed cases (cases \textbf{2}-\textbf{4}) and the surging cases (cases \textbf{6}-\textbf{8}). All these six cases demonstrate more pronounced recoveries in their $\overline{u}_{\mathrm{Disk}}$ profiles compared to their laminar counterparts. Additionally, cases with higher inflow TI demonstrate faster recovery rates, corroborating previous studies \cite{troldborg2009actuator, sarlak2014large}. Furthermore, the $\overline{u}_{\mathrm{Disk}}$ profiles of the fixed and surging cases subjected to turbulent inflow conditions become very similar when sharing the same inflow TI. This observation is in sharp contrast to the laminar cases.

The relative differences of $\overline{u}_{\mathrm{Disk}}$ between the surging and fixed cases for different inflow conditions are given in Table~\ref{tab:crossCompare_TI_SF}. When the inflow conditions are laminar, it can be seen that $\overline{u}_{\mathrm{Disk}}$ for the surging case at $x/D = 5$ and $8$ are $18.8\%$ and $34.7\%$ higher than the fixed case. However, these gains are substantially reduced as the inflow conditions are switched to turbulent, where the gains in $\overline{u}_{\mathrm{Disk}}$ for the surging cases are down to merely around $0.5 \sim 2\%$ at $x/D = 5$ and $8$. Although the gains in $\overline{u}_{\mathrm{Disk}}$ due to surging are relatively mild for turbulent cases, surging cases already exhibit higher values for $\overline{C}_P$ (see Figure~\ref{fig:singleHis_wide}). Notice that both the larger $\overline{u}_{\mathrm{Disk}}$ and the higher $\overline{C}_P$ are advantageous for enhancing the overall power output at the wind farm level.

\begin{table}[htb!]
\centering
\caption{The ratios of $\overline{u}_{\mathrm{Disk}}$ for the surging cases over the fixed cases when subjected to different inflow TI at $x/D = 3$, $5$, and $8$.}
\label{tab:crossCompare_TI_SF}
\adjustbox{max width=\textwidth}{%
\begin{tabular}{r|rrrr}

$\overline{u}_{\mathrm{Disk, surging}}/\overline{u}_{\mathrm{Disk, fixed}}$ & Laminar & TI $= 2.7\%$ & TI $= 5.3\%$ & TI $= 11.6\%$  \\
\hline

$x/D = 3$ & $108.52\%$ & $100.82\%$ & $99.70\%$ & $101.13\%$\\

$x/D = 5$ & $118.83\%$ & $101.41\%$ & $102.84\%$ & $100.63\%$\\

$x/D = 8$ & $134.70\%$ & $102.31\%$ & $100.80\%$ & $100.64\%$\\

\hline
\end{tabular}
}
\end{table}

\paragraph{Impact of surging settings under laminar or turbulent inflow conditions}

$\overline{u}_{\mathrm{Disk}}$ for the twelve cases in \textbf{Group}~$\boldsymbol{b}$ are plotted with Figure~\ref{fig:paper_1_v2_ampFre_con_UAreaAvg}(b1) and Figure~\ref{fig:paper_1_v2_ampFre_con_UAreaAvg}(b2). Specifically, Figure~\ref{fig:paper_1_v2_ampFre_con_UAreaAvg}(b1) plots the six laminar cases (cases \textbf{1}, \textbf{5}, and \textbf{9}-\textbf{12}) and Figure~\ref{fig:paper_1_v2_ampFre_con_UAreaAvg}(b2) plots the six turbulent cases with inflow TI $= 5.3\%$ (cases \textbf{3}, \textbf{7}, and \textbf{13}-\textbf{16}).

In the analysis of the six laminar cases presented in Figure~\ref{fig:paper_1_v2_ampFre_con_UAreaAvg}(b1), it is evident that surging motions facilitate the recovery of $\overline{u}_{\mathrm{Disk}}$. Specifically, cases with a larger $v_{\mathrm{max}}$ exhibit higher $\overline{u}_{\mathrm{Disk}}$ up till $x/D = 6$. This trend suggests a correlation between the $\overline{u}_{\mathrm{Disk}}$ profiles and $v_{\mathrm{max}}$, with larger values of $v_{\mathrm{max}}$ leading to increases in $\overline{u}_{\mathrm{Disk}}$. Interestingly, beyond $x/D \geq 7$, the values of $\overline{u}_{\mathrm{Disk}}$ for these cases appear to converge towards a similar level, a phenomenon also observed and reported by Chen et al. \cite{chen2022modelling}.

In the analysis of the six turbulent cases shown in Figure~\ref{fig:paper_1_v2_ampFre_con_UAreaAvg}(b2), the differences in $\overline{u}_{\mathrm{Disk}}$ between the fixed and surging cases are notably less pronounced compared to the cases subjected to laminar inflow conditions. However, it is observed that values of $\overline{u}_{\mathrm{Disk}}$ are slightly higher in the surging cases than in the fixed cases. Furthermore, there is a tendency for $\overline{u}_{\mathrm{Disk}}$ to be higher with larger $v_{\mathrm{max}}$, indicating that cases with larger $v_{\mathrm{max}}$ have faster wake recovery.

\subsection{Wake structure with contour plots of velocity and TKE fields}

\label{sec:wake_structure}

This subsection provides a comprehensive overview of the impacts of inflow conditions, surging settings, and their interplay on wake structures with contour plots of the selected fields. To maintain focus and clarity, the analysis is confined to four representative cases in \textbf{Group}~$\boldsymbol{c}$, which are cases \textbf{1}, \textbf{3}, \textbf{5}, and \textbf{7}. These four cases represent the four pivotal dimensions of this study, namely fixed-laminar, surging-laminar, fixed-turbulent, and surging-turbulent. The key features discerned from these cases indicate general trends that apply to all 16 cases listed in Table~\ref{tab:cases_Single_rotor}. The fields selected for detailed examination include instantaneous streamwise velocity ($u$), phase-averaged streamwise velocity ($<u>_{0\pi}$), and phase-averaged turbulent kinetic energy ($<$TKE$>_{0\pi}$). By examining these cases and fields, surge-induced periodic coherent structures (SIPeCS) can be identified for cases subjected to both laminar and turbulent inflow conditions.

\paragraph{Instantaneous streamwise velocity fields $u$}

\begin{figure}[t]
\centering
\includegraphics[width=430pt]{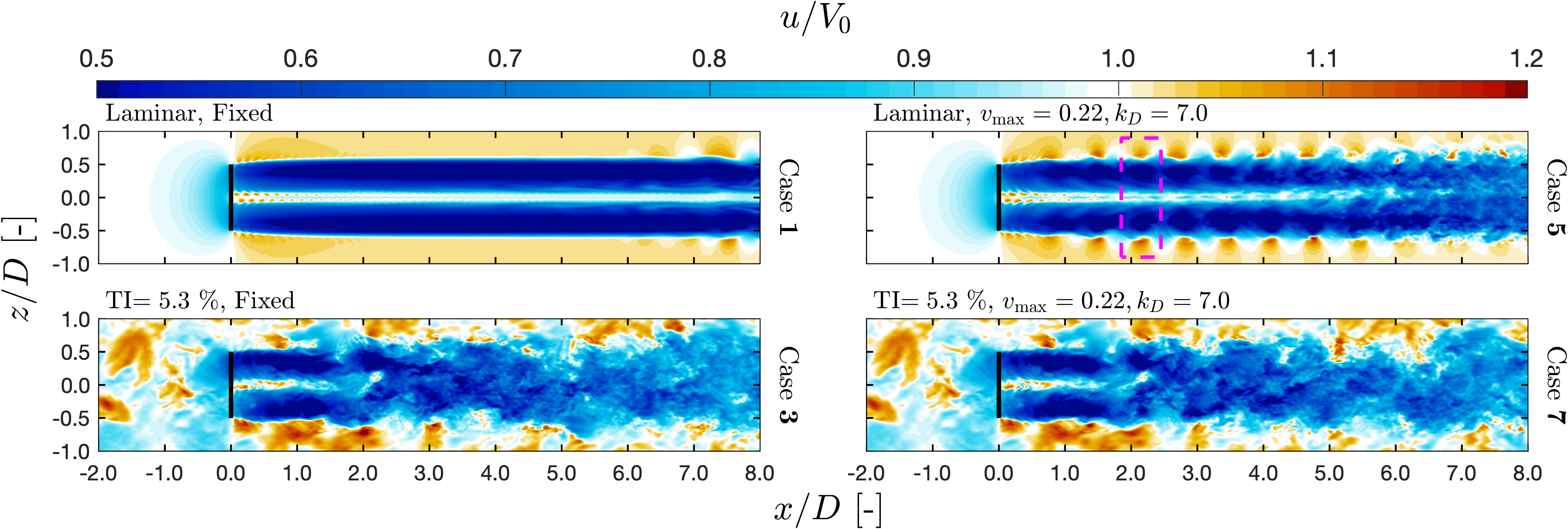}
\caption{Contour plots of instantaneous streamwise velocity fields $u$ for the cases in \textbf{Group}~$\boldsymbol{c}$. Magenta dashed-line box in case~\textbf{5} encloses an example set of SIPeCS. The inflow conditions (TI) and surging settings ($v_{\mathrm{max}}$ and $k_D$) are labeled at the top left of each panel, while the case numbers are labeled on the right.}
\label{fig:paper_1_v2_singleAll_UIns}
\end{figure}

The instantaneous contours of streamwise velocity $u$ for cases of \textbf{Group}~$\boldsymbol{c}$ in Table~\ref{tab:cases_Single_rotor} are displayed in Figure~\ref{fig:paper_1_v2_singleAll_UIns}. All snapshots are captured at $\phi_S = 0.0\pi$, where the rotor moves along the freestream. These snapshots effectively capture the fine structures of wakes, including the detailed influences of tip and root vortices.

\begin{figure}[t]
\centering
\includegraphics[width=430pt]{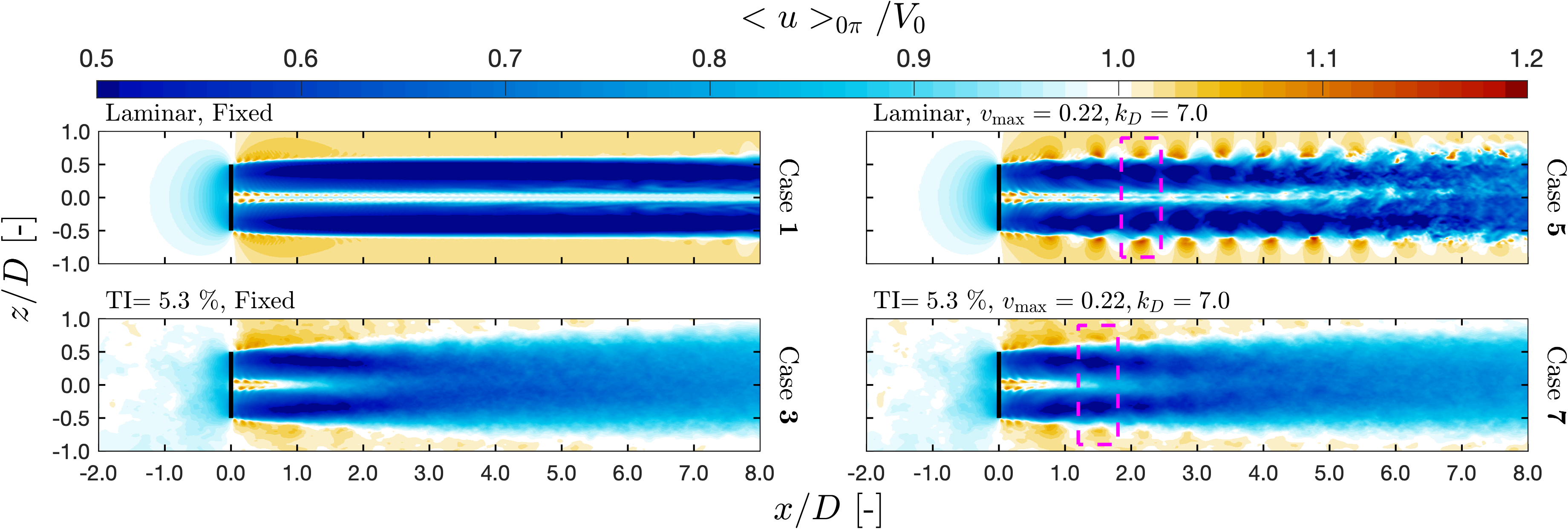}
\caption{Contour plots of phase-averaged streamwise velocity fields $<u>_{0 \pi}$ for the cases in \textbf{Group}~$\boldsymbol{c}$. Magenta dashed-line boxes enclose the example sets of SIPeCS for the surging cases. The inflow conditions (TI) and surging settings ($v_{\mathrm{max}}$ and $k_D$) are labeled at the top left of each panel, while the case numbers are labeled on the right.}
\label{fig:paper_1_v2_singleAll_0000PhaseAvgU}
\end{figure}

Firstly, attention is directed to the two laminar cases depicted in Figure~\ref{fig:paper_1_v2_singleAll_UIns}. The snapshots reveal that the wake structures in the fixed-laminar case are relatively homogeneous in the streamwise direction, and the fluctuations appear only after $x/D = 6.5$. In contrast, the surging-laminar case exhibits noticeable periodic structures within its wake, distinguishing it from the fixed-laminar case. These periodic structures closely resemble those identified in the study by Kleine et al. \cite{kleine2022stability}, where they focused on investigating surging rotors under laminar inflow conditions. We refer to these periodic coherent structures as \emph{Surge-Induced Periodic Coherent Structures} (SIPeCS), with an example set highlighted by the dashed magenta box in Figure~\ref{fig:paper_1_v2_singleAll_UIns}. Note that a set of SIPeCS in case \textbf{5} of Figure~\ref{fig:paper_1_v2_singleAll_UIns} results from a complete surging cycle. SIPeCS can be more effectively visualized and further examined through phase-averaged fields, such as $<u>_{0\pi}$ and $<\omega_y>_{0\pi}$ fields, as will be explored with comprehensive discussions and analysis in the subsequent sections (see Figures~\ref{fig:paper_1_v2_singleAll_0000PhaseAvgU} and \ref{fig:singleAll_8_2_0000PhaseAvgVorticity_y_label}).

Next, the analysis turns to the two turbulent cases. In contrast to the laminar cases, the instantaneous wake systems of the surging-turbulent case (case \textbf{7}) closely resemble those of the fixed-turbulent case (case \textbf{3}), as can be seen in Figure~\ref{fig:paper_1_v2_singleAll_UIns}. Note that this comparison is deemed valid since the inflow conditions are guaranteed to be identical as synthetic turbulent inflow conditions are used (see Section~\ref{sec:setup}). The resemblance between the instantaneous wake structures of fixed-turbulent and surging-turbulent cases suggests that the instantaneous wake structures are predominantly influenced by the inflow turbulence rather than the surging motion of the rotor. Additionally, the general characteristics observed with the time-averaged flow properties of the wakes, particularly in terms of $\overline{u}$ and $\overline{u}_{\mathrm{Disk}}$ reported in Section~\ref{sec:wake_vel}, also exhibit a high degree of similarity between the fixed-turbulent and surging-turbulent cases. This observation reinforces the notion that the impacts of FOWT motions on wake properties are significantly diminished by the presence of inflow turbulence, and these findings are consistent with previous research on FOWT subjected to motions \cite{li2022onset, ramos2022investigationII}.

\paragraph{Phase-averaged streamwise velocity fields $<u>_{0\pi}$}

The phase-averaged streamwise velocity fields as $\phi_S = 0.0\pi$ (and $\phi_{\Omega} = 0.0\pi$), denoted as $<u>_{0\pi}$, for the cases in \textbf{Group}~$\boldsymbol{c}$ are presented in Figure~\ref{fig:paper_1_v2_singleAll_0000PhaseAvgU}. The methodology of phase-averaging and its rationale are elaborated in Section~\ref{sec:phaseLock}.

We have also performed Proper Orthogonal Decomposition (POD) analysis based on phase-locking fields. However, we have concluded that analysis based on phase-averaged fields is sufficient and as representative as the mode shapes from POD analysis, which will be further explained. After performing POD analysis, it is found that the mode energy fractions corresponding to the mean of the phase-locking fields (which are the phase-averaged fields) are significantly predominant. They are about 50 or 5 times larger than the modes with the second highest mode energy fractions for the surging-laminar and the surging-turbulent cases, respectively. Moreover, the energy fractions of subsequent modes gradually and slowly decrease, and the mode shapes related to the phase-averaged fields are the only modes with distinctive patterns in the wake, whereas all the others have relatively disordered patterns. This analysis and observation are detailed in Li \cite{Li2023Numerical}. Consequently, instead of focusing on the mode shapes from the POD analysis, we have chosen to concentrate on the phase-averaged fields for a more straightforward understanding.

Upon examining the contour plots of $<u>_{0\pi}$ for the two laminar cases in Figure~\ref{fig:paper_1_v2_singleAll_0000PhaseAvgU}, it is found that the $<u>_{0\pi}$ fields here and the $u$ fields in Figure~\ref{fig:paper_1_v2_singleAll_UIns} are almost identical. In the fixed-laminar case, both the $u$ and $<u>_{0\pi}$ fields are homogeneous in the streamwise direction. In the surging-laminar case, SIPeCS can be found at exactly the same positions in both $u$ and $<u>_{0\pi}$ field. This striking similarity between the $u$ and $<u>_{0\pi}$ fields suggests that the velocity fields for the laminar cases are highly repeatable with respect to both $\omega_S$ and $\Omega$.

In the $<u>_{0\pi}$ fields for the two turbulent cases appear in Figure~\ref{fig:paper_1_v2_singleAll_0000PhaseAvgU}, the effects of the tip and root vortices are found to become more observable and resemble those of the laminar cases. This observation demonstrates the effectiveness of phase-averaging in revealing repeating flow structures with certain periods by averaging out the noise of inflow turbulence. Particularly noteworthy is the $<u>_{0\pi}$ field for the surging-turbulent case, where SIPeCS are revealed after phase-averaging. Note that SIPeCS are not found in the instantaneous velocity fields of the surging-turbulent case shown in Figure~\ref{fig:paper_1_v2_singleAll_UIns}. The SIPeCS observed in the surging-turbulent case share similar features to those in the surging-laminar case, albeit being more blurred and exhibiting a faster decay. The presence of SIPeCS in the $<u>_{0\pi}$ field of the surging-turbulent case suggests that the effects of surging on wake structures persist when subjected to turbulent inflow conditions, even if not immediately evident in the instantaneous fields. To the best of our knowledge, when subjected to turbulent inflow conditions, we are the first to uncover that the wake structures of a surging rotor are systematically different from those of a fixed rotor. This discovery is important to the offshore wind community, as it indicates that wake interaction among the FOWTs clustered in wind farm scenarios may differ significantly from those of bottom-fixed counterparts. However, in the current work, we do not elaborate on how the downstream FOWTs are influenced by SIPeCS, as we only focus on the isolated rotor scenario. A further analysis of SIPeCS is presented later in Section~\ref{sec:wake_SIPeCS}.

\paragraph{Phase-averaged turbulent kinetic energy fields $<$TKE$>_{0\pi}$}

Various studies have indicated that the surging motion of a FOWT rotor induces instabilities in its wake, promoting the breakdown of larger-scale vortex structures into smaller scales \cite{micallef2021floating, chen2022modelling, kleine2022stability, ramos2022investigationI}. Conventionally, faster breakdown of wake structures caused by this process is believed to lead to an increase of TKE in the wake of a FOWT rotor, thus facilitating quicker wake recovery \cite{chen2022modelling, ramos2022investigationI}. Motivated by this conventional hypothesis, we explore the additional randomness introduced into the wake due to surging motions. Note that randomness is a key essence of turbulent mixing.

\begin{figure}[t]
\centering
\includegraphics[width=430pt]{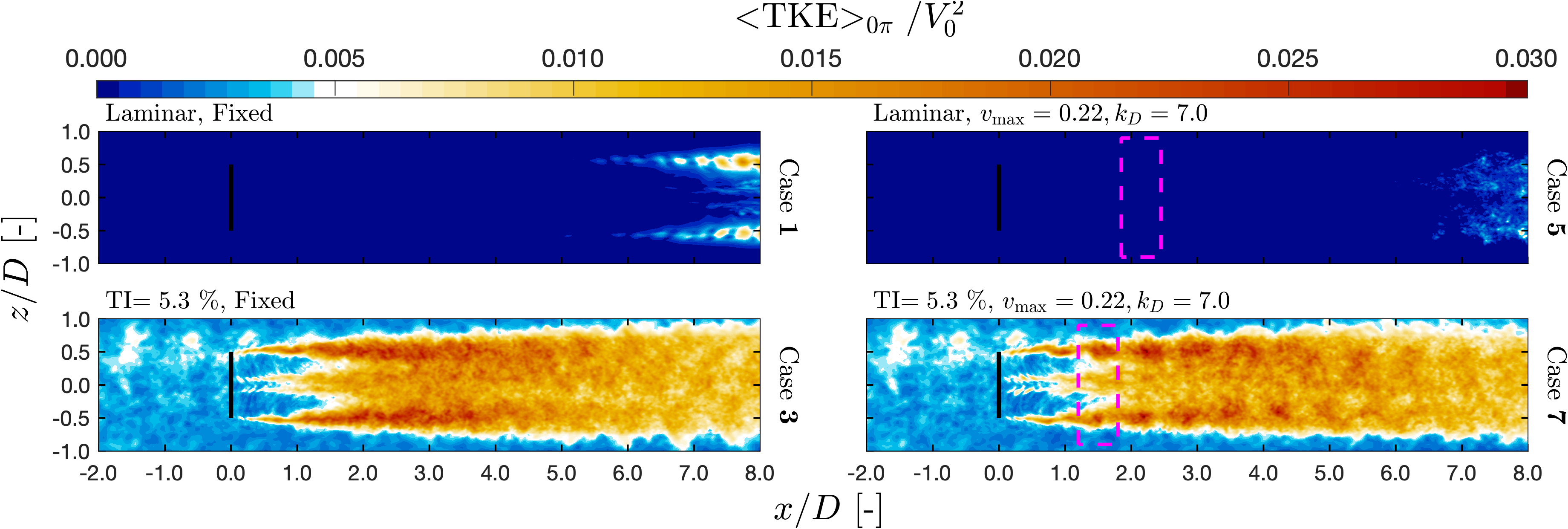}
\caption{Contour plots of phase-averaged turbulent kinetic energy $<$TKE$>_{0 \pi}$ for the cases in \textbf{Group}~$\boldsymbol{c}$. Locations of the magenta dashed-line boxes in the plots are identical to those in Figure~\ref{fig:paper_1_v2_singleAll_0000PhaseAvgU}, which correspond to the positions of the example sets of SIPeCS. The inflow conditions (TI) and surging settings ($v_{\mathrm{max}}$ and $k_D$) are labeled at the top left of each panel, while the case numbers are labeled on the right.}
\label{fig:paper_1_v2_singleAll_0000PhaseAvgTKE_eight}
\end{figure}

To investigate the extent of random fluctuation of velocity fields, we focus on the phase-averaged TKE ($<$TKE$>_{0\pi}$) fields, which are presented in Figure~\ref{fig:paper_1_v2_singleAll_0000PhaseAvgTKE_eight}. The definition of $<$TKE$>_{0\pi}$ is in Section~\ref{sec:phaseLock}. The key idea about using $<$TKE$>_{0\pi}$ instead of conventional TKE is that $<$TKE$>_{0\pi}$ effectively filters out the fluctuations attributable to major periodicities. This ensures that the fluctuations being analyzed are predominantly random. In this study, the significant periodicities include surging motions and the rotation of the FOWT rotors. That is, the periods correspond to $\omega_S$ and $\Omega$.

Firstly, we examined the $<$TKE$>_{0\pi}$ fields of the two laminar cases. Surprisingly, both the fixed and the surging cases exhibit very low $<$TKE$>_{0\pi}$ values, except in the very downstream regions. Note that very low $<$TKE$>_{0\pi}$ fields are also found in the other laminar cases with different surging settings. More interestingly, higher $<$TKE$>_{0\pi}$ values emerge earlier in the $x$-axis for the fixed-laminar case than in the surging-laminar case, following the implications of the results reported by Fang et al. \cite{fang2021effect}. This indicates that despite the complex wake structures observed in the surging-laminar case (see Figures~\ref{fig:paper_1_v2_singleAll_0000PhaseAvgU} and \ref{fig:singleAll_8_2_0000PhaseAvgVorticity_y_label}), our findings show that its flow field is highly periodic and repeatable concerning the surging period. (Note that $<$TKE$>_{0\pi}$ has very low values even around the immediate vicinity of the rotors in both laminar cases, and this is because the simulations are conducted with ALM, where the boundary layers around the rotor geometry are not calculated/modeled and thus missing additional turbulent injections \cite{sarlak2014large}.) Furthermore, $<$TKE$>_{0\pi}$ fields are found to be correlated with larger $v_{\mathrm{max}}$ and $k_D$, where larger $v_{\mathrm{max}}$ and $k_D$ result in lower $<$TKE$>_{0\pi}$ fields. Lower values in $<$TKE$>_{0\pi}$ fields imply that the considered surging motions stabilize the phase-averaged flow fields and delay the wake breakdown, and stronger stabilization effects are observed at larger $v_{\mathrm{max}}$ and higher $k_D$. Crucially, the low $<$TKE$>_{0\pi}$ values in the surging-laminar case clarify that surging motions do not inject additional randomness into the flow fields. Therefore, the mechanism for faster wake recovery in the surging-laminar case is not attributable to increased randomness, and it is proposed that the faster wake recovery is due to the enhanced mixing caused by advection induced by SIPeCS. This will be further elaborated in the later sections (Sections~\ref{sec:wake_SIPeCS} and \ref{sec:phase_energy}).

Next, $<$TKE$>_{0\pi}$ fields for the two turbulent cases are investigated. Generally, $<$TKE$>_{0\pi}$ fields are similar both in patterns and strengths for the two turbulent cases, and the values at regions before the FOWT rotor reflect the inflow turbulence intensity of TI $=5.3\%$ (equivalent to TKE$/V_0^2 = 0.0042$). In both cases, the effects of tip and root vortices are apparent within the $<$TKE$>_{0\pi}$ fields. Moreover, similar to the $<u>_{0\pi}$ fields, periodic structures related to SIPeCS are once again observable in the surging-turbulent case, whereas they are absent in the fixed-turbulent case. It is also noteworthy that when the $<$TKE$>_{0\pi}$ fields are looked alongside the $<\omega_y>_{0\pi}$ fields, as displayed in Figure~\ref{fig:singleAll_8_2_0000PhaseAvgVorticity_y_label} in later section, regions with increased $<$TKE$>_{0\pi}$ coincide with areas with stronger $<\omega_y>_{0\pi}$. This correlation indirectly suggests that inflow turbulence interacts with the vortical structures of SIPeCS.

\subsection{Surge-induced periodic coherent structures (SIPeCS)}

\label{sec:wake_SIPeCS}

This section specifically focuses on delving deeper into the surge-induced periodic coherent structures (SIPeCS) discovered in Section~\ref{sec:wake_structure}. To begin with, the contour plots of the phase-averaged $y$-component vorticity (out-of-plane), $<\omega_y>_{0\pi}$, for all 16 cases in Table~\ref{tab:cases_Single_rotor} are presented and discussed in Section~\ref{sec:SIPeCS_contour}. This discussion explores how various inflow conditions and surging settings, including their interplay, influence the general characteristics of SIPeCS through qualitative visual analysis. Subsequently, in Section~\ref{sec:circ}, we extend our focus to quantifying the impact of SIPeCS on the wake systems of the cases considered. This is achieved by calculating the phase-averaged circulation fields, $<\Gamma>_{0\pi}$, based on the phase-averaged velocity ($<u>_{0 \pi}$ and $<w>_{0 \pi}$). This subsection provides a quantitative methodology for locating the positions of SIPeCS and measuring their strengths. It is important to note that, according to Stokes' theorem, $<\Gamma>_{0\pi}$ and $<\omega_y>_{0\pi}$ are tightly linked.

\subsubsection{Phase-averaged \texorpdfstring{$y$}{text}-component (out-of-plane) vorticity fields}

\label{sec:SIPeCS_contour}

\begin{figure}[t]
\centering
\includegraphics[width=430pt]{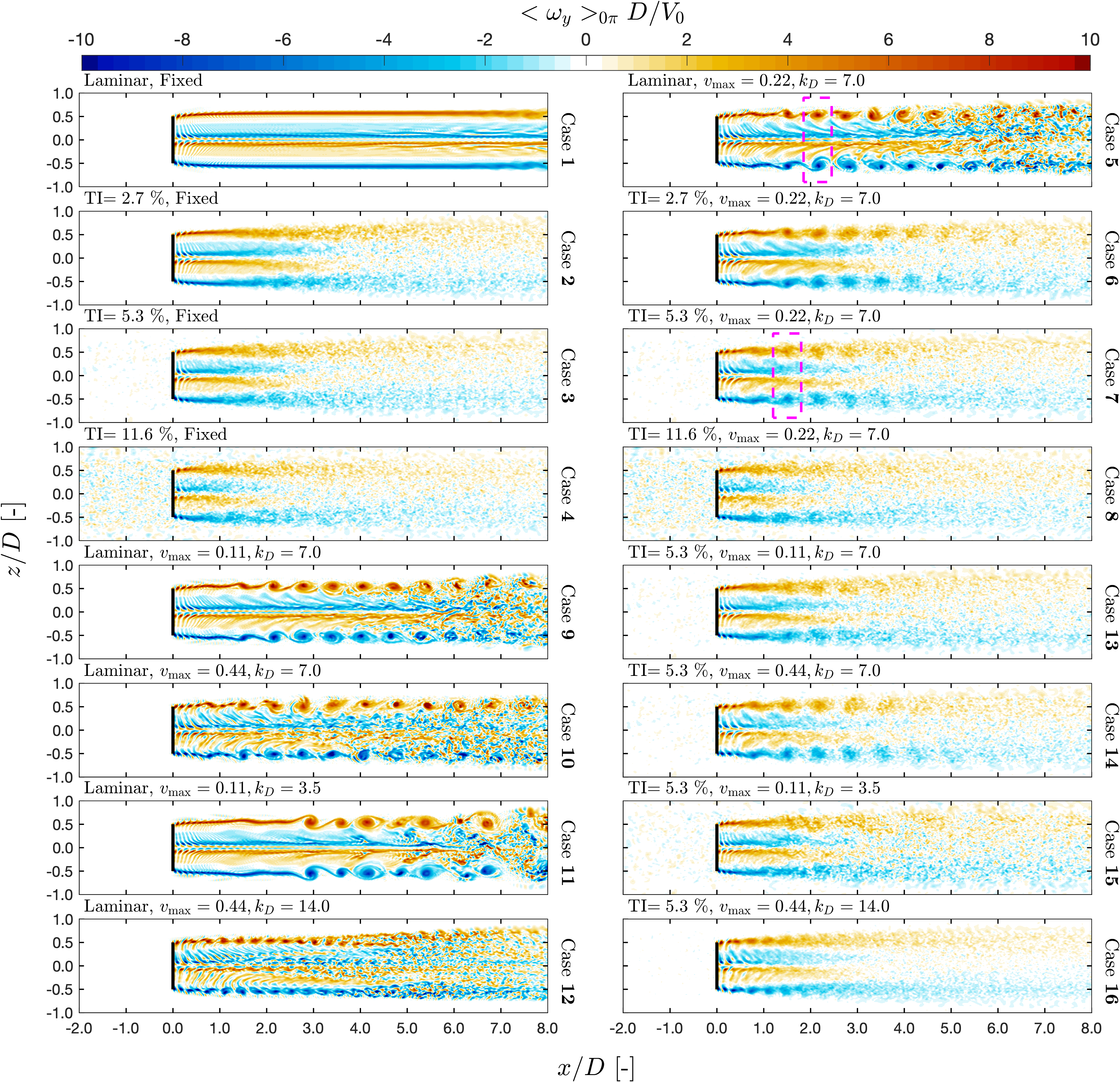}
\caption{Contour plots of phase-averaged $y$-component vorticity $<\omega_y>_{0 \pi}$ for all the 16 cases in Table~\ref{tab:cases_Single_rotor}. Locations of the magenta dashed-line boxes indicating the example sets of SIPeCS in cases \textbf{5} and \textbf{7} are identical with those of Figure~\ref{fig:paper_1_v2_singleAll_0000PhaseAvgU}. The inflow conditions (TI) and surging settings ($v_{\mathrm{max}}$ and $k_D$) are labeled at the top left of each panel, while the case numbers are labeled on the right.}
\label{fig:singleAll_8_2_0000PhaseAvgVorticity_y_label}
\end{figure}

The phase-averaged $y$-component (out-of-plane) vorticity fields $<\omega_y>_{0 \pi}$ are presented in Figure~\ref{fig:singleAll_8_2_0000PhaseAvgVorticity_y_label}. This presentation includes all 16 cases in Table~\ref{tab:cases_Single_rotor} to facilitate comprehensive comparisons. The structure of this subsection for discussing $<\omega_y>_{0 \pi}$ fields is outlined as follows. Initially, the focus is mainly on one case, the surging-laminar case (case~\textbf{5}), to provide an overview of the vortical structures representing SIPeCS. This part of the analysis includes delving into the formation of SIPeCS during a surging cycle, primarily utilizing the vorticity field for the investigation. Next, the analysis proceeds with the four fixed cases in \textbf{Group}~$\boldsymbol{a}$, which are subjected to different inflow TI. Here, the general wake structures and vortex systems of a fixed wind turbine rotor are reviewed, providing a baseline understanding of typical wake structures for a wind turbine rotor. Then, the attention is turned to the four surging cases in \textbf{Group}~$\boldsymbol{a}$. The primary objective here is to explore the influence of inflow TI on SIPeCS. Our findings reveal that SIPeCS dissipates more quickly in environments with higher inflow TI. Subsequently, the laminar cases in \textbf{Group}~$\boldsymbol{b}$ are studied. Observations indicate that under laminar inflow conditions, the patterns of SIPeCS are predominantly influenced by the surging frequency $\omega_S$, with less pronounced effects from surging amplitude $A_S$. Lastly, the turbulent cases in \textbf{Group}~$\boldsymbol{b}$ are examined. The results demonstrate that larger $A_S$ makes SIPeCS patterns more prominent under turbulent inflow conditions. Similarly to the laminar cases, the impacts of $\omega_S$ are primarily observed in spatial repetition rates.

\paragraph{Overviewing SIPeCS with $<\omega_y>_{0\pi}$ fields}

Here, the formation processes and structures of SIPeCS are delved into using the $<\omega_y>_{0\pi}$ field. By examining the contour plot of $<\omega_y>_{0\pi}$ for case \textbf{5} (surging-laminar) in Figure~\ref{fig:singleAll_8_2_0000PhaseAvgVorticity_y_label}, large and strong periodic vortical structures are observed in the wake (see the magenta dashed-line box in the figure). It should be noted that while these structures are depicted in pairs in the $2D$ contour plots, they are in the form of rings in $3D$ space. Due to the clear characteristics of these vortical structures, throughout the rest of this work, the term SIPeCS specifically refers to these ring-like vortical structures. 

The formation of SIPeCS involves the merging of tip vortices through a rolling-up process. In most cases studied here, depending on the surging settings, the tip vortices released within a complete surging cycle merge to form a single set of SIPeCS. This merging process is triggered by the imbalances in the induction fields of consecutive tip vortices since surging motions make the distances between the consecutive tip vortices nonuniform. This tip vortices merging process has previously been reported by Kleine et al. \cite{kleine2022stability}.

The high magnitudes of $<\omega_y>_{0\pi}$ observed for these SIPeCS indicate significant swirling motion around them. The sign of these vortical structures within the SIPeCS indicates that their induction fields are slowing down the flow inside the wake in the streamwise direction while simultaneously accelerating the flow outside the wake. This behavior aligns with the observations from the $<u>_{0\pi}$ fields presented in Figure~\ref{fig:paper_1_v2_singleAll_0000PhaseAvgU}. Furthermore, these vortical structures suggest a dynamic interaction between the flow inside and outside the wake, where the flow outside the wake is being advected into the wake and vice versa. This exchange of flow can be confirmed with the fields of phase-averaged spanwise velocity, denoted as $<w>_{0\pi}$, displayed in Figure~\ref{fig:paper_1_v2_singleAll_0000PhaseAvgW}. Note that this exchanging process is advantageous for wake recovery, as the flow with higher streamwise velocity is brought into the wake, enhancing the momentum exchange and the energy entrainment.

\paragraph{The fixed cases with different inflow TI (\textbf{Group}~$\boldsymbol{a}$)}

For the four fixed cases (cases \textbf{1}-\textbf{4}) in \textbf{Group}~$\boldsymbol{a}$, the distributions of $<\omega_y>_{0 \pi}$ are similar to those reported by previous works \cite{troldborg2009actuator, sarlak2014large}. Regions with higher magnitudes are mainly concentrated at the regions behind the tips and roots, accounting for the released trailing (tip and root) vortices. As observed in Figure\ref{fig:singleAll_8_2_0000PhaseAvgVorticity_y_label}, the tip vortices in these fixed cases tend to smear and merge shortly after being released, forming strips of concentrated $<\omega_y>_{0\pi}$. Moreover, with increasing inflow TI, the strengths of these strips, which are formed by trailing vortices, decay faster. However, even with the highest inflow TI considered, these strips remain evident for at least $2D$ downstream from the rotor.

\paragraph{The surging cases with different inflow TI (\textbf{Group}~$\boldsymbol{a}$)}

For the four surging cases (cases \textbf{5}-\textbf{8}) that share identical surging settings, the $<\omega_y>_{0\pi}$ fields exhibit not only trailing vortices but also the vortical structures of SIPeCS which have been introduced and briefly discussed earlier. Examples of these SIPeCS are highlighted within the magenta dashed-line boxes in Figure~\ref{fig:singleAll_8_2_0000PhaseAvgVorticity_y_label}.

In the surging case subjected to laminar inflow (case \textbf{5}), different from the fixed-laminar case (case \textbf{1}), the released tip vortices right after the rotor is not lined up straight but are distorted. This distortion results from imbalances in the induction forces among consecutive tip vortices. As previously discussed, these imbalances play a crucial role in the merging process of the tip vortices, leading to the formation of SIPeCS.

Moving to the three remaining turbulent cases with a surging rotor, it is apparent that the general features of the $<\omega_y>_{0\pi}$ fields, including the presence of vortical structures of SIPeCS, are similar to those observed in the surging-laminar case (case \textbf{5}). This observation indicates that the mechanisms to form SIPeCS are relatively consistent and remain largely unaffected by the inflow turbulence. However, as expected, the clearness of tip/root vortices and SIPeCS tends to decrease with increased inflow TI. Additionally, the locations where SIPeCS become less discernible shift closer to the rotor with higher inflow TI. This trend reveals that the inflow TI directly influences the persistence of SIPeCS.

\paragraph{The laminar cases with different surging settings (\textbf{Group}~$\boldsymbol{b}$)}

Turning the attention to the impacts of surging settings on SIPeCS, $<\omega_y>_{0\pi}$ fields of the six laminar cases in \textbf{Group}~$\boldsymbol{b}$, encompassing cases \textbf{1}, \textbf{5}, and \textbf{9}-\textbf{12}, are examined. In these cases, with the exception of case \textbf{11}, each set of vortical structures in the $<\omega_y>_{0 \pi}$ fields for the surging laminar cases results from the merging of tip vortices within a complete surging cycle. Consequently, the periodicity of these vortical structures (SIPeCS) aligns exactly with the surging frequency $\omega_S$. However, in the laminar case with $k_D = 3.5$ (case \textbf{11}), the formation of vortical structures deviates from this pattern. Instead of forming a single set of vortical structures per surging cycle, two sets are formed, one stronger (e.g., between $x/D = 2.8$ and $x/D = 3.2$) and another weaker (e.g., between $x/D = 3.5$ and $x/D = 3.8$), highlighting the complexity of the dynamics for SIPeCS formation. Additionally, it is observed that cases with higher $k_D$ tend to form their SIPeCS closer to the rotor.

Regarding the impacts of surging amplitude $A_S$, three laminar cases with the same $k_D$ but varying $v_{\mathrm{max}}$ (identical $\omega_S$ but varying $A_S$) are reviewed (cases \textbf{5}, \textbf{9}, and \textbf{10}). Despite significant differences in rotor performance, the general patterns of SIPeCS in these cases are similar.

\paragraph{The turbulent cases with different surging settings (\textbf{Group}~$\boldsymbol{b}$)}

Shifting the focus toward the six turbulent cases in \textbf{Group}~$\boldsymbol{b}$ (cases \textbf{3}, \textbf{7}, and \textbf{13}-\textbf{16}), we observe that inflow turbulence generally affects SIPeCS like that mentioned in \textbf{Group}~$\boldsymbol{a}$, predominantly by blurring these structures. However, unlike in the laminar cases in {\textbf{Group}}~$\boldsymbol{b}$, where the impacts of variations in $A_S$ are less discernible, the turbulent cases in \textbf{Group}~$\boldsymbol{b}$ reveal that values of $A_S$ have pronounced effects. Specifically, for the turbulent cases with an identical value of $k_D$ but different $v_{\mathrm{max}}$, SIPeCS will appear in more defined forms for the cases with larger $v_{\mathrm{max}}$. This observation suggests that a larger $A_S$ exerts a stronger influence on the wake structures for the six considered turbulent cases.

\subsubsection{Quantification of SIPeCS and its analysis}

\label{sec:circ}

In the previous subsection, SIPeCS are qualitatively evaluated with the contour plots of $<u>_{0\pi}$ and $<\omega_y>_{0\pi}$ fields  (Figures~\ref{fig:paper_1_v2_singleAll_0000PhaseAvgU} and \ref{fig:singleAll_8_2_0000PhaseAvgVorticity_y_label}). In this subsection, the quantitative assessments of SIPeCS about their positions and strengths are provided, which are based on the calculated phase-averaged circulation fields, $<\Gamma>_{0\pi}$. The idea is to characterize SIPeCS by identifying the locations and strengths of ${<\Gamma>_{0\pi, \mathrm{max}}}$, which is the local maximum of $<\Gamma>_{0\pi}$. The calculation of $<\Gamma>_{0\pi}$ is performed using circular rings with the same radius $r_{\Gamma}$. The mathematical formulation of $<\Gamma>_{0\pi}$ is detailed in Equation~\ref{eq:circ_phaseLock}. Utilizing Stokes's theorem, $<\Gamma>_{0\pi}$ can be related to the moving area-sum of $<\omega_y>_{0\pi}$, as described in the latter part of Equation~\ref{eq:circ_phaseLock}. Furthermore, $<\Gamma>_{0\pi}$ is decomposed into two components, $<\Gamma_u>_{0\pi}$ and $<\Gamma_w>_{0\pi}$, as outlined in Equation~\ref{eq:circ_uw_phaseLock}. This decomposition is designed to isolate the strong effects of shear layers \cite{morgan2009vortex}, and it should be noted that $<\Gamma>_{0\pi}$ is the sum of $<\Gamma_u>_{0\pi}$ and $<\Gamma_w>_{0\pi}$. This approach allows for a more detailed understanding of the effects of SIPeCS on wake.

\begin{figure}[t]
\centering
\includegraphics[width=430pt]{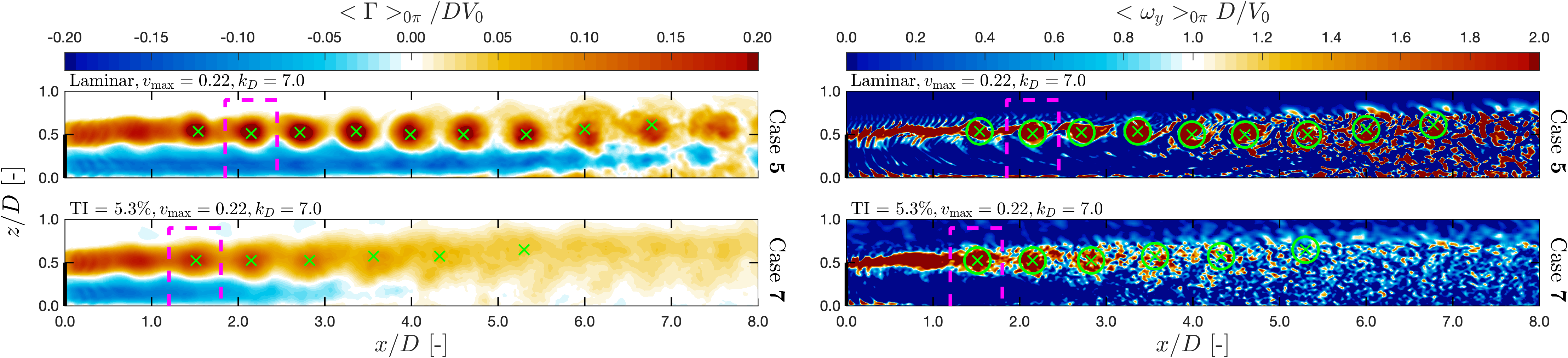}
\caption{(a): Phase-averaged circulation fields $<\Gamma>_{0 \pi}$ calculated based on $<u>_{0\pi}$ and $<w>_{0\pi}$ fields. The path for calculating $<\Gamma>_{0 \pi}$ is circular and the radius $r_{\Gamma}$ is $0.15D$. The locations of local maximums of $<\Gamma>_{0 \pi}$ (${<\Gamma>_{0\pi, \mathrm{max}}}$) are labeled with crosses. (b): The corresponding $<\omega_y>_{0\pi}$ fields of the cases in (a). The crosses in (b) are located at the same positions as (a), where local maximums of $<\Gamma>_{0 \pi}$ are obtained. The circles indicate the paths used to calculate the values of $<\Gamma>_{0 \pi}$ at the crosses, where the radii are $r_{\Gamma}$. Note that not all the values of ${<\Gamma_w>_{0\pi, \mathrm{max}}}$ found at the crosses surpass the threshold in Figure~\ref{fig:paper_1_v2_ampFreqMaxCircValue_All}(c). Locations of the magenta dashed-line boxes in the plots are identical to those in Figure~\ref{fig:paper_1_v2_singleAll_0000PhaseAvgU}, which correspond to the positions of the example sets of SIPeCS.}
\label{fig:paper_1_v2_NREL_plotting_circulation_omegay}
\end{figure}

To calculate the $<\Gamma>_{0\pi}$ fields, we initially select a value for the radius $r_{\Gamma}$ informed by the size of the vortical structures observed in Figure~\ref{fig:singleAll_8_2_0000PhaseAvgVorticity_y_label}. After the $<\Gamma>_{0\pi}$ fields are computed, the locations of SIPeCS are determined at points where local maximums of $<\Gamma>_{0\pi}$ (${<\Gamma>_{0\pi, \mathrm{max}}}$) are identified. Note that the windows for identifying these local maximums are assigned through visual inspection, and we ensure no gaps or overlaps between them. Examples of $<\Gamma>_{0\pi}$ fields and the positions of ${<\Gamma>_{0\pi, \mathrm{max}}}$ are illustrated in Figure~\ref{fig:paper_1_v2_NREL_plotting_circulation_omegay}(a).

Subsequently, we re-evaluate the $r_{\Gamma}$ value to ensure that it satisfies two key criteria. Firstly, circles centered at the locations of ${<\Gamma>_{0\pi, \mathrm{max}}}$ should mostly enclose the vortical structures associated with SIPeCS. Secondly, these circles should not overlap when drawn. If these criteria are not met, adjustments for $r_{\Gamma}$ are made. After a few iterations, a radius of $r_{\Gamma} = 0.15D$ is found to meet both the criteria for all the surging cases in Table~\ref{tab:cases_Single_rotor}. Keeping $r_{\Gamma}$ consistent across cases is advantageous to compare the strength of ${<\Gamma>_{0\pi, \mathrm{max}}}$ between cases, as $<\Gamma>_{0\pi}$ represents the area-sum of $<\omega_y>_{0\pi}$. Examples demonstrating how circles with $r_{\Gamma} = 0.15D$ meet these criteria are provided in Figure~\ref{fig:paper_1_v2_NREL_plotting_circulation_omegay}(b).

For brevity, the analysis of $<\Gamma>_{0\pi}$ only focuses on the regions between $0 \geq y/D \geq 1$ and $0 \geq x/D \geq 8$. Calculations of $<\Gamma>_{0\pi}$ are also performed for the fixed cases for baseline comparisons. The windows selected to identify ${<\Gamma>_{0\pi, \mathrm{max}}}$ in the fixed cases also do not have gaps in between.

\begin{multline}
     <\Gamma>_{0\pi} \Big|_{(x = x_0, y = 0, z = z_0)} \overset{\Delta}{=} \oint_{r = r_{\Gamma}} (<u>_{0\pi}, 0, <w>_{0\pi}) \cdot \mathrm{d} \boldsymbol{l} \\
     = \oint_{r = r_{\Gamma}} (<u>_{0\pi}, <v>_{0\pi}, <w>_{0\pi}) \cdot \mathrm{d} \boldsymbol{l} = \int_{A} \nabla \times (<u>_{0\pi}, <v>_{0\pi}, <w>_{0\pi}) \cdot \mathrm{d} \boldsymbol{A} \\
     = \int_{A} (<\omega_x>_{0\pi}, <\omega_y>_{0\pi}, <\omega_z>_{0\pi}) \cdot \mathrm{d} \boldsymbol{A} = \int_{A} <\omega_y>_{0 \pi} \hspace{2pt} \mathrm{d} A
    \label{eq:circ_phaseLock}
\end{multline}

\begin{multline}
     <\Gamma_u>_{0\pi} \Big|_{(x = x_0, y = 0, z = z_0)} \overset{\Delta}{=} \oint_{r = r_{\Gamma}} (<u>_{0\pi}, 0, 0) \cdot \mathrm{d} \boldsymbol{l}, \\
     <\Gamma_w>_{0\pi} \Big|_{(x = x_0, y = 0, z = z_0)} \overset{\Delta}{=} \oint_{r = r_{\Gamma}} (0, 0, <w>_{0\pi}) \cdot \mathrm{d} \boldsymbol{l}
    \label{eq:circ_uw_phaseLock}
\end{multline}

In the upcoming analysis, we examine the positions of ${<\Gamma>_{0\pi, \mathrm{max}}}$ along with the corresponding values of ${<\Gamma>_{0\pi, \mathrm{max}}}$, ${<\Gamma_u>_{0\pi, \mathrm{max}}}$, and ${<\Gamma_w>_{0\pi, \mathrm{max}}}$ for the selected cases. ${<\Gamma_u>_{0\pi, \mathrm{max}}}$ and ${<\Gamma_w>_{0\pi, \mathrm{max}}}$ represent the
values of $<\Gamma_u>_{0\pi}$ and $<\Gamma_w>_{0\pi}$ calculated at the positions where ${<\Gamma>_{0\pi, \mathrm{max}}}$ are located, respectively. Note that ${<\Gamma>_{0\pi, \mathrm{max}}} = {<\Gamma_u>_{0\pi, \mathrm{max}}} + {<\Gamma_w>_{0\pi, \mathrm{max}}}$. A crucial aspect to emphasize is that although the positions of SIPeCS are identified at the locations of ${<\Gamma>_{0\pi, \mathrm{max}}}$, their strengths are assessed through the values of ${<\Gamma_w>_{0\pi, \mathrm{max}}}$, rather than the values of ${<\Gamma>_{0\pi, \mathrm{max}}}$. One of the reasons is that the values of ${<\Gamma>_{0\pi, \mathrm{max}}}$ are significantly influenced by the strengths of the surrounding shear layers (contributing to the term ${<\Gamma_u>_{0\pi, \mathrm{max}}}$). These shear layers are not directly related to SIPeCS but rather reflect the characteristics of the background flows. Therefore, relying solely on ${<\Gamma>_{0\pi, \mathrm{max}}}$ could potentially misrepresent the strengths of SIPeCS. Another reason is that the strength of ${<\Gamma_w>_{0\pi}}$ serves as a key indicator for the intensity of flow exchange between the inside and outside of the wake. This is based on the fact that ${<\Gamma_w>_{0\pi}}$ are calculated with $<w>_{0\pi}$ fields. Previously, we have pointed out that SIPeCS accelerates the wake recovery rates through this advection process and, therefore, higher values of ${<\Gamma_w>_{0\pi}}$ may indicate more efficient wake recovery. Regarding the above, ${<\Gamma_w>_{0\pi, \mathrm{max}}}$ serves as a better indicator of the strength of SIPeCS compared to ${<\Gamma>_{0\pi, \mathrm{max}}}$.

The analysis of $<\Gamma>_{0 \pi}$ presented later is exclusively about the turbulent cases in \textbf{Group}~$\boldsymbol{b}$ and the surging cases in \textbf{Group}~$\boldsymbol{a}$. (The analysis for the laminar cases in \textbf{Group}~$\boldsymbol{b}$ is presented in Appendix~\ref{app:circ_lam}.) In short, our objective is to locate the positions of SIPeCS based on the locations of ${<\Gamma>_{0\pi, \mathrm{max}}}$ and to determine the strengths of SIPeCS based on the values of ${<\Gamma_w>_{0\pi, \mathrm{max}}}$. With this method, the characteristics of SIPeCS are effectively quantified, and the outcomes closely follow the findings with qualitative analysis in Section~\ref{sec:SIPeCS_contour} by observing the $<\omega_y>_{0\pi}$ fields. Specifically, we found that the spatial repeating rates of SIPeCS are proportional to $\omega_S$, cases with larger $A_S$ (same $k_D$, larger $v_{\mathrm{max}}$) have SIPeCS in stronger strengths (when TI $= 5.3\%$), and SIPeCS is more persistent with lower inflow TI, especially when inflow turbulence is absent. Moreover, according to the set threshold, SIPeCS is generally dissipated after $x/D = 4$ with realistic turbulent inflow conditions (TI $= 5.3\%$).

\paragraph{The turbulent cases in \textbf{Group}~$\boldsymbol{b}$}

\begin{figure}[t]
\centering
\includegraphics[width=430pt]{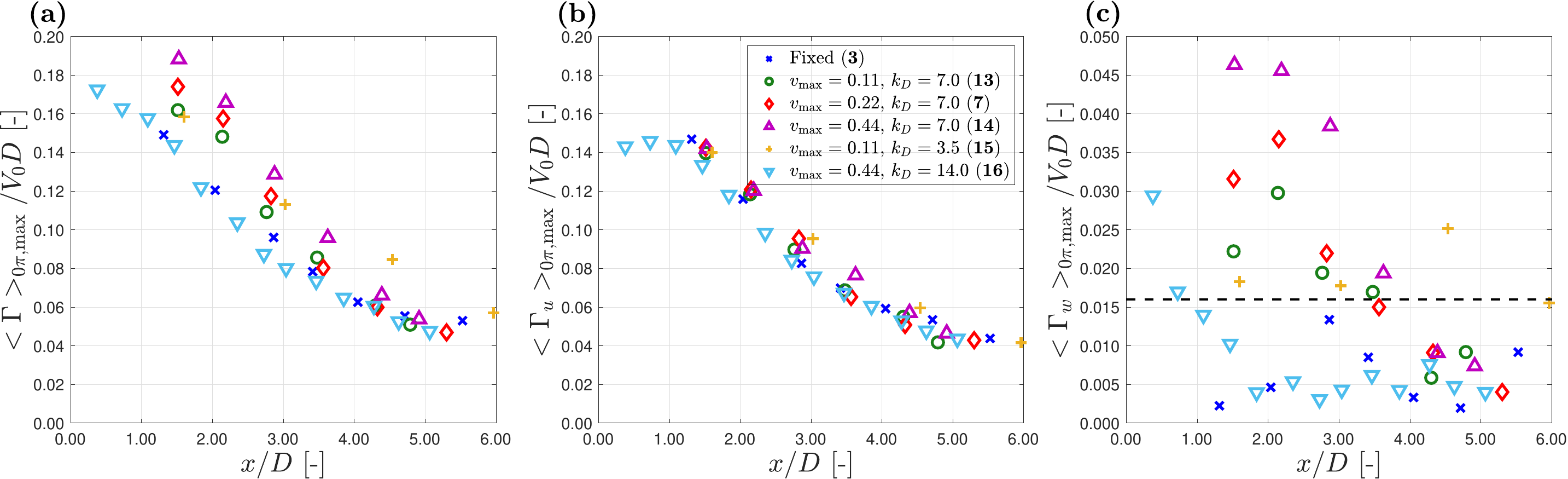}
\caption{Plotting the $x$-positions where the local maximums of phase-averaged circulation $<\Gamma>_{0\pi}$ (${<\Gamma>_{0\pi, \mathrm{max}}}$) are found against the values of ${{<\Gamma>_{0\pi, \mathrm{max}}}}$, ${<\Gamma_u>_{0\pi, \mathrm{max}}}$, and ${<\Gamma_w>_{0\pi, \mathrm{max}}}$ for the turbulent cases (TI $= 5.3\%$) in \textbf{Group}~$\boldsymbol{b}$. ${<\Gamma_u>_{0\pi, \mathrm{max}}}$ and ${<\Gamma_w>_{0\pi, \mathrm{max}}}$ are calculated based on the same paths to obtain ${<\Gamma>_{0\pi, \mathrm{max}}}$, but only considering $<u>_{0 \pi}$ or $<w>_{0 \pi}$. (a): ${<\Gamma>_{0\pi, \mathrm{max}}}$. (b): ${<\Gamma_u>_{0\pi, \mathrm{max}}}$. (c): ${<\Gamma_w>_{0\pi, \mathrm{max}}}$. The dashed line in (c) indicates the threshold for specifying SIPeCS, and the case numbers are labeled in parentheses.}
\label{fig:paper_1_v2_ampFreqMaxCircValue_All}
\end{figure}

Here, the analysis of $<\Gamma>_{0\pi}$ for the six turbulent cases in \textbf{Group}~$\boldsymbol{b}$ are presented. This set of cases represents a surging FOWT rotor with different surging settings when subjected to realistic turbulent inflow conditions (TI $= 5.3\%$). We plot the $x$-positions where ${<\Gamma>_{0\pi, \mathrm{max}}}$ are located against their corresponding values of ${<\Gamma>_{0\pi, \mathrm{max}}}$, ${<\Gamma_u>_{0\pi, \mathrm{max}}}$, and ${<\Gamma_w>_{0\pi, \mathrm{max}}}$, as shown in Figure~\ref{fig:paper_1_v2_ampFreqMaxCircValue_All}.

From the data presented in Figure~\ref{fig:paper_1_v2_ampFreqMaxCircValue_All}(a), it is observed that surging cases generally exhibit higher values of ${<\Gamma>_{0\pi, \mathrm{max}}}$ near the rotor compared to the fixed cases (except for the case with $k_D = 14.0$). As the distance from the rotor increases, the values of ${<\Gamma>_{0\pi, \mathrm{max}}}$ for the surging and fixed cases tend to converge. Furthermore, the strength of ${<\Gamma>_{0\pi, \mathrm{max}}}$ and the positions where they are identified are related to the surging settings. For cases with identical $k_D$ but varying $v_{\mathrm{max}}$ (same $\omega_S$, different $A_S$), the positions of ${<\Gamma>_{0\pi, \mathrm{max}}}$ are similar, yet cases with larger $v_{\mathrm{max}}$ exhibit higher values of ${<\Gamma>_{0\pi, \mathrm{max}}}$. This indicates that SIPeCS are stronger for the cases with a larger $v_{\mathrm{max}}$, which is consistent with the previous findings in this work. However, when comparing cases with different $k_D$, there is no apparent generalized trend in the values of ${<\Gamma>_{0\pi, \mathrm{max}}}$ when fixing $v_{\mathrm{max}}$ (same $A_S$, different $\omega_S$). As for the spatial repetition rates of ${<\Gamma>_{0\pi, \mathrm{max}}}$, they are closely aligned with $k_D$, showing that doubling $k_D$ (equivalent to doubling $\omega_S$) results in doubling the rates.

In Figure~\ref{fig:paper_1_v2_ampFreqMaxCircValue_All}(b), interestingly, despite the discrepancies found with the values of ${<\Gamma>_{0\pi, \mathrm{max}}}$ in Figure~\ref{fig:paper_1_v2_ampFreqMaxCircValue_All}(a), ${<\Gamma_u>_{0\pi, \mathrm{max}}}$ in Figure~\ref{fig:paper_1_v2_ampFreqMaxCircValue_All}(b) for all the six turbulent cases follow the same curve, suggesting the shear layers of these cases possess similar strengths, and the main differences in ${<\Gamma>_{0\pi, \mathrm{max}}}$ lie mainly in ${<\Gamma_w>_{0\pi, \mathrm{max}}}$ (Figure~\ref{fig:paper_1_v2_ampFreqMaxCircValue_All}(c)). Indeed, the trends observed for ${<\Gamma>_{0\pi, \mathrm{max}}}$ are mirrored in ${<\Gamma_w>_{0\pi, \mathrm{max}}}$. Crucially, ${<\Gamma_w>_{0\pi, \mathrm{max}}}$ reflects a key feature of SIPeCS, which is the promotion of flow movement into and out of the wake through their induction fields, and stronger ${<\Gamma_w>_{0\pi, \mathrm{max}}}$ indicates stronger effects, as previously discussed. Another key point is that unlike ${<\Gamma>_{0\pi, \mathrm{max}}}$, values of ${<\Gamma_w>_{0\pi, \mathrm{max}}}$ for the fixed case remain similar along the $x$-direction and remain relatively low, providing a good reference for the background signal. Based on the global maximum of $<\Gamma_w>_{0\pi}$ for the fixed-turbulent case (case \textbf{3}) found in the interested region ($0 \geq y/D \geq 1$ and $0 \geq x/D \geq 8$), we establish a threshold value slightly above it. This threshold serves to identify SIPeCS that are sufficiently strong to be specified.

It is found that all the surging-turbulent cases considered in Figure~\ref{fig:paper_1_v2_ampFreqMaxCircValue_All} have SIPeCS having the strengths surpassing the criteria, highlighting the effectiveness of the implemented method. Also, showing that the surging cases have larger values for ${<\Gamma_w>_{0\pi, \mathrm{max}}}$ compared to the fixed case supports the statement that the SIPeCS-induced advection process is the reason for the faster wake recovery for the surging cases, as larger values of ${<\Gamma_w>_{0\pi, \mathrm{max}}}$ represent stronger flow exchange process due to the advection. 


\paragraph{The surging cases in \textbf{Group}~$\boldsymbol{a}$}

\begin{figure}[t]
\centering
\includegraphics[width=430pt]{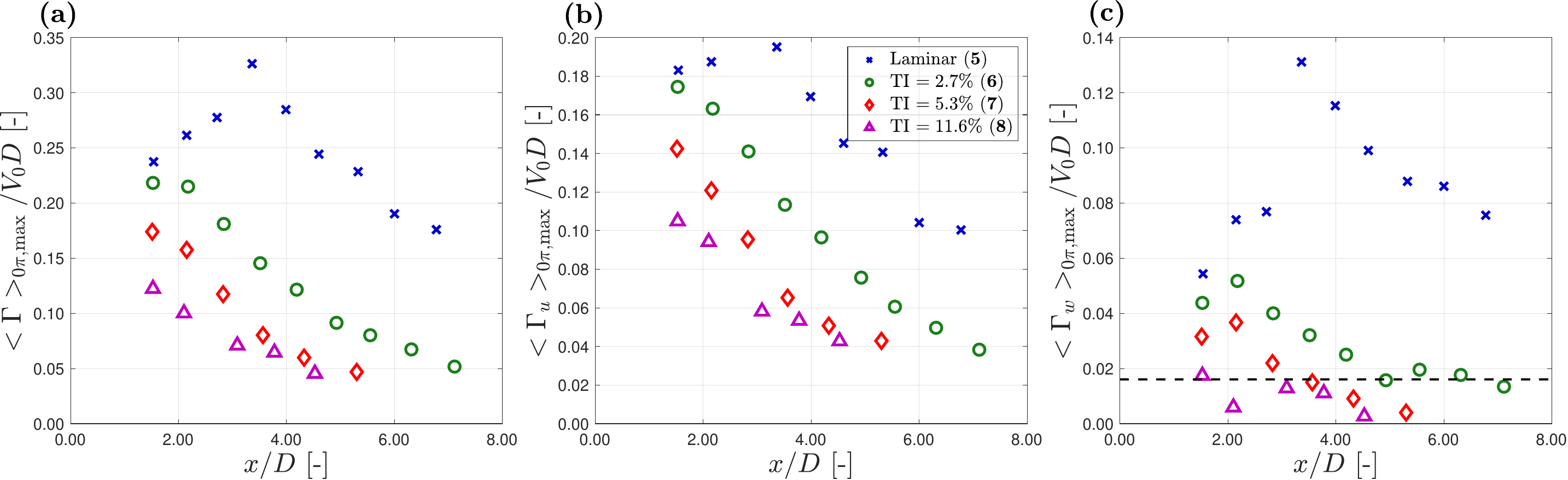}
\caption{Plotting the $x$-positions where ${<\Gamma>_{0\pi, \mathrm{max}}}$ are found against the values of ${<\Gamma>_{0\pi, \mathrm{max}}}$, ${<\Gamma_u>_{0\pi, \mathrm{max}}}$, and ${<\Gamma_w>_{0\pi, \mathrm{max}}}$ for the surging cases in \textbf{Group}~$\boldsymbol{a}$. ${<\Gamma_u>_{0\pi, \mathrm{max}}}$ and ${<\Gamma_w>_{0\pi, \mathrm{max}}}$ are calculated based on the same paths to obtain ${<\Gamma>_{0\pi, \mathrm{max}}}$, but only considering $<u>_{0 \pi}$ or $<w>_{0 \pi}$. (a): ${<\Gamma>_{0\pi, \mathrm{max}}}$. (b): ${<\Gamma_u>_{0\pi, \mathrm{max}}}$.
(c): ${<\Gamma_w>_{0\pi, \mathrm{max}}}$. The dashed line in (c) indicates the threshold for specifying SIPeCS, and the case numbers are labeled in parentheses.}
\label{fig:paper_1_v2_TIMaxCircValue_All}
\end{figure}

After analyzing the impacts of different surging settings on the SIPeCS of the turbulent cases in \textbf{Group}~$\boldsymbol{b}$, our focus shifts to examining the influences of inflow TI on the SIPeCS of the surging cases in \textbf{Group}~$\boldsymbol{a}$. For these cases, ${<\Gamma>_{0\pi, \mathrm{max}}}$, ${<\Gamma_u>_{0\pi, \mathrm{max}}}$, and ${<\Gamma_w>_{0\pi, \mathrm{max}}}$ are plotted and analyzed, as shown in Figure~\ref{fig:paper_1_v2_TIMaxCircValue_All}.

As anticipated, cases with lower TI have higher values in both ${<\Gamma_u>_{0\pi, \mathrm{max}}}$ and ${<\Gamma_w>_{0\pi, \mathrm{max}}}$, especially for the laminar case. Higher values of ${<\Gamma_u>_{0\pi, \mathrm{max}}}$ can be attributed to more preserved shear layers in the wake, as they are less disrupted by ambient turbulence. On the other hand, higher values of ${<\Gamma_w>_{0\pi, \mathrm{max}}}$ result from the vortical structures of SIPeCS being less perturbed by the ambient turbulence, allowing them to sustain for a longer duration before being dissipated. Furthermore, Figure~\ref{fig:paper_1_v2_TIMaxCircValue_All}(c) demonstrates a significant reduction in the strengths of SIPeCS, namely $<\Gamma_w>_{0\pi, \mathrm{max}}$, when transitioning from laminar inflow conditions to turbulent inflow conditions, even at a relatively low level of inflow TI. This observation highlights the profound disruptive impact of inflow turbulence on SIPeCS, emphasizing the strong interactions between SIPeCS and ambient flow. Moreover, this marked decrease in SIPeCS strengths suggests that their influences on wake characteristics for the turbulent cases are much less than those of the laminar cases, as shown in Figure~\ref{fig:paper_1_v2_ampFre_con_UAreaAvg}(a) with the results of $\overline{u}_{\mathrm{Disk}}$.

\subsection{Entrainment of flow kinetic energy with phase-averaged velocity}

\label{sec:phase_energy}

The results based on the phase-averaged circulation ($<\Gamma>_{0\pi}$) in the previous section suggest that the faster wake recoveries of surging cases are due to the enhanced advection process induced by SIPeCS. To verify this hypothesis, a term-by-term analysis of phase-averaged flow kinetic energy entrainment of the wakes is conducted in this subsection based on the phase-averaged velocity data. The analysis is based on the phase-averaged fluxes of flow kinetic energy crossing the edges of the rotor projection area on the $y/D = 0$ plane. That is, the flow kinetic energy that crosses $z/D = \pm 0.5$ on $y/D = 0$ plane. Note that the analysis here is based on the phase-averaged data, as they more precisely reflect the effects of periodic surging motions. For example, properties that vary periodically due to harmonic surging may be considered as ``random'' fluctuations when using time-averaging methods, while they will be treated as phase-averaged quantities when the phase-averaging method is applied. Appendix~\ref{app:time_energy} uses the same analysis in this subsection but replacing phase-averaged data with time-averaged data.

The phase-averaged fluxes of flow kinetic energy are written in Equation~\ref{eq:energyPhaseIn}. The contributions of shear transport (related to $\nu$ and $\nu_T$) are not considered as they are negligible \cite{Li2023Numerical}. The left-hand side of the equation is the dot product of the advection of (specific) flow kinetic energy ($<\boldsymbol{u}(\boldsymbol{u} \cdot \boldsymbol{u}/2)>_{0 \pi}$) and the surface normal vector ($\hat{\boldsymbol{n}}$), while the right-hand side is the decomposition of the left-hand side by letting $\hat{\boldsymbol{n}}$ being parallel to $\pm (0, 0, 1) = \pm \hat{\boldsymbol{z}}$ when $z/D = \mp 0.5$. The decomposition procedure is similar to Reynolds averaging (time-averaging, provided in Appendix~\ref{app:time_energy}), where the relations $u_{0 \pi} = <u>_{0 \pi} + u_{0 \pi}'$ and $<u'>_{0 \pi} = 0$ are used. These are analogous to Reynolds decomposition ($u = \overline{u} + u'$ and $\overline{u'} = 0$).

After decomposing the phase-averaged fluxes of flow kinetic energy, several terms are obtained on the right-hand side of Equation~\ref{eq:energyPhaseIn}, which are the advection of phase-averaged mean kinetic energy($<\mathrm{MKE~advection}>_{0 \pi}$, where $<\mathrm{MKE~advection}>_{0 \pi} = 0.5 <\boldsymbol{u}>_{0 \pi} \cdot <\boldsymbol{u}>_{0 \pi}$), advection of phase-averaged TKE ($<\mathrm{TKE~advection}>_{0 \pi}$), terms related to phase-averaged $2^{\mathrm{nd}}$ order RST (Reynolds stress tensor), and terms related to phase-averaged $3^{\mathrm{rd}}$ order RST. Note that term $<\mathrm{MKE~advection}>_{0 \pi}$ does not relate to fluctuation terms ($<\boldsymbol{u}'>_{0\pi}$), while all the other terms do.

To evaluate the contribution of each term on the left of Equation~\ref{eq:energyPhaseIn} on the energy entertainments of the wakes, their cumulative values are calculated at $z = \pm 0.5$ as written/defined in Equation~\ref{eq:intEnergy}, where the \textbf{energy entraining term} can be substituted to the terms on the right of Equation~\ref{eq:energyPhaseIn}. These \emph{cumulative energy entraining term} are evaluated from $x/D = 1.0$ to $x/D = 8.0$. Regions of very near wake are excluded to avoid the effects of complex aerodynamics of the rotors' tip-vortices, which are not the main focus of this work.

\begin{multline}
    \mathrm{Flux~of~phase-}\mathrm{averaged~kinetic~energy~for}~\hat{n} = \pm \hat{z}~\mathrm{:} \qquad 
    \left<{u_j \left( \frac{u_i u_i}{2} \right) }\right>_{0 \pi} \hat{n}_j 
    \hspace{3pt} = \hspace{3pt}
    \\
    \mp  \left[
    \underbrace{<{w}>_{0 \pi} \left( \frac{ <{u}>_{0 \pi}^2 + <{v}>_{0 \pi}^2 + <{w}>_{0 \pi}^2 }{2} \right)}_{\mathrm{Advection~of~MKE}} \right.
    \\ + \hspace{3pt}
    \underbrace{ <{w}>_{0 \pi} \left( \frac{ <{ \left( u' \right)^2}>_{0 \pi} + <{ \left( v' \right)^2}>_{0 \pi} + <{ \left( w' \right)^2}>_{0 \pi} }{2} \right)}_{\mathrm{Advection~of~TKE}}
    \\
    \hspace{3pt} + \hspace{3pt} \vphantom{ \left[\frac{{u \vphantom{w' w'} }>_{0 \pi} }{2} \right] }
    \underbrace{
    \left(
    <{u \vphantom{w' w'} }>_{0 \pi} \hspace{2pt} <{u' w' \vphantom{w' w'}}>_{0 \pi}
    \hspace{3pt} + \hspace{3pt}
    <{v \vphantom{w' w'} }>_{0 \pi} \hspace{2pt} <{v' w' \vphantom{w' w'}}>_{0 \pi}
    \hspace{3pt} + \hspace{3pt}
    <{w \vphantom{w' w'} }>_{0 \pi} \hspace{2pt} <{w' w' \vphantom{w' w'}}>_{0 \pi}
    \right)
    }_{2^{\mathrm{nd}}~\mathrm{order~RST}}
    \hspace{3pt} \\
    \left. + \hspace{3pt}
    \underbrace{
    \frac{1}{2} \left(
    <{w' u' u'}>_{0 \pi}
    \hspace{3pt} + \hspace{3pt}
    <{w' v' v'}>_{0 \pi}
    \hspace{3pt} + \hspace{3pt}
    <{w' w' w'}>_{0 \pi}
    \right)
    }_{3^{\mathrm{rd}}~\mathrm{order~RST}}
    \right]
    \label{eq:energyPhaseIn}
\end{multline}

\begin{multline}
    \mathrm{Cumulative~\textbf{energy}~\textbf{entraining}~\textbf{term}}(x) \\
    \overset{\Delta}{=}
    \int_{x_0 = 1D}^x 
    \left[ \mathrm{\textbf{energy}~\textbf{entraining}~\textbf{term}}(x_0)\Big|_{z = -0.5D} 
    \hspace{3pt} + \hspace{3pt}
    \mathrm{\textbf{energy}~\textbf{entraining}~\textbf{term}}(x_0)\Big|_{z = 0.5D} 
    \right]
    ~ \mathrm{d} \hspace{1pt} x_0 
    \label{eq:intEnergy}
\end{multline}

\begin{figure}[t]
\centering
\includegraphics[width=430pt]{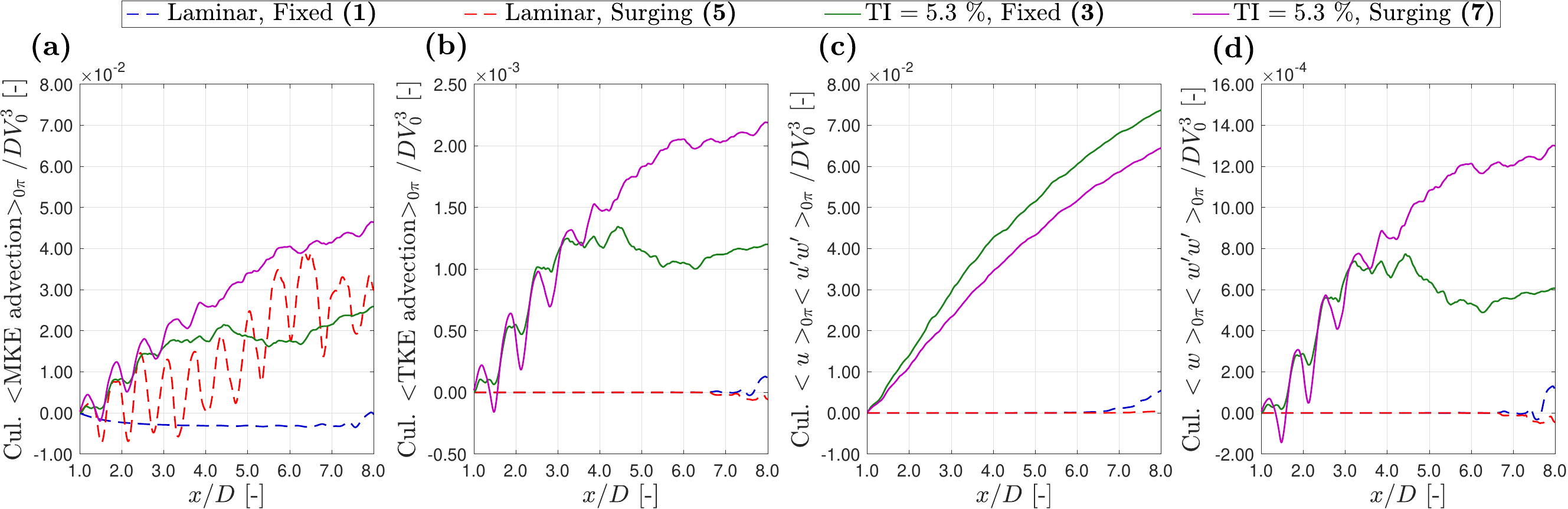}
\caption{Energy entrained into the rotor projection region ($-0.5 \leq z/D \leq 0.5$) on $y/D = 0$ plane with different terms using phase-averaged velocity data. ``Cul.'' is the acronym for ``cumulative''.}
\label{fig:paper_1_v2_pahseLockedEnergyEntrain_cumulative}
\end{figure}

Figures~\ref{fig:paper_1_v2_pahseLockedEnergyEntrain_cumulative}(a) to \ref{fig:paper_1_v2_pahseLockedEnergyEntrain_cumulative}(d) plot out the cumulative contribution of the energy entrained into $-0.5 \leq z/D \leq 0.5$ with the terms of $<\mathrm{MKE~advection}>_{0 \pi}$, $<\mathrm{TKE~advection}>_{0 \pi}$, $<{u}>_{0 \pi} <{u' w'}>_{0 \pi}$, and $<w>_{0 \pi} <w' w'>_{0 \pi}$. The four cases in \textbf{Group}~$\boldsymbol{c}$ are considered, which are cases \textbf{1} (fixed-laminar), \textbf{5} (surging-laminar), \textbf{3} (fixed-turbulent), and \textbf{7} (surging turbulent). Based on the scale of the ordinates in Figure~\ref{fig:paper_1_v2_pahseLockedEnergyEntrain_cumulative}, it is very clear that contributions of $<\mathrm{MKE~advection}>_{0 \pi}$ and $<{u}>_{0 \pi} <{u' w'}>_{0 \pi}$ dominate over those of $<\mathrm{TKE~advection}>_{0 \pi}$ and $<w>_{0 \pi} <w' w'>_{0 \pi}$. Based on this fact, contributions of the terms of $3^{\mathrm{rd}}$ order RST and the out-of-plane term of $2^{\mathrm{nd}}$ order RST ($<{v \vphantom{w' w'} }>_{0 \pi} \hspace{2pt} <{v' w' \vphantom{w' w'}}>_{0 \pi}$) are not presented, as they are smaller or similar to $<w>_{0 \pi} <w' w'>_{0 \pi}$. Additionally, term $<{u}>_{0 \pi} <{u' w'}>_{0 \pi}$ are closely related to $\overline{u \vphantom{u' w'}} \hspace{2pt} \overline{u' w'}$, which is the turbulent mixing term, the main contributor to wind turbine wake recovery when analyzing with time-averaging methods \cite{calaf2010large}.

For the two laminar cases (cases \textbf{1} and \textbf{5}), it can be seen that the fixed-laminar case has almost no energy entrainment. This is reflected on its $\overline{u}_{\mathrm{Dsik}}$ profile presented in Figure~\ref{fig:paper_1_v2_ampFre_con_UAreaAvg}, where wake recovery is not perceivable. As for the surging-laminar case, the energy entrainment of its wake is mainly resulting from $<\mathrm{MKE~advection}>_{0 \pi}$ (the periodic wavy pattern is related to the locations of SIPeCS), while the contribution of term $<{u}>_{0 \pi} <{u' w'}>_{0 \pi}$ is negligible, as hinted by the very low values for the $<$TKE$>_{0 \pi}$ field of case \textbf{5} presented in Figure~\ref{fig:paper_1_v2_singleAll_0000PhaseAvgTKE_eight}. It should be noted that this is in contrast to the common belief that the enhanced wake recovery for a surging FOWT rotor is mainly attributed to the increased turbulence due to the instability \cite{micallef2021floating, chen2022modelling, kleine2022stability, ramos2022investigationI}. Indeed, if the analysis of energy entrainment is performed using time-averaging methods, instead of the increased $\overline{\mathrm{MKE~advection}}$, enhanced $\overline{u \vphantom{u' w'}} \hspace{2pt} \overline{u' w'}$ will become the main contributor to the faster wake recovery of the surging-laminar case (see Appendix~\ref{app:time_energy}). However, treating quantities that can be isolated with the phase-averaging method as turbulence may be misleading, as these quantities are not truly random.

For the two turbulent cases (case \textbf{3} and \textbf{7}), it can be seen that the profiles of cumulative $<{u}>_{0 \pi} <{u' w'}>_{0 \pi}$ are larger than cumulative $<\mathrm{MKE~advection}>_{0 \pi}$ for both the fixed and the surging cases, showing that the effects of turbulence are the main factor for energy entrainment (wake recovery). However, looking closely to Figures~\ref{fig:paper_1_v2_pahseLockedEnergyEntrain_cumulative}(a) and (c), it can be seen that the surging case has higher values for cumulative $<\mathrm{MKE~advection}>_{0 \pi}$ than the fixed case, while having lower values for cumulative $<{u}>_{0 \pi} <{u' w'}>_{0 \pi}$. Moreover, the increase in cumulative $<\mathrm{MKE~advection}>_{0 \pi}$ is larger than the decrease in cumulative $<{u}>_{0 \pi} <{u' w'}>_{0 \pi}$ after $x/D \geq 4.0$. These interesting results indicate two important aspects. The first is that the energy entrained by the wake of the surging-turbulent case is greater than that of the fixed-turbulent case, and this has been reflected by the higher $\overline{u}_{\mathrm{Disk}}$ values for the surging-turbulent cases compared to the fixed-turbulent cases presented in Figure~\ref{fig:paper_1_v2_ampFre_con_UAreaAvg}. The second is that it demonstrates that the enhanced wake recovery of the surging-turbulent case is not mainly related to the increased turbulence level but rather related to the enhanced advection process, again showing disagreements with the common hypotheses. These two aspects are also found when analyzing the time-averaging method, as seen in Appendix~\ref{app:time_energy}.

\section{Conclusions and outlooks}

This paper presented a comprehensive numerical study of a full-scale horizontal-axis Floating Offshore Wind Turbine (FOWT) rotor subjected to harmonic surging motions. The study used high-fidelity CFD simulations, specifically using the Large-Eddy Simulation (LES) coupled with the Actuator Line Model (ALM). The research systematically examined the impacts of various inflow turbulence intensities (TI), surging amplitudes ($A_S$), surging frequencies ($\omega_S$), and their combined interactions. The rotor performance and the wake structures were analyzed and discussed in detail.

The simulation cases in this work were characterized based on two dimensionless parameters, namely $v_{\mathrm{max}}$ and $k_D$. $v_{\mathrm{max}}$ is interpreted as the \emph{magnitude of surging},  defined as the ratio between the \emph{maximum surging velocity of a FOWT rotor} and the \emph{inflow velocity}. $k_D$ represents the \emph{rate of surging}, which is the reduced frequency based on the surging frequency, rotor diameter, and inflow velocity.

Regarding rotor performance, our investigation revealed that inflow turbulence intensity (TI) has minimum impacts on both the time-averaged and cycle-averaged rotor performances. Conversely, rotor performances of a surging rotor are significantly influenced by variations in surging settings, particularly by the values of $v_{\mathrm{max}}$. Notably, in the examined scenarios, before encountering severe (static) stalling, the surging cases with a stronger surging magnitude (larger $v_{\mathrm{max}}$) consistently exhibited a lower time-averaged thrust and a higher time-averaged power compared to the fixed cases. This seemingly contradictory phenomenon was reasoned, potentially providing novel insights for the design of rotors in future FOWT. Furthermore, our research is the first to explicitly show that both $v_{\mathrm{max}}$ and $k_D$ should be considered to adequately characterize the hysteresis effects of a surging FOWT rotor, which is contrary to the previous work, where they showed that only considering $v_{\mathrm{max}}$ should be sufficient \cite{ferreira2022dynamic}. This insight highlights the complexity of aerodynamics for a surging FOWT rotor and suggests the need for future research work to comprehensively grasp the aerodynamic behaviors of surging rotors.

Regarding wake characteristics and structures, our findings affirm that, compared to the cases with a fixed rotor, surging motion significantly promoted wake recovery under laminar inflow conditions \cite{chen2022modelling, kleine2022stability}. However, this enhancement diminished substantially with the presence of inflow turbulence, aligning with previous studies \cite{li2022onset, ramos2022investigationII}. Furthermore, the instantaneous wake structures between fixed and surging cases exhibited limited observable differences with the presence of inflow turbulence, although striking differences were observed when subjected to laminar inflow conditions where Surge-Induced Periodic Coherent Structures (SIPeCS) were found. However, after phase-averaging, SIPeCS became apparent in turbulent cases as well, indicating the effects of surging motions  did exist. Qualitative and quantitative analyses of SIPeCS were conducted to assess their dependency on TI, $A_S$, and $\omega_S$, with consistent agreement between the two approaches. Furthermore, with the phase-averaged quantities, we suggested that the enhanced wake recovery due to surging was mainly due to the advection process rather than the enhanced turbulence level in the wake, where the latter is a previously accepted hypothesis by the FOWT community \cite{micallef2021floating, chen2022modelling, kleine2022stability, ramos2022investigationI}. More evidence was provided based on conducting a term-by-term energy entrainment analysis based on the phase-averaged velocity. The analysis showed that the enhancement in wake recovery (energy entrainment) for the surging cases compared to the fixed cases is due to the increased advection process rather than the increased turbulent mixing for cases with both laminar and turbulent inflows, solidifying the aforementioned claim.

In conclusion, with rich and informative results, our study provides valuable insights into the complex aerodynamics of a surging FOWT rotor. Building upon the foundations established by this research, future investigations can further expand the analysis by including different types of degrees of freedom (DoF) for platform motion, such as pitching and yawing. Complex motions with multiple DoFs combined are also interested in discovering if there are any interplay effects between different DoFs. Additionally, exploring the systems of wake interaction of multiple FOWTs in motions when subjected to realistic turbulent inflow conditions is of great interest, as the effects of motions are observed in the wake of FOWT. Understanding and quantifying the impacts of SIPeCS or other motion-induced coherent wake structures on the downstream FOWTs is critical to the design and operation of floating offshore wind farms, as these effects will likely affect the power outputs and fatigue lifetimes of the downstream rotors.

\section*{Data availability statement}
Data supporting the findings of this study are available upon request.

\section*{Acknowledgement}
We would like to thank Carlos Sim\~{a}o Ferreira for participating in formulating the research questions.

\bibliography{main}%

\appendix

\section{Convergence Tests}

\label{app:convergenceStudy}

\begin{figure}[htb!]
\centering
\includegraphics[width=430pt]{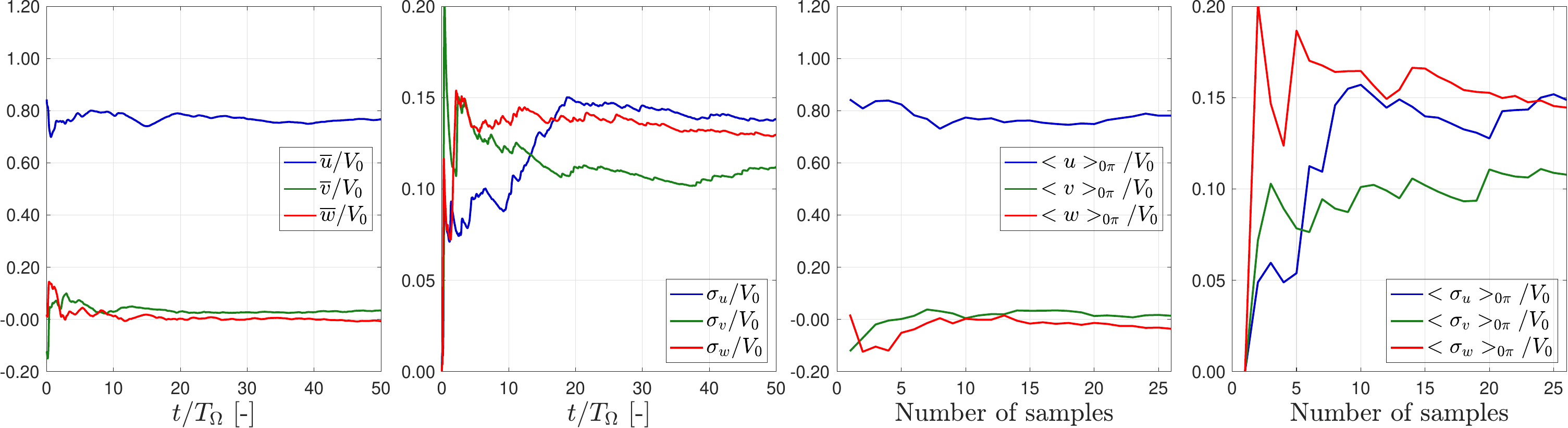}
\caption{Convergence tests of the statistics sampled at $x/D = 3$, $z/D = 0.5$ of case~\textbf{7}. $T_{\Omega}$ is the rotational period for a NREL~5MW rotor operating under its rated conditions.}
\label{fig:convergenceStudyPaper_ConvergencePaper_3}
\end{figure}

Figure~\ref{fig:convergenceStudyPaper_ConvergencePaper_3} displays the results of brief convergence tests for $\overline{u}$, $\sigma_u$, $<u>_{0\pi}$, and $<\sigma_u>_{0\pi}$ based on case~\textbf{7} in Table~\ref{tab:cases_Single_rotor}. The probing point is located at $z/D = 0.5$, and the streamwise position is at $x/D = 3$. The windows for calculating the statistics are described in Section~\ref{sec:Statistics}. See Li \cite{Li2023Numerical} for more information.

\section{Grid independence test}

\label{app:grid_test}

\begin{figure}[t]
\centering
\includegraphics[width=300pt]{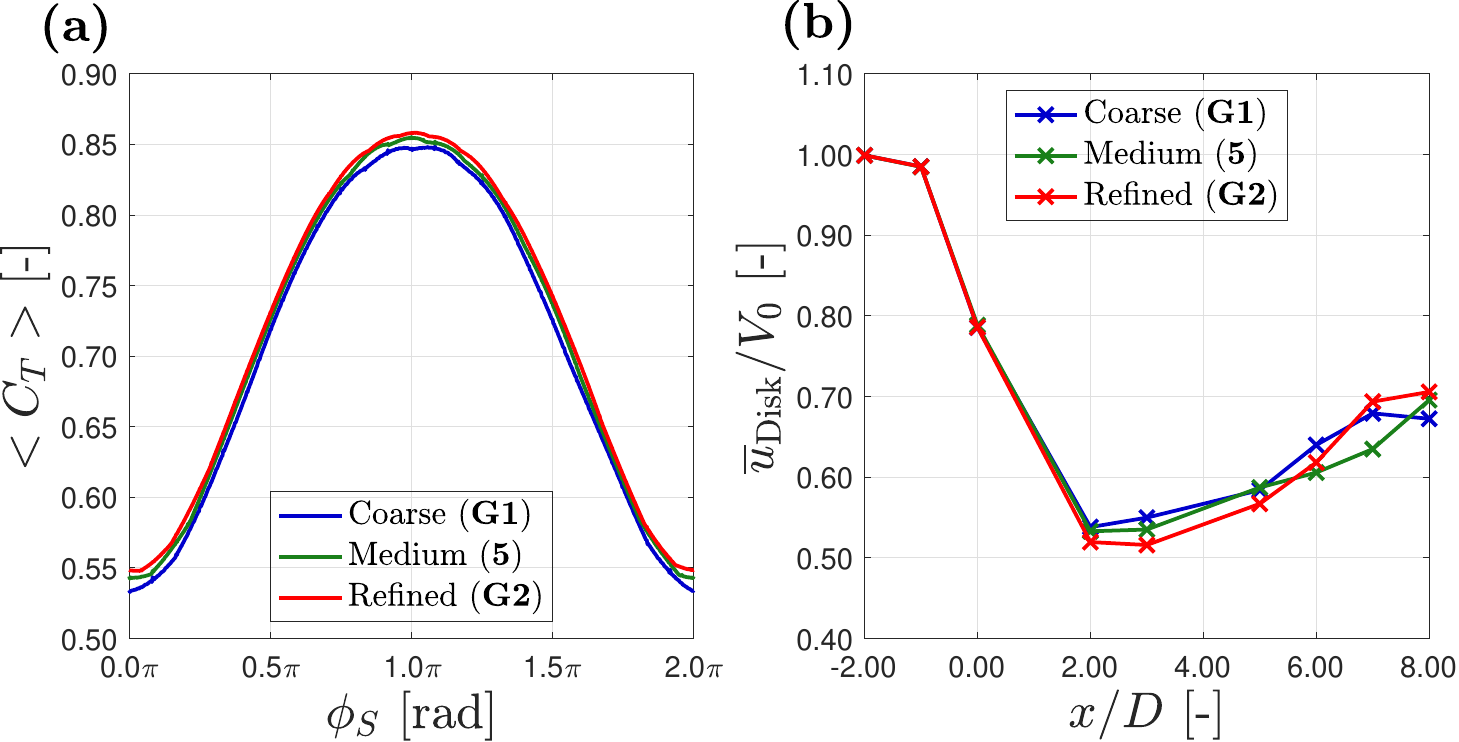}
\caption{Grid sensitivity tests about the cycle-averaged rotor performances ($<C_T>$) and integral wake characteristics ($\overline{u}_{\mathrm{Disk}}$). Numbers in parentheses are the case numbers in Tables~\ref{tab:cases_Single_rotor} and \ref{tab:grid_independent}.}
\label{fig:GridTest}
\end{figure}

A brief grid independence test is performed. It is performed with three resolutions and tested with a surging rotor ($v_{\mathrm{max}} = 0.22$ and $k_D = 7.0$) subjected to laminar inflow conditions. The resolutions for \emph{Level} 4 (see Figure~\ref{fig:paper_mesh}) of the three cases are $\Delta/D = 1.25/80$ (\emph{coarse}, $5.6$M cells, case \textbf{G1} in Table~\ref{tab:grid_independent}), $\Delta/D = 1.0/80$ (\emph{medium}, $10.4$M cells, case \textbf{5} in Table~\ref{tab:cases_Single_rotor}), and $\Delta/D = 0.8/80$ (\emph{fine}, $18.8$M cells, case \textbf{G2} in Table~\ref{tab:grid_independent}). Note that the absolute value of the smoothing factor ($\varepsilon$ in Equation~\ref{eq:ALM}) is kept as $D/40$. Their results are presented in Table~\ref{tab:grid_independent} and Figure~\ref{fig:GridTest}, which show that the results are not sensitive to grid resolution. Though the results of \emph{coarse} are already quite similar to the other configurations, \emph{medium} is chosen as the simulation setup. The choice is based on considering the best practice established by the proceeding works \cite{sarlak2015role, martinez2015large} when using LES with ALM to model the wake of a wind turbine ($\Delta_r < D/70$ and $\Delta_r \simeq \Delta$), and $11$M cells is reasonable with the given computational resources.

\begin{table}[htb!]
\centering
\caption{The basic settings and results for the cases with different grid resolutions of a single surging NREL~5MW rotor.}
\label{tab:grid_independent}
\begin{tabular}{l|ccc|cc}

\textbf{Case} & Number of Cells & $v_{\mathrm{max}}$ & $k_D$ & $\overline{C}_T$ & $\overline{C}_P$ \\
\hline

\textbf{G1} & $5.6$M & $0.22$ & $7.0$ & $0.708$ & $0.512$ \\

\textbf{5} & $10.4$M & $0.22$ & $7.0$ & $0.715$ & $0.523$ \\

\textbf{G2} & $18.8$M & $0.22$ & $7.0$ & $0.720$ & $0.531$ \\

\hline

\end{tabular}
\end{table}

\section{Auxiliary cases with different \texorpdfstring{$V_0$}{text}}

\label{app:auxiliary}

Table~\ref{tab:aux_cases_Single_rotor} lists the basic settings and results for the six auxiliary cases. These auxiliary cases are tested with a fixed configuration and subjected to different inflow velocities $V_0$ under laminar inflow conditions. The values of these $V_0$ are set to meet the maximum and minimum $V_{0, \mathrm{app}}$ for each surging setting appeared in Table~\ref{tab:cases_Single_rotor}. Results of these cases are utilized in Figure~\ref{fig:paper_1_ampFre_concyc_CT}(c) to assess the hysteresis effects (dynamic inflow and unsteady airfoil aerodynamics) due to the surging motions of a FOWT rotor.

\begin{table}[t]
\centering
\caption{The basic settings and results for the auxiliary cases with different $V_0$ but identical rotational frequency ($\Omega = 1.26$~rad/s) of a single fixed NREL~5MW rotor. Note that reference velocity for calculating $\overline{C}_T$ and $\overline{C}_P$ (Equation~\ref{eq:CT_and_CP}) is set to $V_{0, \mathrm{rated}} = 11.4$~m/s.}
\label{tab:aux_cases_Single_rotor}
\begin{tabular}{l|ccc|cc}

\textbf{Case} & TI~[$\%$] & $V_0$~[m/s] & $V_0/V_{0, \mathrm{rated}}$ & $\overline{C}_T$ & $\overline{C}_P$ \\
\hline

\textbf{A1} & Laminar & $6.33$ & $0.56$ & $0.330$ & $0.076$ \\

\textbf{A2} & Laminar & $8.87$ & $0.78$ & $0.538$ & $0.264$ \\

\textbf{A3} & Laminar & $10.13$ & $0.90$ & $0.640$ & $0.388$ \\

\textbf{A4} & Laminar & $12.67$ & $1.11$ & $0.801$ & $0.651$ \\

\textbf{A5} & Laminar & $13.93$ & $1.22$ & $0.859$ & $0.783$ \\

\textbf{A6} & Laminar & $16.47$ & $1.44$ & $0.856$ & $0.846$ \\

\hline

\end{tabular}
\end{table}

\section{Profile of \texorpdfstring{$\alpha_{\mathrm{\MakeLowercase{stall}}}$}{text} for NREL 5MW}

\label{app:alpha_stall}

\begin{figure}[t]
\centering
\includegraphics[width=200pt]{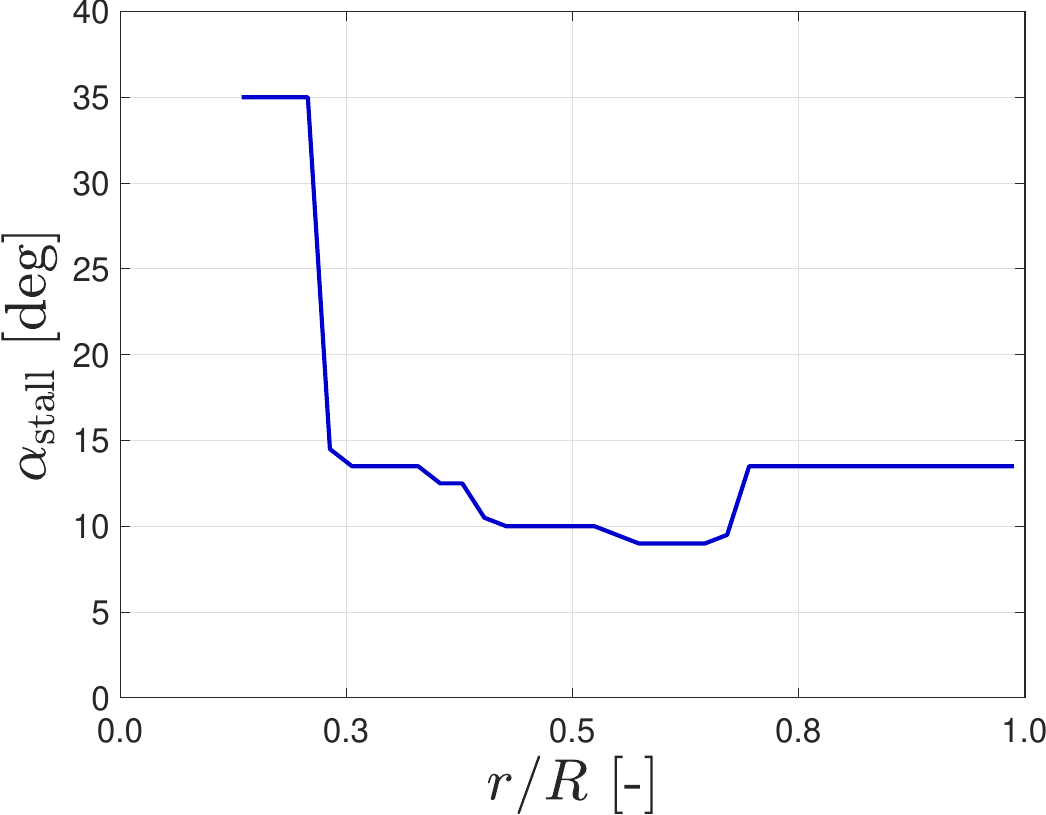}
\caption{Profile of $\alpha_{\mathrm{stall}}$ for NREL 5MW baseline turbine based on the airfoil data given in the documentation \cite{jonkman2009definition}.}
\label{fig:AOA_stallingAngle}
\end{figure}

Figure~\ref{fig:AOA_stallingAngle} presents the $\alpha_{\mathrm{stall}}$ profile along the blade span of NREL~5MW baseline turbine. Note that the airfoil sections are not the same throughout the blade, and thus $\alpha_{\mathrm{stall}}$ varies along the blade.

\section{Phase-averaged spanwise velocity fields \texorpdfstring{ $<\MakeLowercase{w}>_{0\pi}$}{text}}

The phase-averaged spanwise velocity fields at $\phi_S = 0.0\pi$ (and $\phi_{\Omega} = 0.0\pi$), denoted as $<w>_{0\pi}$, for the cases in \textbf{Group}~$\boldsymbol{c}$ are presented in Figure~\ref{fig:paper_1_v2_singleAll_0000PhaseAvgU}. The primary purpose of the figure is to confirm that the induction fields of the SIPeCS exchange the flow inside and outside of the wake for the surging cases, whereas the exchange processes are not detected in the cases with a fixed rotor.

\begin{figure}[t]
\centering
\includegraphics[width=430pt]{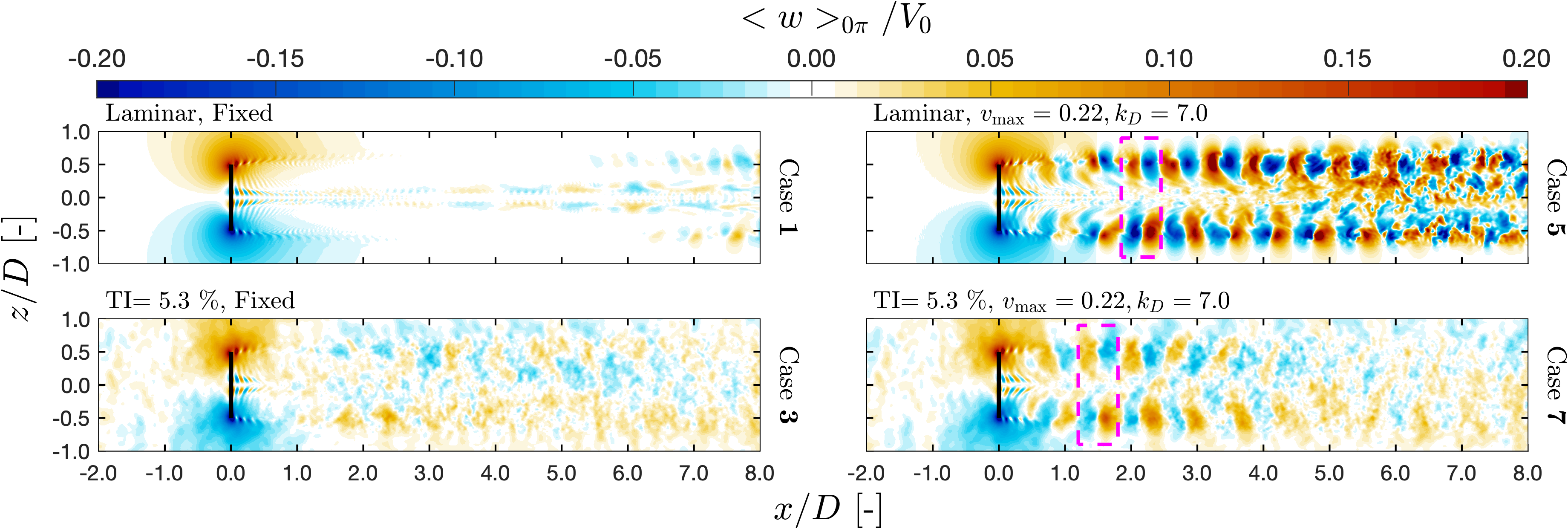}
\caption{Contour plots of phase-averaged spanwise velocity fields $<w>_{0 \pi}$ for the cases in \textbf{Group}~$\boldsymbol{c}$. Locations of the magenta dashed-line boxes in the plots are identical to those in Figure~\ref{fig:paper_1_v2_singleAll_0000PhaseAvgU}, which correspond to the positions of the example sets of SIPeCS. The inflow conditions (TI) and surging settings ($v_{\mathrm{max}}$ and $k_D$) are labeled at the top left of each panel, while the case numbers are labeled on the right.}
\label{fig:paper_1_v2_singleAll_0000PhaseAvgW}
\end{figure}

\section{Quantification of SIPeCS of the laminar cases with different surging settings}

\label{app:circ_lam}

\begin{figure}[t!]
\centering
\includegraphics[width=430pt]{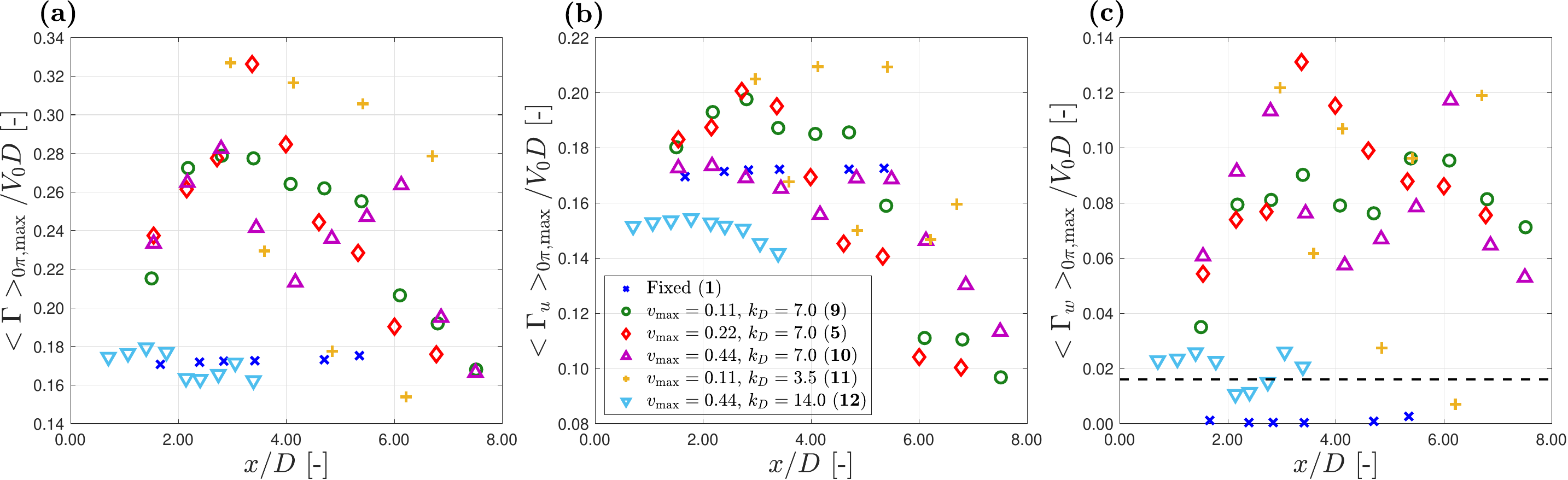}
\caption{Plotting the $x$-positions where ${<\Gamma>_{0\pi, \mathrm{max}}}$ are found against the values of ${<\Gamma>_{0\pi, \mathrm{max}}}$, ${<\Gamma_u>_{0\pi, \mathrm{max}}}$, and ${<\Gamma_w>_{0\pi, \mathrm{max}}}$ for the laminar cases in \textbf{Group}~$\boldsymbol{b}$. ${<\Gamma_u>_{0\pi, \mathrm{max}}}$ and ${<\Gamma_w>_{0\pi, \mathrm{max}}}$ are calculated based on the same paths to obtain ${<\Gamma>_{0\pi, \mathrm{max}}}$, but only considering $<u>_{0 \pi}$ or $<w>_{0 \pi}$. (a): ${<\Gamma>_{0\pi, \mathrm{max}}}$. (b): ${<\Gamma_u>_{0\pi, \mathrm{max}}}$.
(c): ${<\Gamma_w>_{0\pi, \mathrm{max}}}$. The dashed line in (c) indicates the threshold for specifying SIPeCS mentioned in Section~\ref{sec:circ}, and the case numbers are labeled in parentheses.}
\label{fig:paper_1_v2_ampFreqLamMaxCircValue_All}
\end{figure}

The plots of ${<\Gamma>_{0\pi, \mathrm{max}}}$, ${<\Gamma_u>_{0\pi, \mathrm{max}}}$, and ${<\Gamma_w>_{0\pi, \mathrm{max}}}$ against the $x$-positions where ${<\Gamma>_{0\pi, \mathrm{max}}}$ are found for the six laminar cases in \textbf{Group}~$\boldsymbol{b}$ are plotted in Figure~\ref{fig:paper_1_v2_ampFreqLamMaxCircValue_All}. Unlike the turbulent cases in \textbf{Group}~$\boldsymbol{b}$, trends with respect to both $v_{\mathrm{max}}$ and $k_D$ cannot be detected easily. However, despite the irregularity, it can be clearly seen that the strengths of SIPeCS based on ${<\Gamma_w>_{0\pi, \mathrm{max}}}$ for the surging cases are generally much higher than the threshold, except for cases \textbf{12}. Additionally, ${<\Gamma_u>_{0\pi, \mathrm{max}}}$ for the surging cases appear to be lower with larger $x$-positions, showing that the strengths of the shear layer are reduced, unlike the fixed case (case~\textbf{1}). Furthermore, the very similar values of ${<\Gamma>_{0\pi, \mathrm{max}}}$ for the three cases with the same $k_D$ (cases \textbf{5}, \textbf{9}, and \textbf{10}) supports the finding that for the laminar cases, the large and strong vortical structures representing SIPeCS are formed by merging the tip vortices released within a complete surging cycle (except for case \textbf{11}). Note ${<\Gamma>_{0\pi}}$ is the area-sum of $<\omega_y>_{0\pi}$.

\section{Entrainment of flow kinetic energy with time-averaged velocity data}

\label{app:time_energy}

This section briefly covers the energy entrainment process based on the time-averaged data. The analysis procedure is the same as the one used in Section~\ref{sec:phase_energy}. Equation~\ref{eq:energyTimeIn} writes out the decomposition of the fluxes of time-averaged flow kinetic energy, where Reynolds decomposition ($u = \overline{u} + u'$) and Reynolds averaging (\hspace{0.5pt}$\overline{ \overline{u} } = \overline{u}$ and $\overline{u'} = 0$) are utilized. The cumulative fluxes into the rotor projection region ($z/D = \pm 0.5$) starting from $x/D = 1.0$ are calculated based on Equation~\ref{eq:intEnergy}.

\begin{multline}
    \mathrm{Flux~of~time-}\mathrm{averaged~kinetic~energy~for}~\hat{n} = \pm \hat{z}\mathrm{:} \qquad \overline{u_j \left( \frac{u_i u_i}{2} \right) } \hat{n}_j 
    \hspace{3pt} = \hspace{3pt}
    \\
    \mp \left[
    \underbrace{\overline{w} \left( \frac{ \overline{u}^2 + \overline{v}^2 + \overline{w}^2 }{2} \right)}_{\mathrm{Advection~of~MKE}}
    \hspace{3pt} + \hspace{3pt}
    \underbrace{\overline{w} \left( \frac{ \overline{ \left( u' \right)^2} + \overline{ \left( v' \right)^2} + \overline{ \left( w' \right)^2} }{2} \right)}_{\mathrm{Advection~of~TKE}} \right.
    \\ \left.
    \underbrace{
    \hspace{3pt} + \hspace{3pt}
    \left(
    \overline{u \vphantom{w' w'} } \hspace{2pt} \overline{u' w' \vphantom{w' w'}}
    \hspace{3pt} + \hspace{3pt}
    \overline{v \vphantom{w' w'} } \hspace{2pt} \overline{v' w' \vphantom{w' w'}}
    \hspace{3pt} + \hspace{3pt}
    \overline{w \vphantom{w' w'} } \hspace{2pt} \overline{w' w' \vphantom{w' w'}}
    \right)
    }_{2^{\mathrm{nd}}~\mathrm{order~RST}}
    \hspace{3pt} + \hspace{3pt}
    \underbrace{
    \frac{1}{2} \left(
    \overline{w' u' u'}
    \hspace{3pt} + \hspace{3pt}
    \overline{w' v' v'}
    \hspace{3pt} + \hspace{3pt}
    \overline{w' w' w'}
    \right)
    }_{3^{\mathrm{rd}}~\mathrm{order~RST}}
    \right]
    \label{eq:energyTimeIn}
\end{multline}

\begin{figure}[t]
\centering
\includegraphics[width=430pt]{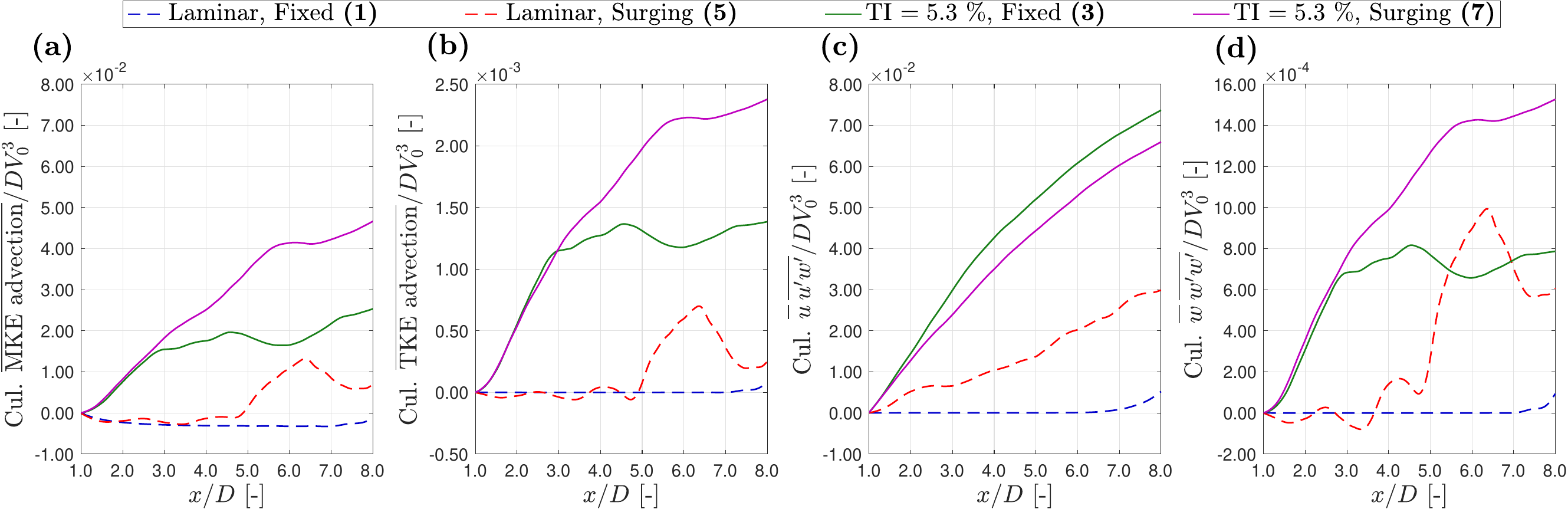}
\caption{Energy entrained into the rotor projection region ($-0.5 \leq z/D \leq 0.5$) on $y/D = 0$ plane with different terms using time-averaged velocity data. ``Cul.'' is the acronym for ``cumulative''. The plots are similar with Figure~\ref{fig:paper_1_v2_pahseLockedEnergyEntrain_cumulative}.}
\label{fig:paper_1_v2_timeAvgEnergyEntrain_cumulative}
\end{figure}

The cumulative values for the cases of \textbf{Group}~$\boldsymbol{c}$ of the same four terms considered in Section~\ref{sec:phase_energy} are plotted in Figure~\ref{fig:paper_1_v2_timeAvgEnergyEntrain_cumulative}. The plots show that the fixed-laminar case entrains almost no energy. While for the surging laminar case, it entrains some energy through term $\overline{u \vphantom{u' w'}} \hspace{2pt} \overline{u' w'}$ but not much from $\overline{\mathrm{MKE}~\mathrm{advection}}$, which is in contrast to the analysis in Section~\ref{sec:phase_energy} with phase-averaged data. For the two turbulent cases, the results are similar to those presented in Section~\ref{sec:phase_energy}, which the surging-turbulent case entrains more energy through $\overline{\mathrm{MKE~advection}}$ but less through the $2^{\mathrm{nd}}$ order RST term ($\overline{u \vphantom{u' w'}} \hspace{2pt} \overline{u' w'}$), while the total entrained energy is more for the surging-turbulent case. Additionally, the wavy patterns found in Figure~\ref{fig:paper_1_v2_pahseLockedEnergyEntrain_cumulative} are lost in Figure~\ref{fig:paper_1_v2_timeAvgEnergyEntrain_cumulative}, showcasing that the periodic effects detected by the phase-averaging method cannot be captured by time-averaged data.

\end{document}